\documentclass[reprint,
 amsmath,amssymb,
 aps,pra, 
 showkeys
]{revtex4-2}

\usepackage{graphicx}
\usepackage{dcolumn}
\usepackage{bm}
\usepackage[version=4]{mhchem}

\usepackage[normalem]{ulem}
\usepackage{color}
\usepackage{cancel}
\definecolor{my_red}{rgb}{.7,0,0}
\definecolor{my_green}{rgb}{0,.7,0}
\definecolor{my_blue}{rgb}{0,0,.7}
\definecolor{my_purple}{rgb}{.7,0,.7}

\begin{document}

\title{Hamiltonian formulation and symplectic split-operator schemes for time-dependent density-functional-theory equations of electron dynamics in molecules}

\author{Fran\c{c}ois Mauger$^{1}$, 
Cristel Chandre$^{2}$,
Mette B.\ Gaarde$^{1}$,
Kenneth Lopata$^{3,4}$, and
Kenneth J.\ Schafer$^{1}$}
\affiliation{%
$^{1}$Department of Physics and Astronomy, Louisiana State University,
    Baton Rouge, Louisiana 70803, USA \\ 
$^{2}$CNRS, Aix Marseille Univ, I2M, Marseille, France \\
$^{3}$Department of Chemistry, Louisiana State University,
    Baton Rouge, Louisiana 70803, USA \\
$^{4}$Center for Computation and Technology, Louisiana State University, Baton Rouge, Louisiana 70803, USA 
}%

\date{\today}

\begin{abstract}
We revisit Kohn-Sham time-dependent density-functional theory (TDDFT) equations and show that they derive from a canonical Hamiltonian formalism. 
We use this geometric description of the TDDFT dynamics to define families of symplectic split-operator schemes that accurately and efficiently simulate the time propagation for certain classes of DFT functionals.
We illustrate these with numerical simulations of the far-from-equilibrium electronic dynamics of a one-dimensional carbon chain. In these examples, we find that an optimized 4$^{\rm th}$ order scheme provides a good compromise between the numerical complexity of each time step and the accuracy of the scheme.
We also discuss how the Hamiltonian structure changes when using a basis set to discretize TDDFT and the challenges this raises for using symplectic split-operator propagation schemes.

\ 

\end{abstract}


\maketitle

\section{Introduction} \label{sec:Introduction}

Since its introduction in the 1980's, time-dependent (TD) density-functional theory (DFT) has been embraced by many science and engineering communities for its ability to systematically and reliably describe the quantum dynamics of systems with a few to many active electrons, up to several thousands~\cite{Marques_2006_book, Marques_2012_book, Maitra_2016, Arpa_2020, Tancogne-Dejean_2020, Smith_2022}.
These systems can range from atoms, to molecules, macro-/bio-molecules and clusters, all the way to condensed matter, both in ordered (solid) and disordered (liquid) phases.
TDDFT is effectively a nonlinear problem and the time propagation of its dynamics is not a trivial task. Traditionally, it is performed using the time-ordered exponential of the DFT Hamiltonian operator~\cite{Castro_2004, Gomez_Pueyo_2018}.
This is computationally challenging because
(i)~computing the exponential of an operator is a known difficult problem without a commonly accepted procedure~\cite{Moler_2003}.
(ii)~Time ordering implies the use of intricate Magnus expansions to obtain schemes beyond second order~\cite{Castro_2004, Gomez_Pueyo_2018, Gomez_Pueyo_2020}.
(iii)~High-order schemes often require storing the solution of intermediate computation steps, potentially making the simulation memory-hungry for systems with many electrons.

The main objective of this article is to use the Hamiltonian formulation of TDDFT to discuss and implement symplectic numerical schemes.
In a generalized formulation of Kohn-Sham TDDFT~\cite{Kohn_1965,Runge_1984}, we carefully detail the canonical Hamiltonian structure by augmenting conventional DFT energy functionals with a Poisson bracket.
Previously, a Hamiltonian structure of TDDFT was revealed  by recasting the dynamics into a set of ordinary differential equations that follows Hamilton's equations~\cite{Gomez_Pueyo_2018}. Instead, we retain the complex-valued partial-differential equation of Kohn-Sham TDDFT in the Hamiltonian formulation and explicit the algebraic structure for its associated Poisson bracket.
As such, our approach is readily compatible with commonly-used TDDFT models and simulations, with only minor and straightforward adjustments required to adapt them to the Hamiltonian formalism.
For grid-based numerical simulations, we show how our Hamiltonian formulation naturally lends itself into using efficient high-order symplectic split-operator propagation schemes.

Specifically, we illustrate possible uses of our TDDFT Hamiltonian formalism along two lines: 
First, we discuss the Hamiltonian dynamics and phase-space structure analyses that are made possible by our approach. 
Second, for some types of DFT functionals, we leverage the geometric description of the TDDFT dynamics to implement symplectic split-operator schemes that have been developed and used in the nonlinear-dynamics community for other applications~\cite{Yoshida_1990, Bandrauk_1992, Blanes_2002, McLachlan_2022, Xiao_2015, He_2015, Xiao_2016, Kraus_2017}.
Notably, the symplectic schemes we propose (i)~are unitary and exactly time reversible, (ii)~easily go beyond second order without extrapolation, and (iii)~do not require saving of intermediate propagation steps, which makes them lean in memory.
Finally, we briefly discuss how the Hamiltonian structure of TDDFT changes when it is described with a time-independent basis, as is commonly done in basis-set quantum chemistry packages~\cite{Arpa_2020, Smith_2022},
and the challenges this raises for using symplectic split-operator propagation schemes.

The paper is organized as follows:
Section~\ref{sec:Model} introduces the TDDFT model and its relationship to the Schr\"{o}dinger-equation description of multi-active-electron molecular systems.
Section~\ref{sec:Hamiltonian_formulation} revisits the TDDFT formalism and shows that it derives from a canonical Hamiltonian framework. In this framework, we show that the kinetic (subsection~\ref{sec:kinetic_functional}), external (\ref{sec:external_functional}), Hartree (\ref{sec:Hartree-energy_functional}), and exchange-correlation (\ref{sec:exchange-correlation-energy_functional}) Hamiltonian functionals all correspond to their standard definition in ground-state DFT. We also discuss how an external driving field can be included in this Hamiltonian formalism (subsection~\ref{sec:External_field_functional}).
Section~\ref{sec:Phase_space} discusses some implications of the phase-space structure associated with the TDDFT Hamiltonian formalism, including continuity equations between the one-body and current densities (subsection~\ref{sec:continuity_equations}) and the derivation of Ehrenfest's theorem for the dipole signal within the Hamiltonian formalism (\ref{sec:Ehrenfest_theorem}).
Section~\ref{sec:Symplectic_schemes} leverages the Hamiltonian formulation of TDDFT to define high-order symplectic schemes that can efficiently propagate the dynamics for certain classes of DFT functionals.
Section~\ref{sec:Basis-set_TDDFT} briefly discusses how the Hamiltonian formulation of TDDFT should be adapted when the system is described with a time-independent spatial basis, as is commonly done in quantum-chemistry packages, and the challenges this raises for implementing symplectic split-operator schemes.
Finally, section~\ref{sec:Conclusion} concludes the paper and discusses the outlook of including a semi-classical description of the nuclear degrees of freedom in so-called Ehrenfest TDDFT.
Throughout the paper, we also discuss computational considerations we use in our numerical simulations; these are marked with ``{\bf Numerical simulations:}'' paragraph headers.
Unless otherwise specified we use atomic units (a.u.).

\section{TDDFT Model} \label{sec:Model}

The quantum dynamics of a system of $N$ interacting electrons is described by the time-dependent Schr\"{o}dinger equation (TDSE)
\begin{equation} \label{eq:TDSE}
    i \partial_t \psi = \hat{\mathcal{H}}_{\rm SE}\psi,
\end{equation}
where $\psi=\psi({\bf x}_1,\ldots,{\bf x}_N;t)$ and the Hamiltonian operator
\begin{equation} \label{eq:TDSE_Hamiltonian_operator}
    \hat{\mathcal{H}}_{\rm SE} =
        \sum_{k=1}^N{-\frac{\Delta_{{\bf x}_k}}{2}} +
        \sum_{k=1}^N{\hat{\mathcal{V}}_{\rm ext}({\bf x}_k)} +
        \sum_{k<l}^N{\hat{\mathcal{V}}_{\rm ee}({\bf x}_k-{\bf x}_l)},
\end{equation}
with each ${\bf x}_k\in\mathbb{R}^d$ in dimension $d$.
The first two terms in the equation respectively correspond to the kinetic and external-potential operators. Together they form the so-called \emph{one-body} interactions.
The last term corresponds to \emph{two-body} interactions, with $\hat{\mathcal{V}}_{\rm ee}$ the electron-electron interaction potential operator. These two-body interactions are the source of the complexity of the TDSE, where the $N$-electron wave function $\psi$ cannot be factorized as an antisymmetrized product of one-electron wave functions, {\it i.e.}, a single-determinant solution.

The non-separability of the TDSE means that the complexity of simulations scales exponentially with the number of electrons. As a result, direct numerical simulations have often been limited to systems with a handful of active electrons, sometimes treated in reduced dimension.
To overcome this limitation, various classes of methods have been proposed.
Staying in the wave-function formalism of the TDSE, one can approximate $\psi$ by a linear combination of antisymmetrized products of one-electron functions. These approaches include Hartree Fock~\cite{Szabo_1996} as well as post-Hartree methods like configuration interaction~\cite{Szabo_1996}, with complete- or restricted-active spaces~\cite{Ishikawa_2015}.
An alternative broad class of simulation methods relinquish finding the $N\times d$-dimensional wave function $\psi$ altogether and, instead, focus on solving for the $d$-dimensional electron density $\rho({\bf x};t)$. Indeed, the Runge-Gross theorem~\cite{Runge_1984, van_Leeuwen_1999} establishes the equivalence between $\psi$ and $\rho$ and is at the heart of time-dependent density functional theory (TDDFT)~\cite{Marques_2004, Marques_2006_book, Marques_2012_book, Ullrich_2014, Maitra_2016}.
In this paper, we focus on the latter in the generalized formulation of \emph{Kohn-Sham} (KS) TDDFT~\cite{Kohn_1965,Runge_1984}.

In the KS formulation of TDDFT, the two-body interactions are recast into a nonlinear effective Hamiltonian operator $\hat{\mathcal{H}}_\text{eff}$ and the time evolution of electron dynamics is instead described by a system of fictitious one-electron KS orbitals $\left\{\phi_k\right\}_k$ as
\begin{equation} \label{eq:TDDFT}
    i\partial_{t} \phi_k \left({\bf x};t\right) = \hat{\mathcal{H}}_\text{eff}\left[\left\{\phi_k\right\}_k\right] \phi_k\left({\bf x};t\right),
\end{equation}
where ${\bf x}\in {\mathbb R}^d$.
Note that, here, the index $k$ labels both the spin and orbital indexes.
Indeed, $\hat{\mathcal{H}}_\text{eff}$ may differ for up- and down-spin KS orbitals ($\hat{\mathcal{H}}_\text{eff}^\uparrow\neq\hat{\mathcal{H}}_\text{eff}^\downarrow$) but is the same for all the orbitals within the same spin channel.
To simplify notation throughout we omit the $^\uparrow$ and $^\downarrow$ labels when referring to indexes and components that cover both spin channels like in the equation above, unless otherwise specified.
For \emph{spin-restricted} systems the up- and down-spin channels are identical, leading to simplified TDDFT equations. We briefly review the spin-restricted equations in Appendix~\ref{app:Spin_restricted}.

Importantly, in the TDDFT formalism the KS orbitals are only used as dynamical variables but do not correspond to physically observable quantities. Instead, the cornerstone observable is the one-body density
\begin{equation} \label{eq:one-body_density}
    \rho = \rho^\uparrow+\rho^\downarrow \quad \text{with} \quad
    \rho^\uparrow\left({\bf x};t\right) = \sum_{k\geq1}{n_k^\uparrow\left|\phi_k^\uparrow\left({\bf x};t\right)\right|^2},
\end{equation}
and likewise for the down-spin density $\rho^\downarrow$. 
The population coefficients $\left\{n_k\right\}_k$ parametrically define the number of electrons in the KS orbitals, with Pauli's exclusion principle imposing $0\leq n_k\leq 1$.
The one-body density is related to the $N$-electron wave function as
$$
    \rho({\bf x};t) = N \int{\left|
        \psi({\bf x}_1,{\bf x}_2,\cdots,{\bf x}_{N-1},{\bf x};t)
    \right|^2 {\rm d}{\bf x}_1 \cdots {\rm d}{\bf x}_{N-1}},
$$
{\it i.e.}, integrating over all the TDSE electronic coordinates except one.
Formally, TDDFT theorems~\cite{Runge_1984, van_Leeuwen_1999} prove that all physical observables are functions or functionals of the one-body density. 
Note that, in theory, this functional dependence involves the full history of the one-body density $\{\rho({\bf x};s)\}_{s\leq t}$ and the initial condition~\cite{Vignale_1997, Maitra_2002, Wijewardane_2005, Marques_2006_book, Lacombe_2020}, which is impractical and hardly ever done in practice. Instead, we assume an instantaneous approximation where the  operator $\hat{\mathcal{H}}_\text{eff}$ only depends on the KS orbitals and density at the same time $t$ it is evaluated in Eq.~\eqref{eq:TDDFT}.

All physical TDDFT Hamiltonians $\hat{\mathcal{H}}_\text{eff}$ in Eq.~\eqref{eq:TDDFT} are Hermitian operators and the KS orbitals form a time-dependent family of orthonormal functions
\begin{equation} \label{eq:KS_are_orthonormal}
    \int{\phi_k^\uparrow\left({\bf x};t\right)^* \phi_l^\uparrow\left({\bf x};t\right) {\rm d}{\bf x}} = \delta_{k,l} \qquad \forall k,l \textrm{ and } \forall t,
\end{equation}
and likewise for the down-spin orbitals.
Following the template of the TDSE of Eq.~\eqref{eq:TDSE_Hamiltonian_operator}, the TDDFT Hamiltonian operator is usually decomposed into its kinetic, external, Hartree, and exchange-correlation components as
\begin{equation} \label{eq:TDDFT_Hamiltonian_decomposition}
    \hat{\mathcal{H}}_\text{eff} = 
        -\frac{\Delta}{2} + \hat{\mathcal{V}}_{\rm ext} +
        \hat{\mathcal{V}}_{H} + 
        \hat{\mathcal{V}}_{\rm XC}.
\end{equation}
The external potential operator $\hat{\mathcal{V}}_{\rm ext}$ may include an explicit time dependence, {\it e.g.}, modeling the interaction with an external driving electric field. In this paper we limit ourselves to dipole-approximation interactions, either in the so-called \emph{length} or \emph{velocity} gauges~\cite{Bandrauk_2013}.
Next, the Hartree potential operator $\hat{\mathcal{V}}_{\rm H}$ corresponds to the classical electrostatic interaction of each KS orbital with the mean-field density generated by all the electrons in the model. 
Finally, the exchange-correlation potential operator $\hat{\mathcal{V}}_{\rm XC}$ collects all the non-classical electron-electron interactions within (exchange) and between (correlation) the up- and down-spin channels. $\hat{\mathcal{V}}_{\rm XC}$ is the only component of $\hat{\mathcal{H}}_\text{eff}$ that may differ for the up- and down-spin channels.
Historically, the exchange-correlation potential was shown to be strictly a functional of the one-body density. It has now been generalized to possibly include an explicit dependence on the KS orbitals~\cite{Seidl_1996}. For instance, this is commonly done when using hybrid DFT functionals (see the discussion at the end of section~\ref{sec:other_exchange-correlation_functionals}).
Interestingly, it was shown that such an explicit dependence on the instantaneous KS orbitals include non-local-in-time effects that are not captured by adiabatic exchange-correlation functionals that solely depend on the one-body density~\cite{Ullrich_1995,Wijewardane_2008}.

\section{Hamiltonian formulation of TDDFT} \label{sec:Hamiltonian_formulation}

In this section, we show that the TDDFT Eq.~\eqref{eq:TDDFT} derive from an infinite-dimensional canonical Hamiltonian formalism, where the Hamiltonian functional matches the energy functional used in conventional ground-state DFT computations~\cite{Gomez_Pueyo_2018}. 
In section~\ref{sec:Basis-set_TDDFT} we will discuss how this Hamiltonian formalism should be adapted when the TDDFT model is represented in a time-independent spatial basis of orbitals.
To simplify notations, in what follows we omit the implicit time dependence of the KS orbitals  $\phi_k({\bf x})\equiv\phi_k({\bf x};t)$ and all other dynamical variables. 

In the Hamiltonian formulation of TDDFT we first define the algebra of functionals $\mathcal{A}$ as the scalar functions of the field variables $\phi_k({\bf x})$ and $\phi_k^*({\bf x})$.
The Hamiltonian structure acts in this space of functionals using a \emph{Poisson bracket} $\{\cdot , \cdot\}$, which is an antisymmetric and bilinear operator in $\mathcal{A}\times\mathcal{A}$
\begin{eqnarray*}
    &\{F_1+\alpha F_2,F_3\}= \{F_1, F_3\} + \alpha \{F_2,F_3\},&\\
    &\{F_1,F_2\} = -\{F_2, F_1\},&
\end{eqnarray*}
and satisfies Leibniz rule and Jacobi identity
\begin{eqnarray*}
    &\{F_1 F_2, F_3\} = F_1\{F_2,F_3\}+ \{F_1, F_3\} F_2, &\\
    &\{F_1,\{F_2,F_3\}\}+\{F_3,\{F_1,F_2\}\}+\{F_2,\{F_3,F_1\}\} = 0,&
\end{eqnarray*}
for all functionals $F_1$, $F_2$, and $F_3$ in $\mathcal{A}$ and all scalar $\alpha$. The dynamics of any functional $F\in\mathcal{A}$ is given by 
\begin{equation} \label{eq:Hamiltonian_flow}
    \dot{F} = \{H, F\},
\end{equation} 
where the overhead dot denotes the temporal derivative and $H\in\mathcal{A}$ is the Hamiltonian -- see Refs.~\cite{Morrison_1998, Morrison_2005, Morrison_2006} for a review on infinite-dimension Hamiltonian systems.

To begin with, we briefly recall the Hamiltonian structure of the TDSE, which we use as a guide for deriving the one for TDDFT. 
The TDSE algebra of functionals is the set of scalar functionals of $\psi$ and $\psi^*$ equipped with the Poisson bracket 
\begin{equation} \label{eqn:TDSEbracket}
    \{F_1, F_2\}_{\rm SE} = \frac{1}{i}\int{  \left(\frac{\delta F_1}{\delta \psi^*} \frac{\delta F_2}{\delta \psi} - \frac{\delta F_1}{\delta \psi} \frac{\delta F_2}{\delta \psi^*} \right)\ {\rm d} {\bf x}_1\cdots {\rm d} {\bf x}_N},
\end{equation}
where $\frac{\delta F}{\delta \psi}$ denotes the functional derivative of $F$ with respect to $\psi$, defined as
$$
    F\left[\psi+\delta\psi\right] = F\left[\psi\right] +
        \int{\frac{\delta F}{\delta \psi} \delta\psi\ {\rm d} {\bf x}_1\cdots {\rm d} {\bf x}_N}
        + \mathcal{O}\left(\delta\psi^2\right).
$$
The TDSE Hamiltonian functional is
$$
    H_{\rm SE} = \int{\psi^* \hat{\cal H}_{\rm SE} \psi\ {\rm d} {\bf x}_1\cdots {\rm d} {\bf x}_N},
$$
from which we recover $\dot{\psi}=\{H_{\rm SE},\psi\}_{\rm SE}=-i\hat{\cal H}_{\rm SE}\psi$ of Eq.~\eqref{eq:TDSE}.

As we establish next, TDDFT also possesses a Hamiltonian structure. 
The TDDFT algebra of functionals is the set of scalar functionals of $\{\phi_k\}_k$ and $\{\phi^*_k\}_k$.
Finding a Hamiltonian structure for Eq.~\eqref{eq:TDDFT} amounts to finding a Poisson bracket and a specific functional $H$ such that
$$
    \dot{\phi}_k = \{H, \phi_k\} \quad {\rm and} \quad
    \dot{\phi}_k^* = \{H, \phi_k^*\} \quad \forall k.
$$
While physical systems are such that the pairs of fields are complex conjugates of each other, $\left(\phi_k\right)^*=\phi_k^*$ for all $k$, in the TDDFT Hamiltonian formalism we treat them as independent canonically-conjugated variables, like in the TDSE example above. Treating $\phi_k$ and $\phi_k^*$ as independent variables ensures that the dynamics has an even number of dynamical field variables, a hallmark of canonical Hamiltonian systems.
Consistent with our choice of dynamical variables, we formally redefine the one-body density functional of Eq.~\eqref{eq:one-body_density} as
\begin{equation} \label{eq:Hamiltonian_one-body_density}
    \rho^\uparrow = 
        \sum_{k}{n_k^\uparrow\ \phi_k^{\uparrow *} \phi_k^\uparrow}
    \quad {\rm and} \quad
    \rho^\downarrow = 
        \sum_{k}{n_k^\downarrow\ \phi_k^{\downarrow *} \phi_k^\downarrow}.
\end{equation}
For any two functionals $F$ and $G$, the TDDFT Poisson bracket is given by
\begin{equation} \label{eq:Poisson_bracket_complex}
    \left\{F,G\right\} = \sum_k{ \frac{1}{n_k} \int{ \frac{1}{i}  \left(
        \frac{\delta F}{\delta \phi_k^*} \frac{\delta G}{\delta \phi_k} - \frac{\delta F}{\delta \phi_k} \frac{\delta G}{\delta \phi_k^*}
        \right) {\rm d} {\bf x}}},
\end{equation}
where the index $k$ runs over both the up- and down-spin variables.
For virtual KS orbitals where $n_k=0$ we take the convention $n_k/n_k=1$, where the population coefficient in the Poisson bracket of Eq.~\eqref{eq:Poisson_bracket_complex} cancels with that of the functional derivative, to match the limit $n_k\to0$.

Following the template for the operator $\hat{\mathcal{H}}_\text{eff}$ in Eq.~\eqref{eq:TDDFT_Hamiltonian_decomposition}, we decompose the TDDFT Hamiltonian functional between a kinetic, external, Hartree, and exchange-correlation components as~\cite{Gomez_Pueyo_2018}
\begin{equation} \label{eq:Hamiltonian_functional}
    H = H_{\rm kin} + H_{\rm ext} + H_{\rm H} + H_{\rm XC}.
\end{equation}
To recover the TDDFT Eq.~\eqref{eq:TDDFT_Hamiltonian_decomposition}, each contributing term in the Hamiltonian should be real valued and satisfy 
$$
    \frac{\delta H_{\rm el}}{\delta \phi_k^*} = n_k \hat{\mathcal{H}}_{\rm el}\ \phi_k,
$$
where $\hat{\mathcal{H}}_{\rm el}$ denotes one of the terms in Eq.~\eqref{eq:TDDFT_Hamiltonian_decomposition}. 
In subsection~\ref{sec:kinetic_functional} through~\ref{sec:External_field_functional} below, we specify the expressions for common components of the TDDFT Hamiltonian functional.
Generally speaking, we show that the Hamiltonian functionals of Eq.~\eqref{eq:Hamiltonian_functional} are obtained by
substituting density components $|\phi_k|^2$ with the product of their associated conjugated phase-space variables $\phi_k^*\phi_k$ in conventional DFT energy functionals.

\subsection{Kinetic-energy functional} \label{sec:kinetic_functional}

In the canonical Hamiltonian framework discussed above, we define the kinetic-energy functional as
\begin{equation} \label{eq:kinetic-energy_functional}
    H_\text{kin} = \sum_k{n_k\int{ \phi_k^* \left(-\frac{\Delta}{2}\right)\phi_k\ \text{d}{\bf x}}},
\end{equation}
with the index $k$ running over both the up- and down-spin variables.
Immediately, we see that 
$$
    \frac{\delta H_\text{kin}}{\delta\phi_k^*} =
        -\frac{1}{2} n_k \Delta\phi_k   ,
$$
and we recover the kinetic-operator component of Eq.~\eqref{eq:TDDFT_Hamiltonian_decomposition}.

\subsection{External-energy functional} \label{sec:external_functional}

For a given static external potential $\mathcal{V}_\text{ext}\left({\bf x}\right)$, the external-energy functional is
\begin{equation} \label{eq:external-energy_functional}
    H_\text{ext} = \int{ \mathcal{V}_\text{ext}\left({\bf x}\right)\rho\left({\bf x}\right)\ \text{d}{\bf x}},
\end{equation}
where $\rho=\rho^\uparrow+\rho^\downarrow$ is the total one-body density of Eq.~\eqref{eq:Hamiltonian_one-body_density}. Typically, $\mathcal{V}_\text{ext}$ corresponds to the atomic or molecular potential with which the electronic degrees of freedom interact.
Once again, we have
$$
    \frac{\delta H_\text{ext}}{\delta\phi_k^*} = n_k\ \mathcal{V}_\text{ext}\phi_k,
$$
and we recover the external-potential-operator component of $\hat{\mathcal{H}}_\text{eff}$ from Eq.~\eqref{eq:TDDFT_Hamiltonian_decomposition}.

\subsection{Hartree-energy functional} \label{sec:Hartree-energy_functional}

The Hartree-energy functional is the first, and often dominant, nonlinear component in the TDDFT framework. It describes the classical part of electron-electron interaction in the system of interest. For a given interaction potential $\mathcal{V}_\text{ee}$, the Hartree-energy functional is
\begin{equation} \label{eq:Hartree-energy_functional}
    H_{\rm H} = \frac{1}{2}\iint \rho({\bf x})\mathcal{V}_\text{ee}({\bf x}-{\bf x}') \rho({\bf x}')\ {\rm d}{\bf x}{\rm d}{\bf x}'.
\end{equation}
Coulomb interactions correspond to $\mathcal{V}_\text{ee}\left({\bf x}\right)=1/\left|{\bf x}\right|$ but an effective potential may also be used in reduced dimension or to avoid the singularity at ${\bf x}={\bf 0}$.
At minimum, physically relevant effective potentials should be isotropic $\mathcal{V}_{\rm ee}({\bf x})=\mathcal{V}_{\rm ee}^{\rm R}(|{\bf x}|)$, where $\mathcal{V}_{\rm ee}^{\rm R}$ denotes the radial part of the potential.

Once again, taking the functional derivative and using that $\mathcal{V}_\text{ee}\left(-{\bf x}\right) = \mathcal{V}_\text{ee}\left({\bf x}\right)$, we get
$$
    \frac{\delta H_\text{H}}{\delta\phi_k^*} = n_k {\mathcal{V}}_\text{H} \phi_k,
$$
with the Hartree potential
\begin{equation} \label{eq:Hartree_potential}
    {\mathcal{V}}_\text{H}\left[\rho\right]\left({\bf x}\right) = \int{
        \mathcal{V}_\text{ee}\left({\bf x}-{\bf x}'\right) \rho({\bf x}')\ {\rm d}{\bf x}' }.
\end{equation}

{\bf Numerical simulations:} From the equation above, we see that the Hartree potential is obtained by the convolution of the (effective) electron-electron interaction potential $\mathcal{V}_\text{ee}$ and the one-body density.
For one-dimensional models, this convolution can be performed numerically either directly or via fast-Fourier transforms. In either cases, because of the long-range Coulomb tail of physical potentials $\mathcal{V}_\text{ee}$, in our simulations we perform the convolution over an extended domain where the density is zero-padded.
For two- and three-dimensional models, in our experience explicit convolutions is impractically slow while the fast-Fourier route remains viable.

\subsection{Exchange-correlation-energy functional} \label{sec:exchange-correlation-energy_functional}

The exchange-correlation-energy functional $H_\text{XC}$ is the second nonlinear component in the TDDFT framework and collects all non-classical electron-electron interaction effects. It is often decomposed into an exchange component, accounting for the indistinguishability of electrons within each spin channel, and a correlation component, including all the other electron interactions. In the following subsections we briefly review common exchange-correlation-energy functionals.

\subsubsection{Exact-exchange-energy functional} \label{sec:exact-exchange-energy_functional}

The \emph{exact-exchange} (XX) energy functional corresponds to the Hartree-Fock exchange term; indeed running a TDDFT computation with only the Hartree- and exact-exchange-energy functionals is equivalent to a TD Hartree-Fock calculation. The exact-exchange-energy functional is
\begin{widetext} \begin{equation} \label{eq:exact-exchange-energy_functional}
    H_\text{XX} = H_\text{XX}^\uparrow + H_\text{XX}^\downarrow \quad \text{with} \quad
    H_{\rm XX}^\uparrow = 
        -\frac{1}{2}\sum_{k,l} n_k^\uparrow n_l^\uparrow \iint 
            \phi_k^{\uparrow *}({\bf x})\phi_l^{\uparrow *}({\bf x}') \mathcal{V}_{\rm ee}({\bf x}-{\bf x}') \phi_k^\uparrow({\bf x}')\phi_l^\uparrow({\bf x}) {\rm d}{\bf x}{\rm d}{\bf x}'. 
\end{equation} 
For simplicity and coherence, in the equation above we use the same electron-electron interaction potential $\mathcal{V}_\text{ee}$ as in the Hartree-energy functional of Eq.~\eqref{eq:Hartree-energy_functional}, but this need not be the case.

Taking the functional derivative of Eq.~\eqref{eq:exact-exchange-energy_functional}, and once again using that $\mathcal{V}_{\rm ee}(-{\bf x})=\mathcal{V}_{\rm ee}({\bf x})$, we find that
$$
    \frac{\delta H_\text{XX}}{\delta\phi_k^{\uparrow *}} = n_k^\uparrow \hat{\mathcal{V}}_\text{XX}^\uparrow \phi_k^\uparrow
    \quad {\rm and} \quad
    \frac{\delta H_\text{XX}}{\delta\phi_k^{\downarrow *}} = n_k^\downarrow \hat{\mathcal{V}}_\text{XX}^\downarrow \phi_k^\downarrow,
$$
with the action of the up-spin exact-exchange operator $\hat{\mathcal{V}}_\text{XX}^\uparrow$ on a given one-electron wave function $\phi$ defined as
\begin{equation} \label{eq:exact-exchange_potential}
    \hat{\mathcal{V}}_\text{XX}^\uparrow\phi  \left({\bf x}\right) = 
        -\sum_{l}{n_l^\uparrow \phi_l^\uparrow\left({\bf x}\right) \int{
        \phi_l^{\uparrow *}\left({\bf x}'\right) \mathcal{V}_\text{ee}\left({\bf x}-{\bf x}'\right)\phi\left({\bf x}'\right) {\rm d}{\bf x}'}},
\end{equation}  \end{widetext} 
and likewise for the down-spin exact exchange.
Note that the exact-exchange potential is a nonlinear operator that mixes (exchanges) the input argument $\phi$ with all the KS orbitals in the same spin channel in the model, weighted by the orbital-population parameters.

{\bf Numerical simulations:} Like for the Hartree potential -- see section~\ref{sec:Hartree-energy_functional} -- the exact exchange involves the computation of convolutions that can be performed using fast-Fourier algorithms (irrespective of the dimension) or with an explicit scheme (one dimension only) over an extended domain. 
In any case, the exact exchange is computationally expensive with a quadratic complexity scaling in the number of KS orbitals in the simulations. Instead, combinations of closed-form functionals like the ones discussed in the following subsections are often preferred for simulations.

\subsubsection{Local-density-approximation Slater exchange} \label{sec:LDA_exchange}

The \emph{local-density-approximation} (LDA) Slater exchange is defined as the exact-exchange for the homogeneous-electron-gas model. In this case, the exchange-energy functional can be explicitly expressed as a functional of the one-body density
\begin{equation}
    H_{\rm X}^{\rm LDA} \left[ \rho^\uparrow,\rho^\downarrow\right] = 
        \frac{1}{2}\left(E_{\rm X}^{\rm LDA} \left[2\rho^\uparrow\right] + E_{\rm X}^{\rm LDA} \left[2\rho^\downarrow\right]\right).
\end{equation}
The homogeneous electron gas is spin unpolarized and the total one-body density $\rho=2\rho^\uparrow=2\rho^\downarrow$ is commonly used in the derivation of the LDA Slater exchange. This convention leads to the factors ``2'' in the equation above, with the energy functional defined as
\begin{equation} \label{eq:LDA_exchange}
    E_{\rm X}^{\rm LDA} \left[\rho\right] = 
        \int{\rho({\bf x})\ \varepsilon_{\rm X}^{\rm LDA}(\rho({\bf x}))\ {\rm d}{\bf x}},
\end{equation}
where $\varepsilon_{\rm X}^{\rm LDA}$ is called the \emph{exchange-energy per particle}. For a given electron-electron-interaction potential $\mathcal{V}_{\rm ee}$, the exchange-energy per particle is given by
\begin{equation}
    \varepsilon_{\rm X}^{\rm LDA}(\rho) = 
        - \int_0^\infty{K(u)\ \mathcal{V}_{\rm ee}^{\rm R}(\alpha(\rho) u)\ {\rm d}u},
\end{equation}
where $\mathcal{V}_{\rm ee}^{\rm R}$ denotes the radial part of $\mathcal{V}_{\rm ee}$, and the kernel $K$ and scaling factor $\alpha(\rho)$ depend on the dimension of the problem as
\begin{subequations}\begin{eqnarray}
    {\rm (1D)}: &\ & 
        K(u) = \frac{\sin^2 u}{\pi u^2} \quad {\rm and} \quad 
        \alpha = \frac{2}{\pi\rho}, \\
    {\rm (2D)}: &\ &
        K(u) = \frac{J_1(u)^2}{u} \quad {\rm and} \quad
        \alpha = \frac{1}{\sqrt{2 \pi \rho}}, \\
    {\rm (3D)}: &\ & 
        K(u) = \frac{3}{\pi} \left(\frac{\cos u}{u} - \frac{\sin u}{u^2}\right)^2,\ 
        \alpha = \frac{1}{\left(3 \pi^2 \rho\right)^{1/3}},
\end{eqnarray} \end{subequations}
with $J_1$ the Bessel function of the first kind.

Taking the functional derivative of Eq.~\eqref{eq:LDA_exchange}, we get that
$$
    \frac{\delta H_\text{X}^\text{LDA}}{\delta\phi_k^{\uparrow *}} = 
        n_k^\uparrow \mathcal{V}_\text{X}^\text{LDA}[2\rho^\uparrow] \phi_k^\uparrow,
$$
with the LDA exchange potential defined as
\begin{equation} \label{eq:LDA_potential}
    \mathcal{V}_{\rm X}^{\rm LDA} \left[\rho\right]({\bf x}) =
        \varepsilon_{\rm X}^{\rm LDA}(\rho({\bf x})) + 
        \rho({\bf x})\ \frac{\partial \varepsilon_{\rm X}^{\rm LDA}}{\partial \rho}(\rho({\bf x})).
\end{equation}

\subsubsection{Other exchange-correlation functionals} \label{sec:other_exchange-correlation_functionals}

Similar to the LDA Slater exchange, LDA correlation energy functionals -- or exchange-correlation if both exchange and correlation effects are grouped together in the functional -- take the generic form
\begin{equation} \label{eq:LDA_correlation}
    H_{\rm C}^{\rm LDA}\left[\rho^\uparrow,\rho^\downarrow\right] = \int{
        \rho({\bf x})\ \varepsilon_{\rm C}^{\rm LDA}\left(
        \rho^\uparrow({\bf x}),\rho^\downarrow({\bf x})
        \right)\ {\rm d}{\bf x}},
\end{equation}
where $\varepsilon_{\rm C}^{\rm LDA}$ is the LDA-correlation energy per particle. Mirroring the derivation for the LDA exchange and for instance looking at the up-spin channel, we find that the correlation potential is
\begin{equation} \label{eq:LDA_correlation_potential}
    \mathcal{V}_{\rm C}^{\rm LDA,\uparrow}[\rho^\uparrow,\rho^\downarrow]({\bf x}) = 
        \varepsilon_{\rm C}^{\rm LDA}(\rho^\uparrow,\rho^\downarrow) +
        \rho\ \frac{\partial \varepsilon_{\rm C}^{\rm LDA}}{\partial \rho^\uparrow},
\end{equation}
where the last term corresponds to the partial derivative of $\varepsilon_{\rm C}^{\rm LDA}$ with respect to the first variable, which should then be evaluated at $\rho^\uparrow({\bf x})$ and $\rho^\downarrow({\bf x})$.

Beyond LDA, a commonly used class of functionals are the so-called \emph{generalized-gradient approximation} (GGA)~\cite{Perdew_1996, Perdew_2001, Marques_2012} which functionally depend on both the one-body density and its gradient. A general expression for GGA exchange-correlation energy functionals is
\begin{eqnarray} 
    && H_{\rm XC}^{\rm GGA}\left[
        \rho^\uparrow,\rho^\downarrow,
        \nabla\rho^\uparrow,\nabla\rho^\downarrow\right] = \nonumber \\ && \quad\quad\quad\quad \int{
        \rho\ \varepsilon_{\rm XC}^{\rm GGA}\left(
        \rho^\uparrow,\rho^\downarrow,
        \nabla\rho^\uparrow,\nabla\rho^\downarrow\right)
        \ {\rm d}{\bf x}}, \label{eq:GGA_exchange-correlation}
\end{eqnarray}
where $\varepsilon_{\rm XC}^{\rm GGA}$ is the GGA exchange-correlation energy per particle.
Once again, taking the functional derivatives with respect to the KS-orbital field variables, we find that
$$
    \frac{\delta H_{\rm XC}^{\rm GGA}}{\delta \phi_k^{\uparrow *}} = 
        n_k^\uparrow \mathcal{V}_{\rm XC}^{\rm GGA, \uparrow} \phi_k^\uparrow
    \ \ {\rm and} \ \ 
    \frac{\delta H_{\rm XC}^{\rm GGA}}{\delta \phi_k^{\uparrow}} = 
        n_k^\uparrow \mathcal{V}_{\rm XC}^{\rm GGA, \uparrow} \phi_k^{\uparrow *},
$$
with the GGA exchange-correlation potential
\begin{equation} \label{eq:GGA_exchange-correlation_potential}
    \mathcal{V}_{\rm XC}^{\rm GGA,\uparrow} = 
        \varepsilon_{\rm XC}^{\rm GGA} + 
        \rho\ \frac{\partial \varepsilon_{\rm XC}^{\rm GGA}}{\partial \rho^\uparrow} - 
        \nabla\cdot\left(\rho\ \frac{\partial \varepsilon_{\rm XC}^{\rm GGA}}{\partial \nabla\rho^\uparrow} \right),
\end{equation}
where each of the partial derivatives are evaluated at $\rho^\uparrow({\bf x})$, $\rho^\downarrow({\bf x})$, $\nabla\rho^\uparrow({\bf x})$, and $\nabla\rho^\downarrow({\bf x})$. 

Beyond the density gradient, one may include higher-order spatial derivatives in analogs to Eq.~\eqref{eq:GGA_exchange-correlation} that lead to more complex, albeight similar in spirit, exchange-correlations potential than Eq.~\eqref{eq:GGA_exchange-correlation_potential}.
In addition to these, so-called \emph{meta-GGA} (mGGA) exchange-correlation adds a functional dependency into (twice) the \emph{kinetic-energy density}~\footnote{The kinetic-energy density is often denoted with the letter $\tau$. Here we use $\kappa$ to distinguish it from the evolution parameter used in the propagation schemes discussed in section~\ref{sec:Symplectic_schemes}.}
\begin{equation} \label{eq:kinetic-energy_density}
    \kappa^\uparrow ({\bf x}) = \sum_k{n_k^\uparrow
        \nabla\phi_k^{\uparrow*}({\bf x})\cdot
        \nabla\phi_k^{\uparrow}({\bf x})},
\end{equation}
and likewise for $\kappa^\downarrow$. Similar to the other exchange-correlation functionals discussed before, the mGGA energy functional is
\begin{equation} \label{eq:mGGA_exchange-correlation}
    H_{\rm XC}^{\rm mGGA}\left[\cdots,\kappa^\uparrow,\kappa^\downarrow\right]= \int{\rho\ \varepsilon_{\rm XC}^{\rm mGGA}(\cdots,\kappa^\uparrow,\kappa^\downarrow)\ {\rm d} {\bf x}},
\end{equation}
where ``$\cdots$'' denotes the one-body densities $\rho^\uparrow$ and $\rho^\downarrow$, their gradients, and any other higher-order spatial derivatives used in the definition of the functional.
Applying similar functional-derivative calculations as above, we find that the mGGA exchange-correlation potential operator is
\begin{equation} \label{eq:mGGA_exchange-correlation_potential}
    \hat{\mathcal{V}}_{\rm XC}^{\rm mGGA,\uparrow}[\cdots,\kappa^\uparrow,\kappa^\downarrow] \phi_k^\uparrow = \cdots - 
    \nabla\cdot\left(\rho\ \frac{\partial \varepsilon_{\rm XC}^{\rm mGGA}}{\partial \kappa^\uparrow} \nabla\phi_k^\uparrow\right),
\end{equation}
where the first part of the potential ``$\cdots$'' on the right-hand side is similar to Eq.~\eqref{eq:GGA_exchange-correlation_potential}.
Note that, unlike its LDA and GGA counterparts, the mGGA exchange-correlation potential explicitly depends on the KS orbitals and not just the one-body density.

As is evident from the LDA, GGA, and mGGA examples above, a wealth of exchange and correlation functionals have been proposed and are routinely used in TDDFT simulations -- {\it e.g.}, see~\cite{Perdew_2001, Marques_2012, Lehtola_2018} for reviews.
Importantly for us, all these functionals are systematically defined through either an energy functional or an energy per particle.
They are thus automatically compatible with the Hamiltonian formalism we lay out in this paper: $H_{\rm XC}$ is simply the DFT exchange-correlation-energy functional and its associated potential is given by
\begin{equation} \label{eq:exchange-correlation_potential}
    \mathcal{V}_{\rm XC}^\uparrow = 
        \frac{\delta H_{\rm XC}}{\delta \rho^\uparrow}
    \quad {\rm and} \quad
    \mathcal{V}_{\rm XC}^\downarrow = 
        \frac{\delta H_{\rm XC}}{\delta \rho^\downarrow}.
\end{equation}

To conclude this section, we briefly mention a special type of exchange-correlation, which are the so-called \emph{hybrid functionals} that mix fractions of exact exchange (Hartree Fock) and DFT. For instance, an exchange global hybrid corresponds to
\begin{equation}
    H_{\rm X}^{\rm GH} = 
        \alpha\ H_{\rm XX} + (1-\alpha)\ H_{\rm X}^{\rm DFT},
\end{equation}
for some $0\leq\alpha\leq1$, and where $H_{\rm XX}$ is defined in Eq.~\eqref{eq:exact-exchange-energy_functional} and $H_{\rm X}^{\rm DFT}$ is some conventional DFT functional (of the one-body density, its gradient, kinetic-energy density, etc.).
Here as well, we immediately see that the hybrid exchange functional is directly compatible with our Hamiltonian formulation and the corresponding potential operator simply reads
\begin{equation}
    \hat{\mathcal{V}}_{\rm X}^{\rm GH} = 
        \alpha\ \hat{\mathcal{V}}_{\rm XX} + 
        (1-\alpha)\ \hat{\mathcal{V}}_{\rm X}^{\rm DFT}.
\end{equation}

\subsection{External driving field} \label{sec:External_field_functional}

The Hamiltonian formulation of TDDFT lends itself very well to the introduction of external driving forces to the electronic dynamics.
Here, we restrict the discussion to TDDFT models with an external driving electric field in the dipole approximation. In the length gauge, such an external field is included in the external potential of Eq.~\eqref{eq:TDDFT_Hamiltonian_decomposition} as
\begin{equation} \label{eq:length-gauge_external_potential}
    {\mathcal{V}}_\text{ext}\left({\bf x},t\right) = 
        {\mathcal{V}}_\text{ext}\left({\bf x}\right) + 
        {\bf E}\left(t\right)\cdot {\bf x},
\end{equation}
where the first term corresponds to the time-independent atomic or molecular potential and, in the second term, ${\bf E}$ is the spatially-homogeneous electric field.

\subsubsection{Autonomization of the Hamiltonian system} \label{sec:autonomization}

As is standard in driven Hamiltonian models, we first autonomize the system by treating time as a dynamical variable and by adding its canonically conjugated energy variable $\xi$ in an extended Hamiltonian
\begin{equation} \label{eq:Autonomized_Hamiltonian}
    \overline{H}\left[\phi_k,\phi_k^*,t,\xi\right] = H\left[\phi_k,\phi_k^*,t\right] + \xi,
\end{equation}
together with the extended Poisson bracket
\begin{equation} \label{eq:Autonomized_Poisson_bracket}
    \overline{\left\{F,G\right\}} = \left\{F,G\right\} + 
        \frac{\partial F}{\partial \xi}\frac{\partial G}{\partial t} -\frac{\partial F}{\partial t} \frac{\partial G}{\partial \xi},
\end{equation}
for all functionals $F$ and $G$ depending functionally on $\phi_k$, $\phi_k^*$, $t$, and $\xi$. In this way, from Eq.~\eqref{eq:Hamiltonian_flow}, we have
$$
    \dot{t}=\overline{\{\overline{H}, t\}}= 1
    \quad {\rm and} \quad
    \dot{\xi} = \overline{\{\overline{H}, \xi\}}=
        -\frac{\partial H}{\partial t},
$$
such that $\xi$ is related to the amount of energy brought in and out of the system by the time-dependent driving. 
In what follows we implicitly assume that the Hamiltonian and the Poisson bracket are autonomized and remove the overlines for simplicity. 

\subsubsection{Length-gauge formulation} \label{sec:length-gauge_formulation}

Following the template of section~\ref{sec:external_functional}, the length-gauge external-energy functional is
\begin{equation} \label{eq:driven_external-energy_functional}
    H_\text{ext} = 
        \int{ \mathcal{V}_\text{ext}\left({\bf x}\right)\rho\left({\bf x}\right)\ \text{d}{\bf x}} + 
        \int{ {\bf E}\left(t\right)\cdot {\bf x}\ \rho\left({\bf x}\right)\ \text{d}{\bf x}},
\end{equation}
with its functional derivative
$$
    \frac{\delta H_\text{ext}}{\delta\phi_k^*({\bf x})} = 
        n_k \mathcal{V}_\text{ext}({\bf x},t)\phi_k
$$
and $\mathcal{V}_{\rm ext}({\bf x},t)$ defined in Eq.~\eqref{eq:length-gauge_external_potential}.

Aside from the driven external potential, it is also interesting to look at the time evolution of the phase-space variable $\xi$ that tracks the amount of energy injected into, or withdrawn from, the electronic degrees of freedom by the field. 
From Eqs.~\eqref{eq:Hamiltonian_flow} and~\eqref{eq:Autonomized_Poisson_bracket}, we directly deduce that 
\begin{equation}
    \dot{\xi} = \left\{\overline{H},\xi \right\} = 
        -\frac{\partial H_\text{ext}}{\partial t} =
        -\dot{\bf E}\left(t\right)\cdot \int{  {\bf x}\ \rho\left({\bf x}\right)\ \text{d}{\bf x}},
\end{equation}
where the integral term corresponds to the electric dipole.
For example, in transient-absorption spectroscopy, the spectral composition of $\dot{\xi}$ yields the frequency-dependent absorption and emission of light by the target~\cite{Gaarde_2011}.

{\bf Numerical simulations:} Solving for the energy coordinate $\xi$, along with the KS orbitals, can be used to check the conservation of the Hamiltonian functional
$$
    \dot{\overline{H}} = 
        \left\{\overline{H},\overline{H}\right\}=0,
$$
as a proxy to evaluate the accuracy of numerical simulations. 
Otherwise, the dynamics for $\xi$ has no effect on the KS orbital or the one-body density and can be ignored.

\subsubsection{Velocity-gauge formulation} \label{sec:velocity-gauge_formulation}

The velocity gauge is obtained by performing the change of variables 
$$
    (\phi_k,\phi_k^*,\xi,t) \mapsto
        (\tilde{\phi}_k,\tilde{\phi}_k^*,\tilde{\xi},t)
$$
defined as
\begin{subequations} \label{eq:length_vs_velocity-gauge_change_of_variables} \begin{eqnarray}
    \tilde{\phi}_k({\bf x}) &=& 
        {\rm e}^{-i {\bf A}(t) \cdot {\bf x}}\phi_k({\bf x}),\\
    \tilde{\xi} &=& 
        \xi +\int{ {\bf E}(t) \cdot {\bf x}\  \rho({\bf x})\ {\rm d}{\bf x}},  
\end{eqnarray} \end{subequations}
where $\dot{\bf A}=-{\bf E}$.
This transformation ensures that
$$
    \{\tilde{\phi}_k^*,\tilde{\phi}_k\}=-i/n_k
    \quad {\rm and} \quad
    \{\tilde{\phi}_k,\tilde{\xi}\}=0,
$$
so that the expression for the Poisson bracket of Eq.~\eqref{eq:Poisson_bracket_complex} is unchanged and the change of variables is canonical. 
In the end, the velocity-gauge Hamiltonian functional becomes 
$$
    \tilde{H} = \tilde{H}_{\rm kin} + \tilde{H}_{\rm ext} + H_{\rm H} + H_{\rm XC} + \tilde{\xi},
$$
with the kinetic- and external-energy functional respectively changed to
\begin{eqnarray}
    \tilde{H}_{\rm kin} &=& 
        \sum_k{ n_k \int \tilde{\phi}_k^* \frac{\left(-i\nabla +{\bf A}(t)\right)^2}{2} \tilde{\phi}_k\ {\rm d}{\bf x}}, \\
    \tilde{H}_{\rm ext} &=& 
        \int{ \mathcal{V}_\text{ext}({\bf x}) \tilde{\rho}({\bf x})\ {\rm d}{\bf x}}, \nonumber
\end{eqnarray}
where the first sum runs over both up- and down-spin components and only the time-independent part of the external potential is retained in the second equation.
Once again, we get the dynamic for the energy variable 
\begin{eqnarray}
    \dot{\tilde{\xi}} & = & 
        \left\{\tilde{H},\tilde{\xi}\right\} \nonumber \\
    & = & {\bf E}\left(t\right)\cdot \sum_{k}{ n_k \int{\tilde{\phi}_k^* \left(-i\nabla+{\bf A}\left(t\right)\right) \tilde{\phi}_k\ {\rm d}{\bf x}}}, \nonumber \\
    & = & {\bf E}\left(t\right)\cdot\int{{\bf j}\left({\bf x},t\right)\ {\rm d}{\bf x} },
\end{eqnarray}
where the current density ${\bf j}={\bf j}^\uparrow+{\bf j}^\downarrow$ is defined in Eq.~\eqref{eq:velocity-gauge_current_density} below.

\subsection{Self-interaction correction} \label{sec:SIC}

A known issue in TDDFT simulations is the so-called self-interaction of each KS orbital with itself, through the total one-body density. For standard functionals, this leads to an improper tail of the KS potential as compared to the expected Coulomb shape.
Remedies to this issue are generally referred to as \emph{self-interaction correction} (SIC). 
A version of SIC that is easily applicable to the Hamiltonian formalism discussed above is the \emph{average-density SIC} (ADSIC)~\cite{Legrand_2002}. ADSIC is obtained by a global parametric scaling of the one-body densities used in the Hartree and exchange-correlation functional
\begin{widetext}
\begin{equation} \label{eq:ADSIC_Hamiltonian_functionals}
    H_{\rm H}^{\rm ADSIC} \left[\rho^\uparrow,\rho^\downarrow\right] = 
        \frac{N-1}{N} H_{\rm H} \left[\rho^\uparrow,\rho^\downarrow\right]
    \ {\rm and} \
    H_{\rm XC}^{\rm ADSIC} \left[\rho^\uparrow,\rho^\downarrow\right] =
        H_{\rm XC} \left[\rho^\uparrow,\rho^\downarrow\right] - 
        N^\uparrow H_{\rm XC} \left[\frac{\rho^\uparrow}{N^\uparrow},0\right]  -
        N^\downarrow H_{\rm XC} \left[0,\frac{\rho^\downarrow}{N^\downarrow}\right],
\end{equation}
\end{widetext}
where $N^\uparrow=\sum_{k}{n_k^\uparrow}=\int{\rho^\uparrow({\bf x})\ {\rm d}{\bf x}}$, likewise for $N^\downarrow$, and $N=N^\uparrow+N^\downarrow$.
ADSIC yields the scaled potentials
\begin{subequations} \begin{eqnarray}
    \mathcal{V}_{\rm H}^{\rm ADSIC} \left[\rho^\uparrow,\rho^\downarrow\right] & = &
        \frac{N-1}{N} \mathcal{V}_{\rm H} \left[\rho^\uparrow,\rho^\downarrow\right], \\
    \mathcal{V}_{\rm XC}^{\uparrow {\rm ADSIC}}\left[\rho^\uparrow,\rho^\downarrow\right] &=&
        \mathcal{V}_{\rm XC}^{\uparrow}\left[\rho^\uparrow,\rho^\downarrow\right] - 
        \mathcal{V}_{\rm XC}^{\uparrow}\left[\frac{\rho^\uparrow}{N^\uparrow},0\right], \\
    \mathcal{V}_{\rm XC}^{\downarrow {\rm ADSIC}}\left[\rho^\uparrow,\rho^\downarrow\right] &=&
        \mathcal{V}_{\rm XC}^{\downarrow}\left[\rho^\uparrow,\rho^\downarrow\right] - 
        \mathcal{V}_{\rm XC}^{\downarrow}\left[0,\frac{\rho^\downarrow}{N^\downarrow}\right].
\end{eqnarray} \end{subequations}
Note that the exact exchange of section~\ref{sec:exact-exchange-energy_functional} does not suffer self interaction and no SIC is applied.

\section{Hamiltonian dynamics and structure of phase space} \label{sec:Phase_space}

In the previous section, we established that the Poisson bracket of Eq.~\eqref{eq:Poisson_bracket_complex} combined with the Hamiltonian functional of Eq.~\eqref{eq:Hamiltonian_functional} generate a Hamiltonian flow equivalent to the TDDFT system of Eq.~\eqref{eq:TDDFT}.
In this section, we briefly comment on properties of the TDDFT Hamiltonian flow that will be useful in the remainder of the paper.
We also illustrate the use of the Hamiltonian formalism to recover continuity equations and Ehrenfest's theorem for the dipole signal.

\subsection{Physical vs. dynamical variables}

As mentioned in section~\ref{sec:Model} above, KS orbitals \emph{do not} correspond to physical observables, {\it i.e.}, one generally cannot define an operator that maps the TDSE wave function $\psi$ of Eq.~\eqref{eq:TDSE} onto the TDDFT orbitals $\{\phi_k\}_k$ of Eq.~\eqref{eq:TDDFT}. Instead, all the physical information is contained in the one-body densities $\rho^\uparrow$ and $\rho^\downarrow$.
For instance, in recent studies we showed that density-based analyses of attosecond time scale electron dynamics in molecules can be very enlightening~\cite{Bruner_2017,Tuthill_2020,Folorunso_2021} while orbital-based interpretations may be misleading~\cite{Mauger_2022}.
In this section we briefly discuss some of the difficulties that arise in dynamical analyses because of the KS-orbitals {\it vs.} one-body-density distinction.

The first and most obvious source of difficulties is in the way coherence is accounted for in TDDFT. Within each KS orbital, coherence appears explicitly in a similar fashion as for the TDSE. On the other hand, since orbitals are summed incoherently in the one-body density of Eq.~\eqref{eq:one-body_density}, between them coherence is treated implicitly through the nonlinearity of the TDDFT Hamiltonian operator. This second source of coherence is rather intricate owing to the complexity of the various DFT components discussed above.

Second, intrinsically correlated processes are structurally difficult to interpret in the density-based TDDFT framework.  
A good example of this is sequential double ionization, where an atom or molecule is independently stripped of two electrons. While there is no temporal coherence between the two ionization events, double ionization means that both electrons are concomitantly far away from the system's core. This type of spatial coherence is not readily accessible in the one-body density and is hard to distinguish from, {\it e.g.}, competing single ionization from the same channels.

The third source of difficulty for nonlinear analyses resides in the effective dimension of phase space.
Parsing through the various DFT Hamiltonian operators above we notice that, within each spin channel, any unitary transformation (linear mixing) between orbitals that have the same occupation leads to identical one-body densities and have equivalent dynamics -- in other words, the TDDFT Hamiltonian operator commutes with the group formed by the unitary-transform matrices.
This suggests that the physical phase space is obtained by taking the quotient of TDDFT phase space with those unitary transforms, which is obviously not a simple task.
Anecdotally, in Ref.~\cite{Mauger_2022} we observed that the effective dimension of the TDDFT dynamics in physical space can be further drastically reduced to just a handful, or even one, apparent degrees of freedom.

For at least some of these issues, we hope that the Hamiltonian formalism we laid out in the previous section provides a path forward to addressing them by bringing tools and methods from nonlinear dynamics and Hamiltonian systems.

\subsection{Continuity equations} \label{sec:continuity_equations}

For the first illustration of how the TDDFT Hamiltonian formalism discussed in section~\ref{sec:Hamiltonian_formulation} can be used, we recover the continuity equation that links the electronic one-body and current densities.

The external-energy functional $H_{\rm ext}$ of Eq.~\eqref{eq:external-energy_functional} and Hartree energy  $H_{\rm H}$ of Eq.~\eqref{eq:Hartree-energy_functional} are both functionally dependent on the one-body density. Thus, since the Poisson bracket is antisymmetric, they commute with it
%
\begin{equation}  \label{eq:density_Poisson_bracket_kernel_1}
    \{H_{\rm ext},\rho^\uparrow\} = 
    \{H_{\rm H},\rho^\uparrow\} = 0,
\end{equation}
and likewise for $\rho^\downarrow$. Per the same argument, LDA-type exchange-correlations also Poisson-commute with the density. Actually, we find this to be a general property of exchange-correlation energy functionals that solely depend on the one-body density, and potentially any of its spatial derivatives, such as the GGA of Eq.~\eqref{eq:GGA_exchange-correlation}
\begin{equation}  \label{eq:density_Poisson_bracket_kernel_2}
    \{H_{\rm XC}^\text{LDA},\rho^\uparrow\} =
    \{H_{\rm XC}^\text{GGA},\rho^\uparrow\} =
    \ldots = 0.
\end{equation}
In a similar fashion, using the symmetry of the electron-electron-interaction potential $\mathcal{V}_{\rm ee}(-{\bf x})=\mathcal{V}_{\rm ee}({\bf x})$, we find that the one-body density Poisson commutes with the exact-exchange energy $H_{\rm XX}$ of Eq.~\eqref{eq:exact-exchange-energy_functional}
\begin{equation}  \label{eq:density_Poisson_bracket_kernel_3}
    \{H_{\rm XX},\rho^\uparrow\} = \{H_{\rm XX},\rho^\downarrow\} = 0.
\end{equation}
%

From the Poisson-commutation property of Eqs.~(\ref{eq:density_Poisson_bracket_kernel_1}-\ref{eq:density_Poisson_bracket_kernel_3}), we deduce that currents in the TDDFT formalism are solely generated by kinetic functionals. This leads to the continuity equation
\begin{equation} \label{eq:continuity_equation}
    \dot{\rho}^\uparrow  =  \{ H, \rho^\uparrow\} = 
    -\nabla \cdot {\bf j}^\uparrow.
\end{equation}
The \emph{current density} generated by the kinetic-energy functional of Eq.~\eqref{eq:kinetic-energy_functional} is
\begin{equation} \label{eq:current_density}
    {\bf j}^\uparrow = \{ H_{\rm kin}, \rho^\uparrow\}=
        \frac{1}{i}\sum_{k}{\frac{n_k^\uparrow}{2} \left(\phi_k^{\uparrow *} \nabla \phi_k^{\uparrow} - \phi_k^{\uparrow} \nabla \phi_k^{\uparrow *} \right)},
\end{equation}
or, in the velocity gauge variables -- see section~\ref{sec:velocity-gauge_formulation}
\begin{equation} \label{eq:velocity-gauge_current_density}
    {\bf j}^\uparrow =  
        \frac{1}{i}\sum_{k}{\frac{n_k^\uparrow}{2} \left(\tilde{\phi}_k^{\uparrow *} \nabla \tilde{\phi}_k^{\uparrow} - \tilde{\phi}_k^{\uparrow} \nabla \tilde{\phi}_k^{\uparrow *} \right)}
        + {\bf A}(t) \rho^\uparrow,
\end{equation}
and likewise for the down-spin density and current.
For physical systems where $\phi_k^*=\left(\phi_k\right)^*$, the expression for the current density simplifies to
\begin{equation} \label{eq:current_density_in_simulations}
    {\bf j}^\uparrow = 
        \sum_{k}{n_k^\uparrow {\rm Im}\left(\phi_k^{\uparrow *} \nabla \phi_k^{\uparrow}\right)} = 
        \sum_{k}{n_k^\uparrow {\rm Im}\left(\tilde{\phi}_k^{\uparrow *} \nabla \tilde{\phi}_k^{\uparrow}\right)} + {\bf A}(t) \rho^\uparrow.
\end{equation}
Note that, as a physical observable, the current density is real valued.

For mGGA functionals, the dependency on the  kinetic energy density $\kappa$ of Eq.~\eqref{eq:kinetic-energy_density} prevents the  Poisson commutation with the one-body density, leading to a current-like term. For instance, the mGGA of Eq.~\eqref{eq:mGGA_exchange-correlation} gives
\begin{equation} \label{eq:mGGA_current}
    \{H_{\rm XC}^\text{mGGA},\rho^\uparrow\} = 
        - \nabla\cdot\left(2\rho\ \frac{\partial \varepsilon_{\rm XC}^{\rm mGGA}}{\partial \kappa^\uparrow}\ {\bf j}^\uparrow_\text{kin}\right),
\end{equation}
with the current density ${\bf j}^\uparrow_\text{kin}$ as defined in Eqs.~(\ref{eq:current_density}-\ref{eq:current_density_in_simulations}). 
As such, the expression for the current density becomes
\begin{equation}
    {\bf j}^\uparrow = \left(1+2\rho\ \frac{\partial \varepsilon_{\rm XC}^{\rm mGGA}}{\partial \kappa^\uparrow}\right) {\bf j}^\uparrow_\text{kin}.
\end{equation}

Unsurprisingly, the TDDFT continuity equation and current density are reminiscent of those for the TDSE with the $N$-body density $\rho_{\rm SE}({\bf x}_1,\ldots,{\bf x}_N;t)$ 
\begin{eqnarray*}
    \dot{\rho_{\rm SE}} & = & 
        \{H_{\rm SE},\rho_{\rm SE}\}_{\rm SE} \\ & = & 
        \frac{1}{2i}\left( \psi \Delta \psi^*-\psi^*\Delta \psi \right) =
        -\nabla \cdot {\bf j}_{\rm SE},
\end{eqnarray*}
from the Poisson bracket of Eq.~\eqref{eqn:TDSEbracket}, $\Delta=(\Delta_{{\bf x}_1},\ldots,\Delta_{{\bf x}_N})$, and
$$
    {\bf j}_{\rm SE}({\bf x}_1,\ldots,{\bf x}_N;t) = \frac{1}{2i}\left( \psi^* \nabla \psi-\psi\nabla \psi^* \right).
$$

\subsection{Dipole signal dynamics} \label{sec:Ehrenfest_theorem}

As a second illustration of TDDFT Hamiltonian manipulations, we look at Ehrenfest's theorem which links the dipole, dipole-velocity, and dipole-acceleration signals associated with the system's electronic dynamics.
For each, we show how the Hamiltonian flow $\dot{F}=\{H,F\}$ of Eq.~\eqref{eq:Hamiltonian_flow} often provides closed-form equations for the dipole successive time derivatives. 
We note that, on-the-fly computation of these quantities are staples of many quantum-chemistry codes since they define the optical-response properties of the system.
Beyond recovering the well-established Ehrenfest's theorem equations, this section illustrates how one can use the Hamiltonian formalism of section~\ref{sec:Hamiltonian_formulation} to systematically obtain the dynamics of relevant observables.

First, the \emph{dipole} observable is
\begin{equation} \label{eq:dipole}
    {\bf d}(t) = \int{{\bf x} \left(\rho^\uparrow({\bf x})+\rho^\downarrow({\bf x})\right) \ {\rm d}{\bf x} }.
\end{equation}
From there, we simply obtain the \emph{dipole-velocity} observable  $\dot{\bf d}=\{H,{\bf d}\}$
\begin{equation} \label{eq:dipole_velocity}
    \dot{\bf d} (t)=
        \int{{\bf j}^\uparrow({\bf x})+{\bf j}^\downarrow({\bf x})
        \ {\rm d}{\bf x} }, 
\end{equation}
using Eq.~\eqref{eq:continuity_equation} and an integration by parts.

In the absence of mGGA functionals, one can also obtain a closed-form formula for the \emph{dipole-acceleration} observable
$
    \ddot{\bf d}(t) = \{ H, \dot{{\bf d}}\},
$
with the functional derivatives
$$
    \frac{\delta \dot{\bf d}}{\delta \phi_k^\uparrow} = 
        i n_k^\uparrow \nabla \phi_k^{\uparrow *} \quad {\rm and} \quad
    \frac{\delta \dot{\bf d}}{\delta \phi_k^{\uparrow *}} = 
        -i n_k^\uparrow \nabla \phi_k^\uparrow,
$$
and likewise for the down-spin channel.
The equation for the dipole acceleration becomes
\begin{eqnarray}
    \ddot{\bf d} (t) & = & \int{ \left(
        \frac{\delta H}{\delta \phi_k^\uparrow}\nabla \phi_k^\uparrow + 
        \frac{\delta H}{\delta \phi_k^{\uparrow *}}\nabla \phi_k^{\uparrow *}
        \right)\ {\rm d}{\bf x}} + \nonumber \\ && \int{ \left(
        \frac{\delta H}{\delta \phi_k^\downarrow}\nabla \phi_k^\downarrow + 
        \frac{\delta H}{\delta \phi_k^{\downarrow *}}\nabla \phi_k^{\downarrow *}
        \right)\ {\rm d}{\bf x}}. \label{eq:dipole_acceleration_general_equation}
\end{eqnarray}
From there, we find that the kinetic-energy functional of Eq.~\eqref{eq:kinetic-energy_functional} does not contribute to the dipole acceleration since 
$$
    \frac{\delta H_{\rm kin}}{\delta \phi_k^*}=-\frac{1}{2}n_k \Delta \phi_k.
$$
Aside from an eventual exact-exchange piece, the external, Hartree, and exchange-correlation functionals all define local potentials with
$$
    \frac{\delta H_{\rm el}}{\delta \phi_k^\uparrow} = 
        n_k^\uparrow \mathcal{V}_{\rm el}  \phi_k^{\uparrow *}
    \quad {\rm and} \quad
    \frac{\delta H_{\rm el}}{\delta \phi_k^{\uparrow *}} = 
        n_k^\uparrow \mathcal{V}_{\rm el}  \phi_k^\uparrow.
$$
We combine these with the previous derivatives and, after integrating by parts, we find
\begin{eqnarray}
    \ddot{\bf d} (t) & = & \int{
        (\rho^\uparrow + \rho^\downarrow)
        (-\nabla \mathcal{V}_{\rm ext}-\nabla \mathcal{V}_{\rm H})
        \ {\rm d}{\bf x}}  \nonumber \\ &&
    +\int{\left(-\rho^\uparrow \nabla\mathcal{V}_{\rm XC}^\uparrow 
        -\rho^\downarrow \nabla\mathcal{V}_{\rm XC}^\downarrow \right)\ {\rm d}{\bf x}}  \nonumber \\ &&
         - {\bf E}(t) N, \label{eq:dipole_acceleration}
\end{eqnarray}
where ${\bf E}(t)$ appears from the dipole-approximation external-electric-field term of Eq.~\eqref{eq:driven_external-energy_functional} and recall that $N$ is the total charge. Note that the added electric-field contribution ${\bf E}(t) N$ is included in both the length- and velocity-gauge formulations of TDDFT and it has the same spectral composition as the driving field.

Unlike other common exchange-correlation energy functionals, the exact-exchange of Eq.~\eqref{eq:exact-exchange-energy_functional} does not contribute to the dipole acceleration. For example, from Eq.~\eqref{eq:dipole_acceleration_general_equation}, the up-spin exact-exchange contribution to the dipole acceleration is
\begin{widetext}
\begin{equation}
    \ddot{\bf d}_{\rm XX}^\uparrow (t)  =   \int{\left(
        \frac{\delta H_{\rm XX}}{\delta \phi_k^\uparrow}\nabla \phi_k^\uparrow +
        \frac{\delta H_{\rm XX}}{\delta \phi_k^{\uparrow *}}\nabla \phi_k^{\uparrow *}
        \right)\ {\rm d}{\bf x}}  =  
    - \sum_{k,l}{n_k^\uparrow n_l^\uparrow \iint{
        \phi_k^{^\uparrow *}({\bf x})\phi^{^\uparrow *}_l({\bf x}')
        \nabla {\cal V}_{\rm ee}({\bf x}-{\bf x}')
        \phi_k^\uparrow({\bf x}')\phi_l^\uparrow({\bf x})
        \ {\rm d}{\bf x}'{\rm d}{\bf x}}}.
\end{equation}
\end{widetext}
For electron-electron interaction potentials where $\mathcal{V}_{\rm ee}$ is isotropic, the sum above vanishes.

Finally, for TDDFT simulations that \emph{do include} some mGGA functional one may apply a similar procedure to derive a closed-form formula for the dipole acceleration. 
However, the practical use of such a closed form may be limited. 
Looking at Eq.~\eqref{eq:dipole_velocity}, we see that the mGGA-specific term induces second-order derivatives of $\varepsilon_{\rm XC}^{\rm mGGA}$ in $\ddot{\bf d}$, with no straightforward cancellations.
In such cases, it might be easier to obtain the acceleration signal by taking the numerical derivative of the dipole of Eq.~\eqref{eq:dipole} or dipole velocity of Eq.~\eqref{eq:dipole_velocity}.

\section{Symplectic split-operator propagation schemes} \label{sec:Symplectic_schemes}

In this section, we leverage the Hamiltonian structure of TDDFT introduced above to lay out symplectic propagation schemes that efficiently propagate the dynamics while preserving the underlying structure of the Hamiltonian flow.
Specifically we focus on the class of split exponential-operator schemes~\cite{Yoshida_1990, Blanes_2002, McLachlan_2022} and local LDA-type exchange-correlation functionals (see discussion in sections~\ref{sec:LDA_exchange} and~\ref{sec:other_exchange-correlation_functionals}). 
Implicitly, here we assume a discretization of the TDDFT problem on a grid and over a domain large enough that the KS orbitals, and all their spatial derivative, effectively vanish at the edges of the domain, {\it i.e.}, we ignore boundary-condition effects.
We touch on the difficulties for defining symplectic split-operator schemes when using  a spatial basis-set discretization of the TDDFT problem in the next section.

\subsection{Split exponential-operator schemes}

For any functional $F$, the formal solution to the Hamiltonian flow of Eq.~\eqref{eq:Hamiltonian_flow} is 
\begin{equation} \label{eq:Hamiltonian_flow_solution}
    \dot{F} = \mathcal{L}_H F \quad \Rightarrow \quad
        F\left(\tau_0+\tau\right) = {\rm e}^{\tau \mathcal{L}_H}F\left(\tau_0\right),
\end{equation}
where $\tau$ is the evolution parameter and
\begin{equation} \label{eq:Liouville_operator}
    \mathcal{L}_H = \left\{H,\cdot\right\},
\end{equation}
is the \emph{Liouville operator}.
Note that the formal solution above does \emph{not} involve the time-ordering operator, which is already accounted for in the exponential of the Liouville operator.
Applied to the autonomization of Eqs.~\eqref{eq:Autonomized_Hamiltonian} and~\eqref{eq:Autonomized_Poisson_bracket}, 
we get the evolution of the time variable
$$
    t\left(\tau_0+\tau\right) = 
        {\rm e}^{\tau \mathcal{L}_H}t\left(\tau_0\right) = 
        t\left(\tau_0\right) +\tau,
$$
given that $\mathcal{L}_H t = 1$ and $\mathcal{L}_H^2 t = 0$. 
This ensures that the evolution parameter is synchronized with the time variable in driven systems $t=\tau$.
Apart from trivial cases, computing ${\rm e}^{\tau \mathcal{L}_H}F$ is challenging if not outright unfeasible.

To approximate the solution of Eq.~\eqref{eq:Hamiltonian_flow_solution}, symplectic split-operator schemes decompose the Hamiltonian into pieces for which the computation of the operator exponential is facilitated.
We consider the example of a split operator with two components $H=H_1+H_2$ and note that if $H_1$ and $H_2$ Poisson commute the exponential-operator split can be performed exactly
$$
    \{H_1,H_2\} = 0 \quad \Rightarrow \quad
    {\rm e}^{\tau \mathcal{L}_{H_1+H_2}} = 
        {\rm e}^{\tau \mathcal{L}_{H_1}}
        {\rm e}^{\tau \mathcal{L}_{H_2}},
$$
as a result of the commutation of the two Liouville operators $\mathcal{L}_{H_1}$ and $\mathcal{L}_{H_2}$, from the Jacobi identity. 
When the two components do not commute, we instead use the expansion
\begin{equation} \label{eq:symplectic_split_operator}
    {\rm e}^{\tau \mathcal{L}_{H_1+H_2}} = \prod_{k=1}^K{
        {\rm e}^{a_k \tau \mathcal{L}_{H_1}}
        {\rm e}^{b_k \tau \mathcal{L}_{H_2}}}
        + \mathcal{O}\left(\tau^{p+1}\right),
\end{equation}
where $p$ is the order of the scheme and for some suitably chosen coefficients $a_k$ and $b_k$. This split leads to the computation of up to  $2K$ exponentials per time step. The specific choice of $p$ and $K$ results from a compromise between the accuracy of the numerical scheme and the number of operations to be performed. The procedure to determine the relevant coefficients $\{a_k\}_k$ and $\{b_k\}_k$ heavily relies on the Baker-Campbell-Hausdorff formula~\cite{Serre_1992_book}.
Finally, we note that palindromic split schemes --- where $a_K=0$, $a_k=a_{K-k}$, and $b_k=b_{K+1-k}$ for all $k$ --- are exactly time reversible.

Many split schemes for Eq.~\eqref{eq:symplectic_split_operator} have been proposed and are routinely used to propagate Hamiltonian flows~(see, {\it e.g.}, Ref.~\cite{McLachlan_2022} and references therein).
Here we consider the conventional 2$^{\rm nd}$ order Strang splitting (a.k.a.\ Verlet)~\cite{Strang_1968,Yoshida_1990,Blanes_2002} and the 4$^{\rm th}$ order Forest Ruth~\cite{Forest_1990,Yoshida_1990}. We compare them to Blanes and Moan optimized 4$^{\rm th}$ order ${\rm BM}_{6}(4)$ and 6$^{\rm th}$ order ${\rm BM}_{10}(6)$ schemes~\cite{Blanes_2002} that were designed to significantly reduce numerical errors in the propagation.
Note that Strang, Forest Ruth, and Blanes and Moan splittings are all general-purpose algorithms for evaluating the exponential of two operators that do not commute.
Here we leverage the Hamiltonian structure of TDDFT equations laid out in the previous section to obtain a symplectic version of each propagator.
We provide details about the split-operator coefficients for each scheme in appendix~\ref{app:symplectic_split_operator} and note that they are all exactly time reversible.

{\bf Numerical simulations:} In all numerical simulations we consider physical systems for which the canonically-conjugated KS-orbital field variables are complex conjugate from each other: $\phi_k^* = \left(\phi_k\right)^*$ for all $k$. Therefore we only need to propagate the $\left\{\phi_k\right\}_k$ orbitals and ignore the $\left\{\phi_k^*\right\}_k$.
We also stress two important features of the symplectic split-operator schemes we discuss in this section:
(i) Because we compute the operator exponentials in Eq.~\eqref{eq:symplectic_split_operator} without relying on their Taylor-series expansion (see below), all the schemes are systematically unitary, at machine precision, irrespective of the propagation time step.
(ii) Each operator exponentials in Eq.~\eqref{eq:symplectic_split_operator} is computed independently of the others, meaning that one does not need to store the orbitals or the KS potential between different iteration within, or between, each propagation time step. Concretely it means that these symplectic split-operator schemes have relatively low memory requirements for the propagation -- actually all schemes, irrespective of their order, have the same memory footprint.

\subsection{Field-free dynamics} \label{sec:Symplectic_field_free}

We begin with the simplest case of a field-free configuration where the TDDFT Hamiltonian of Eq.~\eqref{eq:Hamiltonian_functional} has no explicit time dependence. A natural splitting is between the kinetic and potential parts of the Hamiltonian:  $H_1=H_\text{kin}$ and $H_2=H_\text{ext}+H_\text{H}+H_\text{XC}$ in Eq.~\eqref{eq:symplectic_split_operator}. 

The first type of components in the split operator is for $\mathcal{L}_{H_1}$ with
$$
    \mathcal{L}_{H_1} \phi_k = 
        i \frac{\Delta}{2} \phi_k, \quad
    \mathcal{L}_{H_1}^2 \phi_k = 
        \left(i \frac{\Delta}{2}\right)^2 \phi_k, \quad \ldots
$$
such that 
$
    {\rm e}^{\tau \mathcal{L}_{H_1}} = 
        {\rm e}^{i \tau \frac{\Delta}{2}}.
$
We note that the Laplacian is diagonal in Fourier space and we therefore compute the exponential operator as
\begin{equation}
    {\rm e}^{\tau \mathcal{L}_{H_1}} \phi_k = 
        \mathcal{F}^{-1}\ {\rm e}^{-i \tau |{\bf p}|^2/2 }\ \mathcal{F}\ \phi_k,
\end{equation}
where $\mathcal{F}$ is the Fourier transform and ${\bf p}$ is the $d$-dimensional coordinate in Fourier-space.

The second type of split component in Eq.~\eqref{eq:symplectic_split_operator} is for $\mathcal{L}_{H_2}$ with
$
    \mathcal{L}_{H_2} \phi_k = 
        -i \left(\hat{\mathcal{V}}_{\rm ext} + \hat{\mathcal{V}}_{\rm H} + \hat{\mathcal{V}}_{\rm XC}\right) \phi_k.
$
For simplicity, here we restrict the discussion to local LDA-type exchange-correlation functionals such that the total DFT potential
$
    \hat{\mathcal{V}}_{\rm DFT}[\rho] = 
        \hat{\mathcal{V}}_{\rm ext} + 
        \hat{\mathcal{V}}_{\rm H} [\rho] + 
        \hat{\mathcal{V}}_{\rm XC}[\rho]
$
is a multiplicative operator that functionally depends on the one-body density only. Thus
\begin{eqnarray}
    \mathcal{L}_{H_2}^2 \phi_k & = &
        \mathcal{L}_{H_2} \left(-i
            \mathcal{V}_{\rm DFT}[\rho] \phi_k
        \right), \nonumber \\ & = &
       -i \left(\mathcal{L}_{H_2} \mathcal{V}_{\rm DFT}[\rho] \right) \phi_k -i 
        \mathcal{V}_{\rm DFT}[\rho] \mathcal{L}_{H_2} \phi_k, \nonumber \\
    & = & \left( -i \mathcal{V}_{\rm DFT}[\rho]\right)^2  \phi_k,
\end{eqnarray}
using the Leibniz rule property of the Poisson bracket and given that potential terms do not generate currents with $\mathcal{L}_{H_2} \rho = 0$ from Eqs.~(\ref{eq:density_Poisson_bracket_kernel_1}-\ref{eq:density_Poisson_bracket_kernel_3}), such that $\mathcal{L}_{H_2} \mathcal{V}_{\rm DFT}[\rho]=0$. More precisely, to prove that $\{H_2,\mathcal{V}_{\rm DFT}\}=0$, we use the following two functional derivatives
\begin{widetext}
\begin{equation}
    \frac{\delta H_2}{\delta \phi_k^{\uparrow *}} = 
        n_k^\uparrow ({\cal V}_{\rm ext}+{\cal V}_{\rm H}+{\cal V}_{\rm XC})\phi_k^\uparrow
    \quad\quad {\rm and} \quad\quad
    \frac{\delta {\cal V}_{\rm XC}({\bf x}')}{\delta \phi_k^\uparrow({\bf x})} = 
        n_k^\uparrow \left(\frac{\partial \varepsilon_{\rm XC}}{\partial \rho}\delta({\bf x}-{\bf x}')-\nabla \cdot \left(\frac{\partial \varepsilon_{\rm XC}}{\partial \nabla \rho}\delta({\bf x}-{\bf x}') \right) \right) \phi_k^{\uparrow *}.
\end{equation}
\end{widetext}
The fact that one quantity is proportional to $\phi_k$ and the other one to $\phi_k^*$ makes their pairing in the Poisson bracket vanish. 
The result above can trivially be extended an any power of $\mathcal{L}_{H_2}$ and thus the exponential operator simply is
\begin{equation}
    {\rm e}^{\tau \mathcal{L}_{H_2}} \phi_k = 
        {\rm e}^{-i \tau \left( \mathcal{V}_{\rm ext} + \mathcal{V}_{\rm H} [\rho] + \mathcal{V}_{\rm XC}[\rho] \right) } \phi_k,
\end{equation}
where $\rho\left[\left\{\phi_k \right\}_k\right]$ need be evaluated right before the exponentiation.

\begin{figure}
    \centering
    \includegraphics[width=\linewidth]{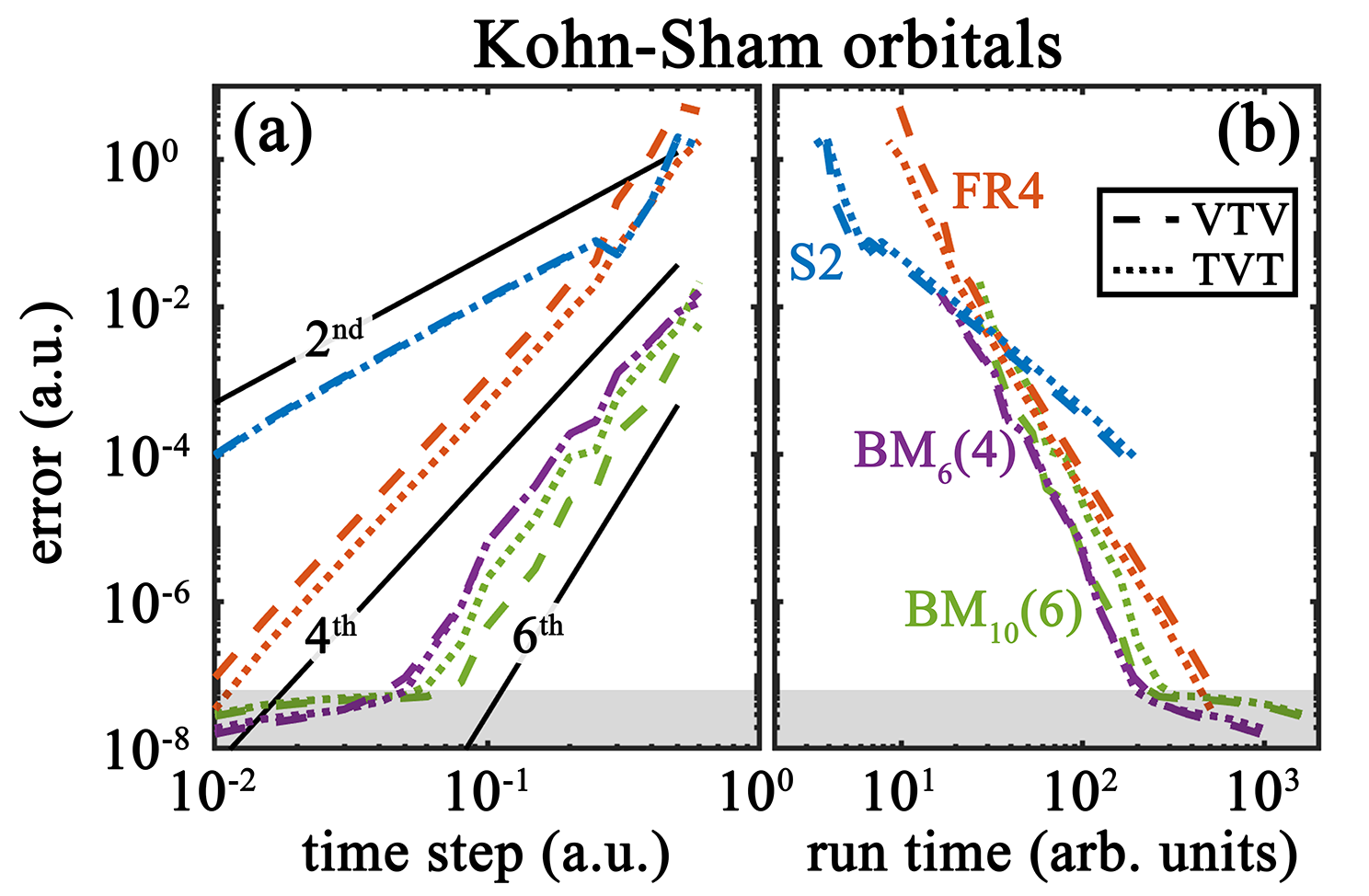}
    \includegraphics[width=\linewidth]{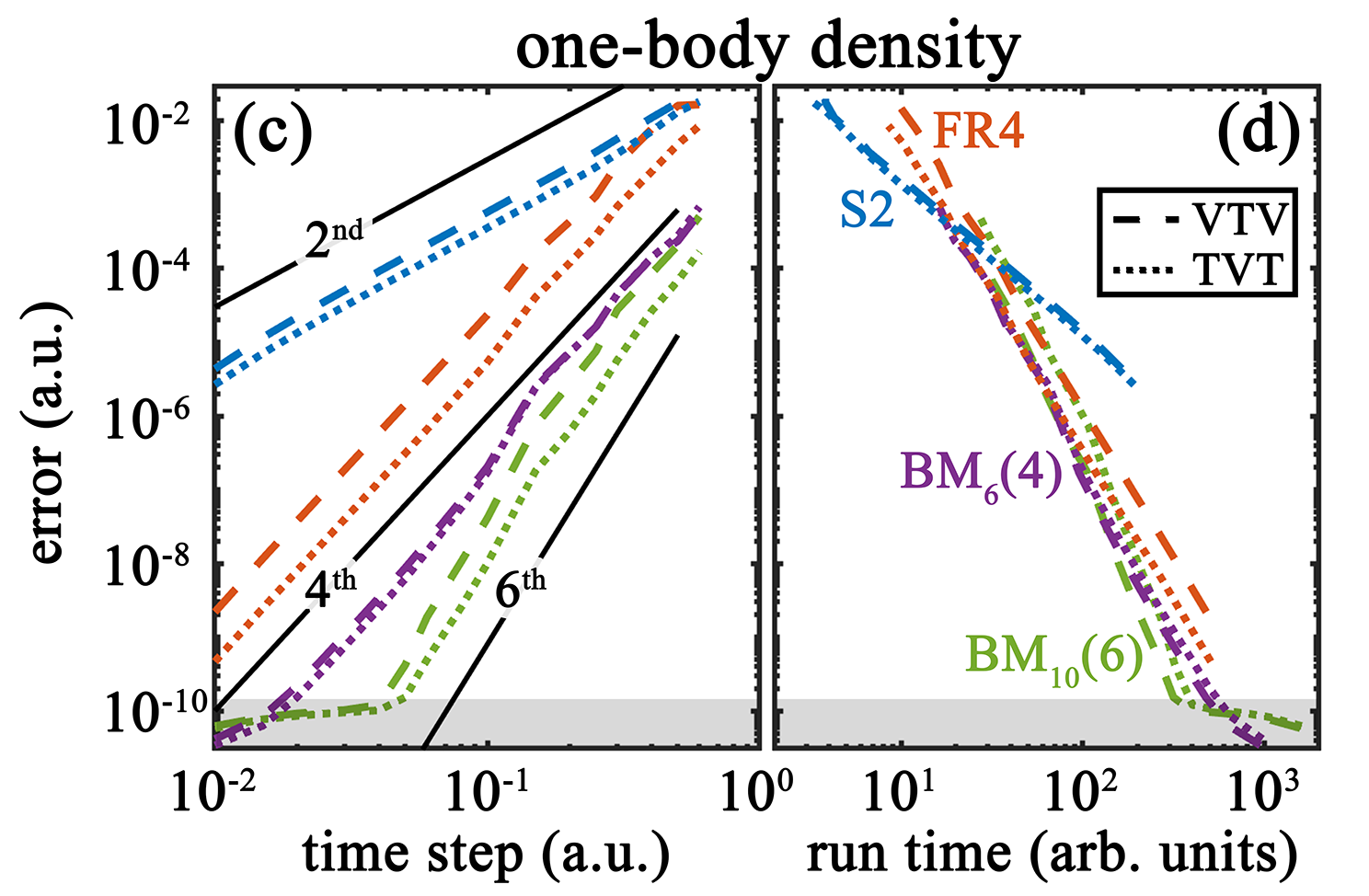}
    \includegraphics[width=\linewidth]{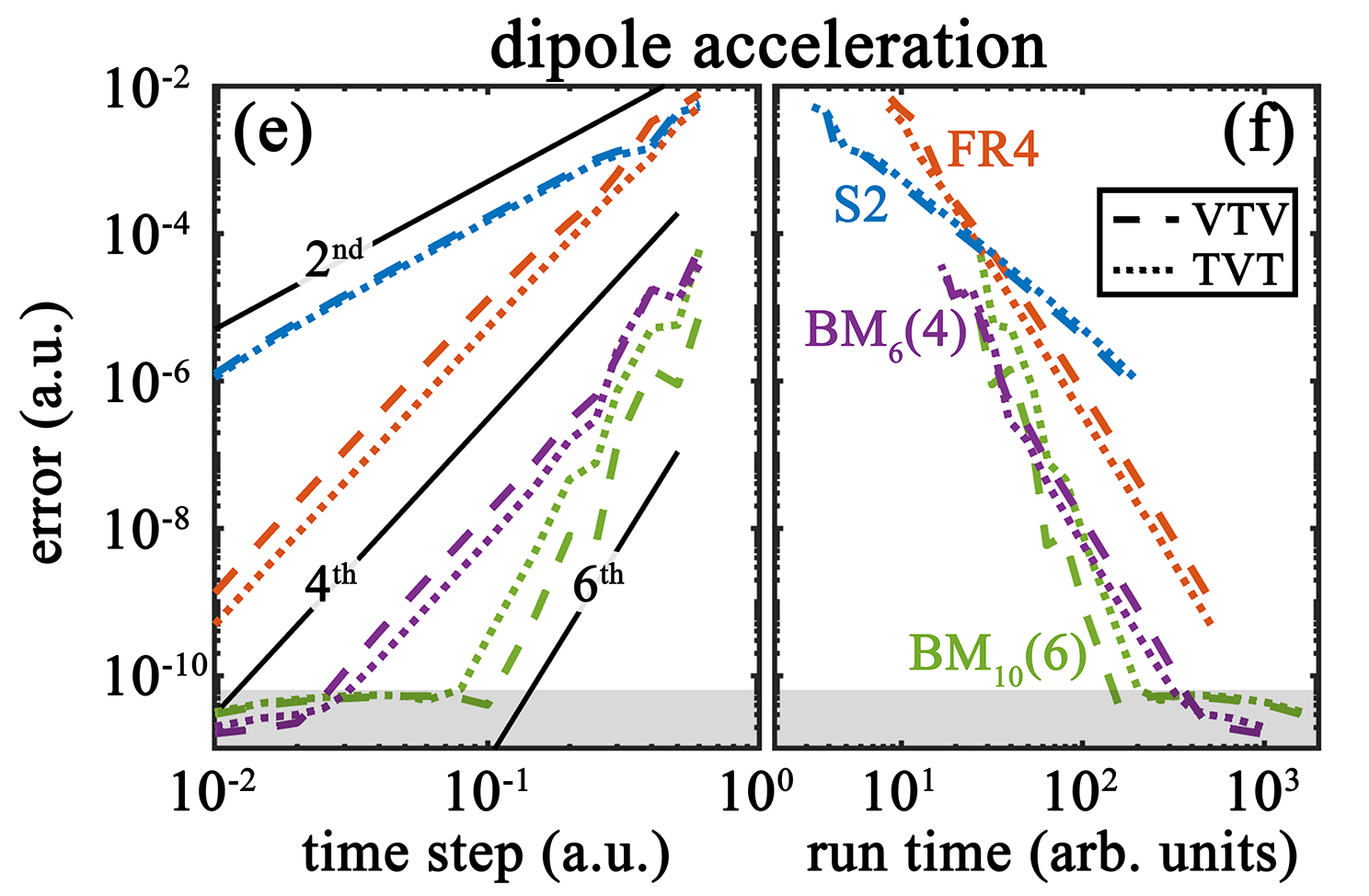}
    \caption{\label{fig:compare_order-FF-SP}
    Comparison of the (a,c,e) accuracy and (b,d,f) efficacy of the (S2) Strang-splitting, (FR4) Forest-Ruth, and Blanes and Moan optimized 4$^{\rm th}$ order ${\rm BM}_6(4)$ and 6$^{\rm th}$ order ${\rm BM}_{10}(6)$ schemes for field-free TDDFT simulations.
    For each scheme we compare the results when we first apply the exponential of the (VTV) potential or (TVT) kinetic term in the split operator of Eq.~\eqref{eq:symplectic_split_operator}.
    Panels (a,b) show the ${\rm L}^2$-norm error in the KS orbitals defined in Eq.~\eqref{eq:KSO_error}, (c,d) the error in the one-body density of Eq.~\eqref{eq:density_error}, and (e,f) that of the dipole acceleration of Eq.~\eqref{eq:dipole-acceleration_error}.
    In each panel, solid lines mark expected scaling trends, with the order printed on them, and the shaded area at the bottom highlights where simulation errors are visibly dominated by extraneous effects to the propagation scheme.
    }
\end{figure}

{\bf Numerical simulations:} In figure~\ref{fig:compare_order-FF-SP} we compare the accuracy (panels a, c, e) and efficacy (b, d, f) of four symplectic split-operator schemes in propagating the field-free TDDFT equations. For the comparison we consider the far-from-equilibrium dynamics of a localized one-electron hole introduced in one of the spin channels of a one-dimensional carbon chain similar to~\cite{Mauger_2022} -- see appendix~\ref{app:numercial_simulation_model} for details about the simulation model we use and how we compute the error.

In all the accuracy plots of Fig.~\ref{fig:compare_order-FF-SP} (left panels), we clearly see that the scaling of the simulation-time-step dependent errors for each scheme matches its predicted order -- compare the colored dashed and dotted curves with solid black guidelines.
For the most accurate results, we remark that the schemes eventually reach a plateau where the error scaling abruptly degrades, marked with gray shading in each panel. We attribute this to roundoff effects associated with the discrete arithmetic in numerical simulations (our code uses double-precision numbers with 16 significant digits). 
The precision at which these appear, though, is beyond what most simulations require. 

Outside of the saturation regime we see that, for a given propagation time step, the $2^{\rm nd}$-order (S2) Strang splitting scheme is the least accurate. Among the two $4^{\rm th}$ order schemes, the ${\rm BM}_{6}(4)$ is at least an order of magnitude more accurate than the (FR4) Forest Ruth~(see, {\it e.g.}, Ref.~ \cite{McLachlan_2022}). Of all, with its higher $6^{\rm th}$-order, the ${\rm BM}_{10}(6)$ is the most accurate.
For each scheme, the relative advantage of starting the split of Eq.~\eqref{eq:symplectic_split_operator} with the exponential of the potential (VTV) or that of the kinetic (TVT) component is inconsistent~\cite{Fasso_2003}. 
For computationally-intensive analyses that run many TDDFT simulations a prior profiling of the VTV {\it vs.} TVT pattern might thus be worth considering.
More generally, our approach provides a straightforward template for implementing high-order symplectic split-operator schemes developed in the nonlinear community to TDDFT simulations.

For all practical purposes, the accuracy of each scheme should be balanced with its efficacy since increased accuracy generally comes at the price of more computations.
In the right panels of Fig.~\ref{fig:compare_order-FF-SP} we compare the efficacy of each scheme, defined as the amount of time it takes to reach a given error in the simulation.
Aside from the crudest result, S2 is the slowest and higher-order schemes should always be preferred.
Between the two $4^{\rm th}$ order schemes, the advantage of the ${\rm BM}_{6}(4)$ compared to FR4 is situational. For the KS orbitals in panel~(b) we observe a slight advantage to ${\rm BM}_{6}(4)$ while the two schemes are about on par with the one-body density in panel~(d).
We note that the TDDFT dynamics we use to test the various schemes is far from equilibrium and should probably be seen as a stress test.
To illustrate this, in Fig.~\ref{fig:compare_order-FF-SP}~(f) we compare each scheme's efficacy in computing the dipole-acceleration signal of Eq.~\eqref{eq:dipole_acceleration} whose main contributions come from the dynamics close to the molecular core where the gradient of the KS potential is largest. In this case, the ${\rm BM}_{6}(4)$ scheme has a decisive edge with over an order of magnitude gain in efficacy as compared to FR4.
Finally, for our test case, the increased complexity of the ${\rm BM}_{10}(6)$ scheme does not justify reaching $6^{\rm th}$ order. Only in the dipole acceleration signal does it eventually prove more efficient than  ${\rm BM}_{6}(4)$ but for accuracy levels generally beyond what one needs.

In the following two subsections, we discuss how symplectic split-operator schemes should be modified to account for an external electric field in the dipole approximation, both in the length and velocity gauges. 

\subsection{Length-gauge driven dynamics} \label{sec:Symplectic_length_gauge}

Compared to the previous section, adding an external driving field requires proper handling of the conjugated time $t$ and energy $\xi$ variables.
In the length gauge and the split-operator of Eq.~\eqref{eq:symplectic_split_operator}, we group $\xi$ with $H_{\rm kin}$, with
$
    H_1 = H_{\rm kin} + \xi,
$
while the expression for $H_2$ is unchanged, albeit picking an explicit time dependence through the external potential $\mathcal{V}_{\rm ext}({\bf x},t)$. 

The updated expression for the split operator $H_1$ gives
$$
    \mathcal{L}_{H_1} \phi_k = i \frac{\Delta}{2} \phi_k \quad {\rm and} \quad
    \mathcal{L}_{H_1} t = 1.
$$
The first equation means that, like in the field-free case, the exponential kinetic-operator can be computed efficiently in Fourier space, where it is diagonal.
The second equation tells us that, in the split expansion of Eq.~\eqref{eq:symplectic_split_operator}, the time variable should be updated after each ${\rm e}^{a \tau \mathcal{L}_{H_1}}$. In other words
\begin{equation}
    {\rm e}^{b \tau \mathcal{L}_{H_2}} {\rm e}^{a \tau \mathcal{L}_{H_1}} \phi_k (t) =
        {\rm e}^{-i b \tau \mathcal{V}_{\rm KS}[\rho(t+a\tau)](t+a\tau) } \phi_k(t+a\tau),
\end{equation}
where
the scalars $a$ and $b$ correspond to the exponential expansion coefficients of Eq.~\eqref{eq:symplectic_split_operator},
$\phi_k(t+a\tau)={\rm e}^{a \tau \mathcal{L}_{H_1}} \phi_k (t)$ and $\mathcal{V}_{\rm KS}=\mathcal{V}_{\rm ext}+\mathcal{V}_{\rm H}+\mathcal{V}_{\rm XC}$, all evaluated with the updated one-body density $\rho(t+a\tau)=\rho\left[\left\{\phi_k(t+a\tau)\right\}_k\right]$.

As in the field-free case, ${\rm e}^{b\tau \mathcal{L}_{H_2}} \phi_k$ is efficiently computed in real space owing the same expression for $H_2$.
Finally, if tracking the energy coordinate $\xi$, we notice that $\mathcal{L}_{H_2}^2\xi=0$ such that
\begin{equation}
    {\rm e}^{b \tau \mathcal{L}_{H_2}} \xi = 
        \xi + b \tau \mathcal{L}_{H_2} \xi = 
        \xi -b \tau\ \dot{\bf E}\cdot \int{  {\bf x}\ \rho\left({\bf x}\right)\ \text{d}{\bf x}}.
\end{equation}

\begin{figure}
    \centering
    \includegraphics[width=\linewidth]{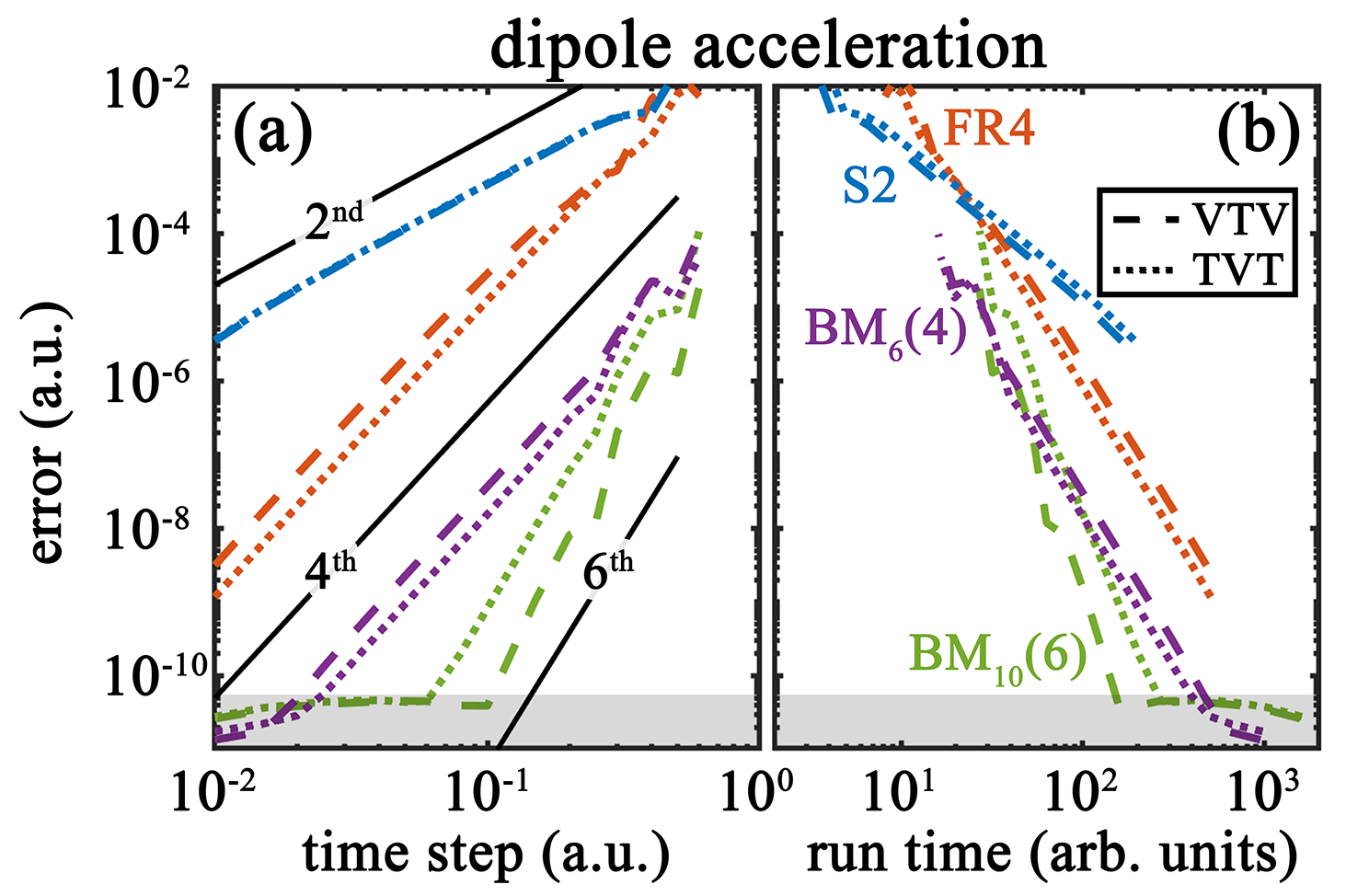}
    \caption{\label{fig:compare_order-LD-SP}
    Like in figure~\ref{fig:compare_order-FF-SP}~(e,f), comparison of the (a) accuracy and (b) efficacy of different symplectic split-operator schemes in simulating length-gauge laser-driven TDDFT dynamics.
    }
\end{figure}

{\bf Numerical simulations:} In figure~\ref{fig:compare_order-LD-SP} we compare the (a) accuracy and (b) efficacy of symplectic split-operator schemes when the TDDFT dynamics is driven by an external electric field in the length gauge -- see appendix~\ref{app:numercial_simulation_model} for details.
Qualitatively the results are very similar to the field-free configuration and, for brevity, we only show the results for the dipole acceleration. We  refer to the discussion in the previous subsection for an analysis of the relative strength of each scheme.
Interestingly, we notice that the simulation time is barely affected by adding the electric-field driving. This can be understood with the simpler form for, and thus negligible computational cost of, the laser external potential in Eq.~\eqref{eq:length-gauge_external_potential} as compared to, {\it e.g.}, the kinetic or Hartree-potential terms.

\subsection{Velocity-gauge driven dynamics} \label{sec:Symplectic_velocity_gauge}

Compared to the length gauge of the previous section, in the velocity gauge the explicit time dependence is moved to the kinetic energy.
To balance this, in the split operator we move the $\xi$ term to $H_2=H_{\rm ext}+H_{\rm H}+H_{\rm XC}+\xi$ and $H_1=H_{\rm kin}$.
Overall, in the split operator propagation, this results in swapping the steps when the time and energy variables should be updated, respectively with $H_2$ and $H_1$.

{\bf Numerical simulations:} Within our driven TDDFT test-case dynamics we find that the accuracy and efficacy of velocity-gauge simulations are almost identical to those of the length-gauge case shown in Fig.~\ref{fig:compare_order-LD-SP}.
This suggests that the choice of the length- {\it vs.} velocity-gauge model should not be directed by the propagation scheme but rather be made based on the analysis framework.

\section{Basis-set TDDFT} \label{sec:Basis-set_TDDFT}

In this section, we briefly discuss how the TDDFT Hamiltonian formalism detailed above changes when the dynamics is restricted to a time-independent spatial basis of orbitals $\left\{\Theta_k\right\}_k$. This configuration is commonly encountered in basis-set TDDFT simulations, with the orbital basis built from linear combinations of atomic orbitals positioned around the various molecular centers in the system.
We also discuss the difficulty engendered by such basis-set models for defining symplectic schemes.

\subsection{Hamiltonian formulation of basis-set TDDFT} \label{sec:Hamiltonian_basis-set_TDDFT}

We derive the Hamiltonian formulation of basis-set TDDFT from that of section~\ref{sec:Hamiltonian_formulation} using a sub-algebra reduction approach:
We define $\mathcal{P}$ the orthogonal projector on the vector space spanned by the orthonormal basis $\left\{\Theta_k\right\}_k$ -- if the basis set is implicitly defined from atom-centered atomic orbitals one can, for instance, obtain the basis $\left\{\Theta_k\right\}_k$ by diagonalizing the overlap matrix between the atomic orbitals. The projector acts on a wavefunction $\phi$ as
\begin{equation}
    {\mathcal P}\phi ({\bf x}) = \sum_l{\Theta_l({\bf x}) \int{
        \Theta_l^*({\bf x}') \phi({\bf x}')
        \ {\rm d}{\bf x}'}}. 
\end{equation}
Implicitly, to be well-posed, the number of orbital-basis elements -- {\it i.e.}, the dimension of the range of $\mathcal{P}$ -- must be larger than the number of KS orbitals in each spin channel.

Using the decomposition of the identity operator
$
    1 = \mathcal{P} + \left(1 - \mathcal{P}\right),
$
we introduce the change of variables
\begin{equation}
   \phi_{\parallel k} = \mathcal{P}\phi_k
    \quad {\rm and} \quad
    \phi_{\perp k} = \left(1-\mathcal{P}\right) \phi_k,
\end{equation}
and, given that the projector $\mathcal{P}$ is self-adjoint $\mathcal{P}^\dagger=\mathcal{P}$,
$$
    \phi_{\parallel k}^* = 
        \mathcal{P}\phi_k^* \quad {\rm and} \quad \phi_{\perp k}^* = \left(1-\mathcal{P}\right) \phi_k^*,
$$
where the index $k$ covers both the KS-orbital and spin labels. 
Next, with a simple chain rule we get, {\it e.g.},
$$
    \frac{\delta F}{\delta \phi_{k}^*} = 
        \mathcal{P}\frac{\delta F}{\delta \phi_{\parallel k}^*} + 
        \left(1-\mathcal{P}\right)\frac{\delta F}{\delta \phi_{\perp k}^*},
$$
so that the Poisson bracket~\eqref{eq:Poisson_bracket_complex} becomes
\begin{widetext}
$$
    \left\{F,G\right\} = \sum_k{ \frac{1}{n_k} \int{ \frac{1}{i}  \left(
         \frac{\delta F}{\delta \phi_{\parallel k}^*} 
         \mathcal{P} \frac{\delta G}{\delta \phi_{\parallel k}} -  \frac{\delta F}{\delta \phi_{\parallel k}} 
         \mathcal{P} \frac{\delta G}{\delta \phi_{\parallel k}^*} +
         \frac{\delta F}{\delta \phi_{\perp k}^*}
         (1-\mathcal{P}) \frac{\delta G}{\delta \phi_{\perp k}} -  \frac{\delta F}{\delta \phi_{\perp k}} 
         (1-\mathcal{P}) \frac{\delta G}{\delta \phi_{\perp k}^*}
    \right) {\rm d} {\bf x}}}.
$$
\end{widetext}
Finally, we obtain the reduction by restricting the algebra of functionals to those of the form $F[\phi_{\parallel k},\phi_{\parallel k}^*]$ so that the Poisson bracket becomes
\begin{equation} \label{eq:basis_set_Poisson_bracket}
    \left\{F,G\right\}_\parallel = \sum_k{ \frac{1}{n_k} \int{ \frac{1}{i}  \left(
         \frac{\delta F}{\delta \phi_{\parallel k}^*} 
         \mathcal{P} \frac{\delta G}{\delta \phi_{\parallel k}} -  \frac{\delta F}{\delta \phi_{\parallel k}} 
         \mathcal{P} \frac{\delta G}{\delta \phi_{\parallel k}^*}
    \right) {\rm d} {\bf x}}}.
\end{equation}

The basis-set reduction leads to similar TDDFT equations as throughout section~\ref{sec:Hamiltonian_formulation}, where each operator $\hat{\mathcal{H}}_{\rm el}$ is substituted with
$
    \mathcal{P} \hat{\mathcal{H}}_{\rm el} \mathcal{P}.
$
Note that if $\hat{\mathcal{H}}_{\rm el}$ is Hermitian, so is $\mathcal{P}\hat{\mathcal{H}}_{\rm el}\mathcal{P}$ such that the orthonormality property for the KS orbitals of Eq.~\eqref{eq:KS_are_orthonormal} is automatically retained in the basis-set formulation.
On the other hand, the basis-set reduction removes the Poisson-bracket commutation properties of the one-body density with the external, Hartree, and exchange-correlation potentials of Eqs.~(\ref{eq:density_Poisson_bracket_kernel_1}-\ref{eq:density_Poisson_bracket_kernel_3}) as
\begin{widetext}
\begin{equation} \label{eq:basis-set_spurious_currents}
    \left\{H_{\rm ext},\rho \right\}_\parallel = \sum_{k}{\frac{n_k}{i} \left(
        (\mathcal{P}\ \mathcal{V}_{\rm ext}\phi_{\parallel k})\phi_{\parallel k}^*-
        (\mathcal{P}\ \mathcal{V}_{\rm ext}\phi_{\parallel k}^*)\phi_{\parallel k}
    \right)} \neq 0 \quad {\rm likewise}\ 
    \left\{H_{\rm H},\rho \right\}_\parallel \neq 0 \quad {\rm and}\ 
    \left\{H_{\rm XC},\rho \right\}_\parallel \neq 0.
\end{equation}
\end{widetext}
In turn, this breaks the continuity equation~\eqref{eq:continuity_equation}.
Intuitively, Eq.~\eqref{eq:basis-set_spurious_currents} can be thought of as akin to spurious current contributions induced by the basis-set reduction. For numerical simulations, the non-vanishing terms of Eq.~\eqref{eq:basis-set_spurious_currents} engenders serious challenges for defining symplectic split-operator schemes, as discussed in the next section.
Similarly, we note that the Ehrenfest theorem detailed in section~\ref{sec:Ehrenfest_theorem} does not hold in the basis-set because of the additional terms that appear when evaluating $\dot{\bf d}=\{H,{\bf d}\}_\parallel$.

To conclude this subsection, we note that one can recover Hamilton's equations for TDDFT detailed in Ref.~\cite{Gomez_Pueyo_2018} by considering two canonical changes of variables.
The first change of variables explicitly introduces the complex-valued linear-expansion coefficients of the KS orbitals into the projection basis
$$
    \{\phi_k,\phi_k^*\}_{k} \to \{c_{k,l},c_{k,l}^*\}_{k,l} \quad {\rm with} \quad
    \phi_k = \sum_{l}{c_{k,l} \Theta_{l} },
$$
and likewise for $\phi_k^*$. With this change of variables, the Poisson bracket~\eqref{eq:basis_set_Poisson_bracket} becomes
\begin{equation}
    \left\{F,G\right\}_\parallel = \sum_{k,l}{\frac{1}{i n_k} \left(
        \frac{\partial F}{\partial c_{k,l}^*} \frac{\partial G}{\partial c_{k,l}} -
        \frac{\partial F}{\partial c_{k,l}} \frac{\partial G}{\partial c_{k,l}^*}
        \right)},
\end{equation}
owing to the fact that the expansion basis is orthonormal. Note that the functional derivatives of Eq.~\eqref{eq:basis_set_Poisson_bracket} are replaced with partial derivatives with respect to the scalar expansion coefficients. Similarly to their parent KS-orbital field variables, the pairs of $(c_{k,l},c_{k,l}^*)$ are treated as independent, canonically conjugated, dynamical variables while, for physical systems, they remain complex conjugate of each other.
The second change of variables $\{c_{k,l},c_{k,l}^*\}_{k,l}\to\{q_{k,l},p_{k,l}\}_{k,l}$ introduces the real and imaginary parts of the population coefficients with 
$$
    q_{k,l} = \frac{1}{\sqrt{2}}\left(c_{k,l}+c_{k,l}^* \right)
    \quad {\rm and} \quad
    p_{k,l} = \frac{1}{i\sqrt{2}}\left(c_{k,l}-c_{k,l}^* \right),
$$
leading to the canonical Poisson bracket
\begin{equation}
    \left\{F,G\right\}_\parallel = \sum_{k,l}{\frac{1}{n_k} \left(
        \frac{\partial F}{\partial p_{k,l}} \frac{\partial G}{\partial q_{k,l}} -
        \frac{\partial F}{\partial q_{k,l}} \frac{\partial G}{\partial p_{k,l}}
        \right)},
\end{equation}
from which Hamilton's equations of Ref.~\cite{Gomez_Pueyo_2018} are recovered.

\subsection{Symplectic schemes} 

Intuitively, the non-vanishing contribution of potential terms to the current density in Eq.~\eqref{eq:basis-set_spurious_currents} can be understood as follows:
With a basis set discretization, all operators are effectively non local. This is because they operate over the entire support of each basis element.
This can lead to intricate features in spatial areas where several basis vectors overlap.
For all practical purposes this means that the naive implementation of the split-operators of section~\ref{sec:Symplectic_schemes}, taking ${\rm e}^{\tau \mathcal{L}_{H_2}}\equiv{\rm e}^{\tau \hat{\mathcal{V}}_{\rm DFT}}$, with basis-set TDDFT models systematically lead to first-order schemes.
This underpins the difficulty of leveraging the symplectic structure of TDDFT equations when expanding the KS orbitals into a basis set like in Ref.~\cite{Gomez_Pueyo_2018}.

Going back to the definition of the exponential operator
$$
    {\rm e}^{\tau \mathcal{L}_{H_2}} F = \sum_{k\geq0}{\frac{\tau^k }{k!} \mathcal{L}_{H_{2}}^k F},
$$
we see that, because of non-vanishing Eq.~\eqref{eq:basis-set_spurious_currents}, each order in the truncated sum may bring an ever growing number of terms.
This is a major difficulty for defining split-operator symplectic schemes similar to those of section~\ref{sec:Symplectic_schemes} but with basis sets. Yet, symplectic schemes could provide superior results, especially when long propagation times are considered~\cite{McLachlan_2022}.
We note that basis-set TDDFT codes are ubiquitous to high-performance quantum-chemistry simulation. Thus progress with defining symplectic, or more generally geometric, propagation schemes would likely prove very valuable.
To finish, we mention commutator-free Magnus expansion schemes~\cite{Gomez_Pueyo_2018,Gomez_Pueyo_2020,Gomez_Pueyo_2020_2} that have shown excellent results but  are implicit and, {\it e.g.},  require extrapolation and keeping the results of multiple previous time steps. We hope that the Hamiltonian formalism we lay out in this paper can help with further development of propagation schemes.

\section{Summary, conclusions, and outlook} \label{sec:Conclusion}

In summary, we have shown that generalized Kohn-Sham TDDFT derives from a canonical Hamiltonian formalism with the Hamiltonian functional matching the expression for standard ground-state DFT~\cite{Gomez_Pueyo_2018}.
We illustrated the use of the Hamiltonian formalism to recover continuity equations and Ehrenfest's theorem linking the electron density, current, dipole, dipole-velocity, and dipole-acceleration observables.
We then leveraged the Hamiltonian structure to define symplectic split-operator schemes for local exchange-correlation potentials. These symplectic schemes can accurately and efficiently propagate the TDDFT equations with a minimal memory footprint, while being exactly unitary and time reversible. 
In the test cases we use to compare these propagation schemes, we find that an error-optimized $4^{\rm th}$ order method BM$_6(4)$ provides the best trade-off between simulation error and run time.
Finally, we showed how the canonical Hamiltonian framework should be adapted when using a spatial basis-set discretization, and the challenges this brings for defining symplectic split-operator schemes.


Throughout, we have considered external potentials with a fixed molecular geometry.
One can easily extend the Hamiltonian formalism we have laid out to include a semi-classical treatment of the nuclear degrees of freedom, corresponding to so-called \emph{Ehrenfest TDDFT}~\cite{Marques_2012_book, Castro_2013}. 
In this formalism, the nuclear degrees of freedom are treated classically and interact with the electrons through the mean-field one-body density. 
Unsurprisingly, Ehrenfest TDDFT also possesses a Hamiltonian structure~\cite{Gomez_Pueyo_2020}.
As an outlook, in appendix~\ref{app:Ehrenfest_TDDFT}, we briefly lay out how the Hamiltonian structure of section~\ref{sec:Hamiltonian_formulation} can be modified to include semi-classical nuclear motion.

\begin{acknowledgments}
This work was supported by the U.S.\ Department of Energy, Office of Science, Office of Basic Energy Sciences, under Award No.~DE-SC0012462.
CC acknowledges funding from the European Union’s Horizon 2020 research and innovation program under the Marie Sk\l odowska-Curie Grant Agreement No. 734557.
\end{acknowledgments}

\ 

{\bf CRediT:}
    Conceptualization, Methodology, Investigation, and Writing - Original Draft: F.M. and C.C.;
    Software and Validation: F.M.;
    Writing - Review \& Editing: M.B.G., K.L., and K.J.S.

\appendix

\section{Spin restricted models} \label{app:Spin_restricted}

Spin-restricted models correspond to systems for which both the population parameters $\left\{n_k\right\}_k$ and the KS orbitals $\left\{\phi_k\right\}_k$ are identical for the up- and down-spin channels ($n_k^\uparrow=n_k^\downarrow$ and $\phi_k^\uparrow=\phi_k^\downarrow$ for all $k$).
In such a case, the TDDFT formalism can be simplified to account for the symmetry. We introduce the joint population coefficients $n_k=n_k^\uparrow+n_k^\downarrow=2n_k^\uparrow=2n_k^\downarrow$. It is also sufficient to follow half of the KS-orbital field variables and we omit the spin label, which is now redundant.
For the sake of brevity, we only review equations that change from the spin-polarized case laid out in the main text.

The spin-channel-resolved one-body densities of Eq.~\eqref{eq:Hamiltonian_one-body_density} are replaced with the total one-body density
\begin{equation} \label{eq:spin_restricted_Hamiltonian_one-body_density}
    \rho({\bf x}) = 
        \sum_k{n_k \phi_k^*({\bf x})\phi_k({\bf x})} = 
        2\rho^\uparrow({\bf x}) = 2\rho^\downarrow({\bf x}),
\end{equation}
while the expression for the (canonical) Poisson bracket~\eqref{eq:Poisson_bracket_complex} is unchanged.
The expressions for the kinetic, external, Hartree, and exchange-correlation functionals are straightforwardly obtained by substituting any $\rho^\uparrow$ and $\rho^\downarrow$ by $\rho/2$ throughout section~\ref{sec:Hamiltonian_formulation}.
Note that the exact-exchange energy functional and potential of Eq.~\eqref{eq:exact-exchange-energy_functional} and Eq.~\eqref{eq:exact-exchange_potential} respectively becomes
\begin{widetext}
\begin{equation} \label{eq:spin_restricted_exact-exchange-energy_functional}
    H_{\rm XX} = -\frac{1}{4}\sum_{k,l}{n_k n_l\iint{
        \phi_k^{*}({\bf x})\phi_l^{*}({\bf x}') \mathcal{V}_{\rm ee}({\bf x}-{\bf x}') \phi_k({\bf x}')\phi_l({\bf x}) {\rm d}{\bf x}{\rm d}{\bf x}',
    }}
\end{equation}
\begin{equation}
    \hat{\mathcal{V}}_{\rm XX} \phi = -\frac{1}{2}\sum_{l}{n_l
        \phi_l({\bf x}) \int{\phi_l^*({\bf x}') \mathcal{V}_{\rm ee}({\bf x}-{\bf x}') \phi({\bf x}')\ {\rm d}{\bf x}'
    }},
\end{equation}
\end{widetext}
where the $1/2$ factor in the potential compensates for $n_l$ being twice the spin-resolved charge for the orbital $l$.

\begin{figure}[b]
    \centering
    \includegraphics[width=\linewidth]{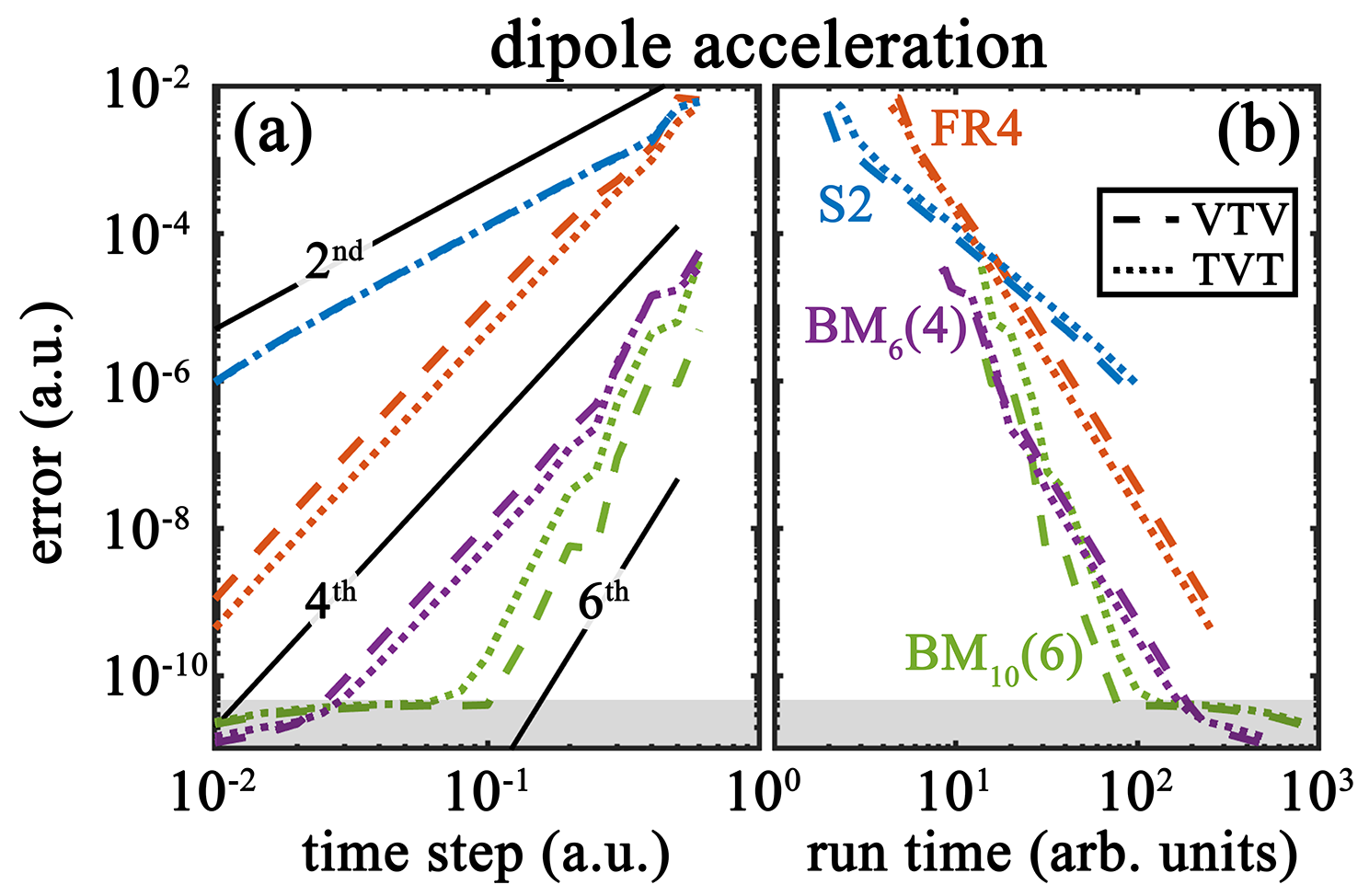}
    \caption{\label{fig:compare_schemes-spin_restricted}
    Like in figure~\ref{fig:compare_order-FF-SP}~(e,f), comparison of the (a) accuracy and (b) efficacy of different symplectic split-operator schemes in simulating field-free spin-restricted TDDFT dynamics.
    }
\end{figure}

{\bf Numerical simulations:} Figure~\ref{fig:compare_schemes-spin_restricted} compares the accuracy and efficacy of symplectic split-operator schemes in propagating field-free spin-restricted TDDFT dynamics.
Once again, we find qualitatively very similar results to spin-polarized simulations. 
Quantitatively, the reduced number of orbitals being propagated in spin restricted models leads to a speedup in the efficacy results -- compare the y axes in Fig.~\ref{fig:compare_schemes-spin_restricted}~(b) and Fig.~\ref{fig:compare_order-FF-SP}~(f).
We also observe similar results and conclusions when adding an external driving field in the length- or velocity-gauge formalism -- not shown.

\begin{widetext}
\section{Symplectic split-operator schemes} \label{app:symplectic_split_operator}

The split-operator coefficients of Eq.~\eqref{eq:symplectic_split_operator} for each of the symplectic schemes we consider are
\begin{itemize}
    \item Strang splitting (a.k.a.\ Verlet; 2$^{\rm nd}$ order)~\cite{Strang_1968,Yoshida_1990,Blanes_2002}: ${\bf a}=\left\{1, 0\right\}$ and ${\bf b}=\left\{\frac{1}{2}, \frac{1}{2}\right\}$ 
    \item Forest Ruth (4$^{\rm th}$ order)~\cite{Forest_1990,Yoshida_1990}: ${\bf a}=\left\{\theta, 1-2\theta,\theta,0\right\}$ and ${\bf b}=\left\{\frac{\theta}{2}, \frac{1-\theta}{2}, \frac{1-\theta}{2}, \frac{\theta}{2}\right\}$, with $\theta= 1/(2 - 2^{1/3})$.
    \item Blanes and Moan optimized ${\rm BM}_{6}(4)$ (4$^{\rm th}$ order)~\cite{Blanes_2002}: ${\bf a} = \{$0.209515106613362, -0.143851773179818, 0.434336666566456, 0.434336666566456, -0.143851773179818, 0.209515106613362, 0$\}$ and ${\bf b} = \{$0.0792036964311957, 0.353172906049774, -0.0420650803577195, 0.2193769557534996, -0.0420650803577195, 0.353172906049774, 0.0792036964311957$\}$.
    \item Blanes and Moan optimized ${\rm BM}_{10}(6)$ (6$^{\rm th}$ order)~\cite{Blanes_2002}: ${\bf a} = \{$0.148816447901042, -0.132385865767784, 0.067307604692185, 0.432666402578175, -0.016404589403618, -0.016404589403618, 0.432666402578175, 0.067307604692185, -0.132385865767784, 0.148816447901042, 0$\}$ and ${\bf b} = \{$0.0502627644003922, 0.413514300428344, 0.0450798897943977, -0.188054853819569, 0.54196067845078, -0.7255255585086897, 0.54196067845078, -0.188054853819569, 0.0450798897943977, 0.413514300428344, 0.0502627644003922$\}$.
    %
\end{itemize}

\end{widetext}
\section{Numerical simulation models} \label{app:numercial_simulation_model}

To numerically test for the accuracy and efficacy of the various split-operator schemes we discuss in this paper, we consider a one-dimensional molecular model and in-house TDDFT code written in MATLAB similar to Ref.~\cite{Mauger_2022}.
Briefly, we set a series of four pairs of atom-like centers that emulates conjugated (doubly- or triply-bounded) carbon chains, each contributing one electron to the neutral compound.
Because of the reduced dimension, we substitute all Coulomb interactions with a softened version of the potential~\cite{Javanainen_1988}.
Like in Ref.~\cite{Mauger_2022}, we use LDA Slater exchange (see section~\ref{sec:LDA_exchange}) and correlation functionals with ADSIC (see section~\ref{sec:SIC}).

\begin{figure}
    \centering
    \includegraphics[width=\linewidth]{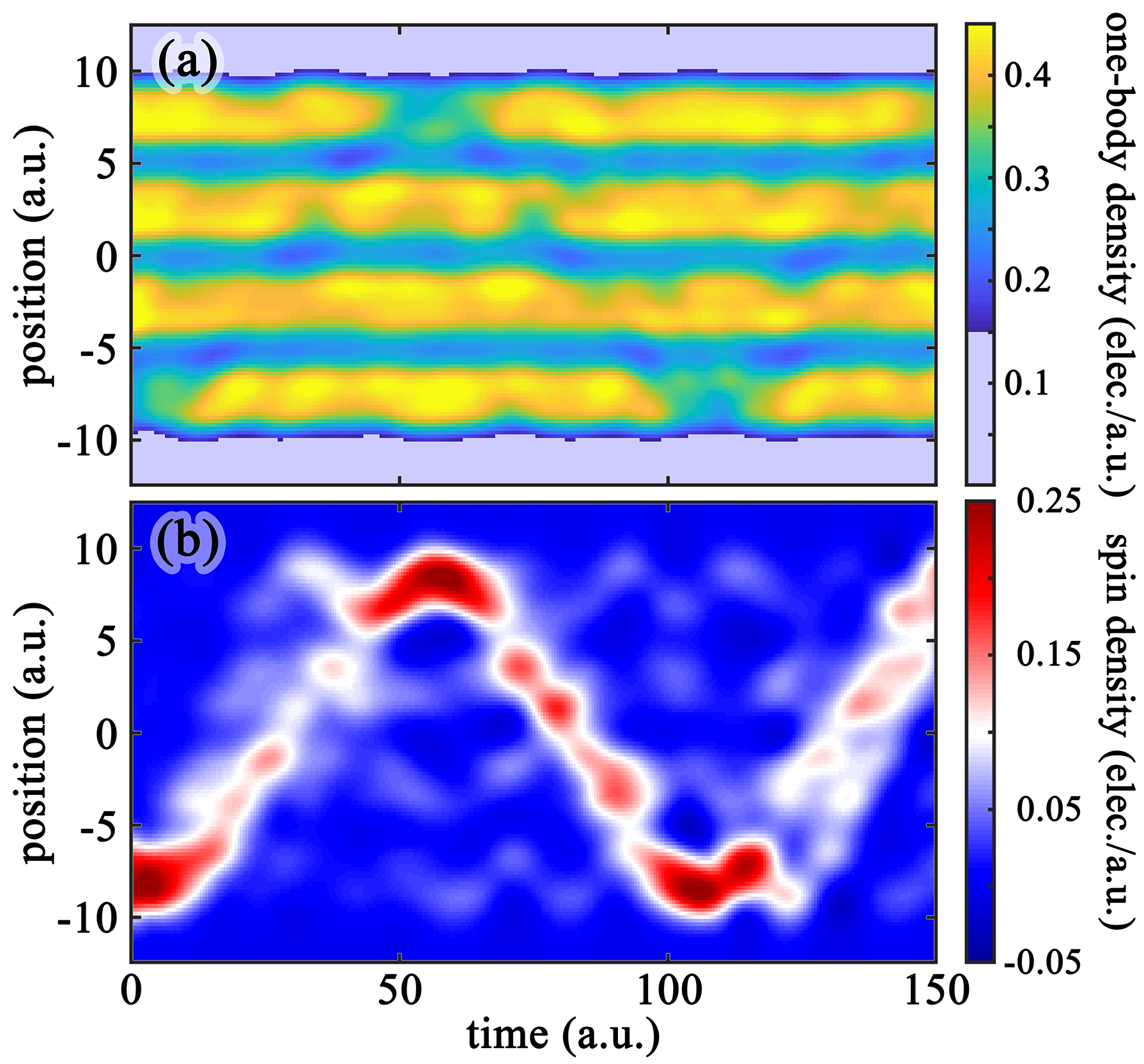}
    \caption{\label{fig:TDDFT_dynamic_example}
    TDDFT dynamic we use to evaluate the accuracy and efficacy of the symplectic split-operator schemes we discuss in the Paper. Panel (a) shows the total one-body density $\rho^\uparrow+\rho^\downarrow$ and (b) the spin density $\rho^\uparrow-\rho^\downarrow$.}
\end{figure}

In the cation -- {\it i.e.}, having a total of 7 active electrons in the (TD)DFT model -- we initialized a dynamics far from the ground-state equilibrium by forcing a one-electron hole to be localized on one end of the chain. The subsequent field-free TDDFT dynamics results in a hole that remains localized while periodically traveling back and forth along the molecular backbone.
This is clearly visible in Fig.~\ref{fig:TDDFT_dynamic_example}, in both the (a) one-body density $\rho^\uparrow+\rho^\downarrow$ and (b) spin density $\rho^\uparrow-\rho^\downarrow$ dynamics.
For laser-driven simulations, we use the same molecular model and initial condition and simply add a strong laser field corresponding to the vector potential
$$
    A(t) = 0.07 \times \sin(0.1 t) \times \sin(0.005\pi t)^2.
$$
We select this field such that (i) it visibly alter the KS-orbitals and one-body density dynamics, as well as the dipole-acceleration signal but (ii) without overwhelming them, such that the computation of propagation errors remain meaningful.
We have checked that length- and velocity-gauge simulations yield consistent results with the change of variables of Eq.~\eqref{eq:length_vs_velocity-gauge_change_of_variables}.

In all simulation results, we propagate the TDDFT dynamics for a total of 150~a.u. ($\approx$3.6~f.s.).
For the first 60~a.u.\ we solely solve for the KS orbitals' dynamics of Eq.~\eqref{eq:TDDFT} and record the time elapsed doing so. This is the time we quote in all efficacy plots.
Between 60 and 150~a.u.\ we restart the TDDFT dynamics (from the 60~a.u. configuration) but this time also compute the dipole acceleration of Eq.~\eqref{eq:dipole_acceleration}.
At the end of the simulations, we define three types of errors:
(i)~the $L^2$-norm error on the KS orbitals
\begin{equation} \label{eq:KSO_error}
    {\rm Err}(\{\phi_k\}_k) = \sum_{k}{\sqrt{\int{\left|
        \phi_k\left({\bf x}\right) - 
        \phi_k^{\rm ref}\left({\bf x}\right)
    \right|^2\ {\rm d} {\bf x} }}},
\end{equation}
where the sum runs over both the up- and down-spin channels and the error is computed at time $t=150$; (ii)~the $L^2$-norm error on the one-body density
\begin{equation} \label{eq:density_error}
    {\rm Err}(\rho) = \sqrt{\int{\left(
        \rho\left({\bf x}\right) - \rho^{\rm ref}\left({\bf x}\right)
    \right)^2\ {\rm d} {\bf x} }},
\end{equation}
again taken at $t=150$; and (iii)~the $L^2$-norm error on the dipole/velocity/acceleration signal
\begin{equation} \label{eq:dipole-acceleration_error}
    {\rm Err}(\ddot{\bf d}) = \sqrt{\int_{60}^{150}{\left\Vert
        \ddot{\bf d}\left(t\right) - \ddot{\bf d}^{\rm ref}\left(t\right)
    \right\Vert^2\ {\rm d}t }}.
\end{equation}
In all errors, the ``${\rm ref}$'' corresponds to the result of a TDDFT simulation, with the same scheme, and a time step of $5\times10^{-3}$~a.u.\ ($1/2$ of the smallest time step reported in the figures).
For the KS orbitals and density error, we restrict the spatial integrals to $-200\leq x\leq200$~a.u.\ to reduce boundary effects. Integrating over the entire simulation domain leads to similar results with slightly worse accuracy/efficacy (most apparent for the two BM schemes).
We also have compared and checked that we obtain the same TDDFT dynamics between different schemes -- withing the numerical error associated with each scheme -- and choices of gauges.
\section{Ehrenfest TDDFT} \label{app:Ehrenfest_TDDFT}

Ehrenfest TDDFT describes the interaction between the electronic and nuclear degrees of freedom with a semi-classical treatment of nuclear motion: nuclei are treated classically and interact with the (quantum) electrons through the mean-field one-body density.
To reveal the Hamiltonian structure of Ehrenfest TDDFT, we first expand the external (molecular) potential as the sum of its atomic constituents
\begin{equation} \label{eq:Ehrenfest_molecular_potential}
    \mathcal{V}_{\rm ext}[\{{\bf r}_k\}_k]({\bf x}) = 
        \sum_{k=1}^{M}{\mathcal{V}_{k}^{a}({\bf x} - {\bf r}_k)},
\end{equation}
where $\mathcal{V}_k^a$ is the $k^{\rm th}$ (effective) atomic potential with position ${\bf r}_k\in\mathbb{R}^d$ and $M$ is the number of atomic centers in the molecule. The corresponding external-energy functional is unchanged from Eq.~\eqref{eq:external-energy_functional}.
Next, the interaction between the atomic centers is given by the nuclear potential-energy functional
\begin{equation} \label{eq:Ehrenfest_nuclear_potential}
    H_{\rm pot}^{\rm n} = \sum_{k<l}^{M}{\mathcal{V}_{kl}^{\rm n}\left({\bf r}_k - {\bf r}_l\right)},
\end{equation}
where $\mathcal{V}_{kl}^{\rm n}$ is the interaction between the $k^{\rm th}$ and $l^{\rm th}$ atomic centers, generally taken as a scaled potential
$
    \mathcal{V}_{kl}^{\rm n}({\bf r}) = Z_k Z_l \mathcal{V}_{\rm n} (|{\bf r}|),
$
with $Z$ the (effective) bare charge of each atomic center.
Finally, the kinetic energy of the nuclear degrees of freedom is
\begin{equation} \label{eq:Ehrenfest_nuclear_kinetic}
    H_{\rm kin}^{\rm n} = \sum_{k=1}^{M}{\frac{|{\bf p}_k|^2}{2 m_k}},
\end{equation}
where $m_k$ and ${\bf p}_k$ are respectively the mass and canonical momentum of the $k^{\rm th}$ atomic center.

All put together, the Ehrenfest Hamiltonian functional is
\begin{equation} \label{eq:Ehrenfest_Hamiltonian_functional}
    H_{\rm E} = 
        H_{\rm kin}^{\rm n} + H_{\rm pot}^{\rm n} +
        H_{\rm ext} +
        H_{\rm kin} + H_{\rm H} + H_{\rm XC},
\end{equation}
with the extended canonical Poisson bracket
\begin{equation} \label{eq:Ehrenfest_Poisson_bracket}
    \left\{F,G\right\}_{\rm E} = \sum_{k=1}^{M}{
        \frac{\partial F}{\partial {\bf p}_k} \cdot \frac{\partial G}{\partial {\bf r}_k} - \frac{\partial F}{\partial {\bf r}_k} \cdot \frac{\partial G}{\partial {\bf p}_k}
    } + \left\{F,G\right\},
\end{equation}
for all functionals $F$ and $G$ and where $\{\cdot,\cdot\}$ is the bracket of Eq.~\eqref{eq:Poisson_bracket_complex}.
With this, we recover the dynamics for the $k^{\rm th}$ nuclear coordinate
\begin{equation} \label{eq:Ehrenfest_TDDFT_nuclear_dynamics}
    \dot{\bf r}_k = \frac{{\bf p}_k}{m_k}
    \quad {\rm and} \quad
    \dot{\bf p}_k = -\int{\nabla_k\mathcal{V}_{\rm ext} ({\bf x}) \rho({\bf x})\ {\rm d}{\bf x}} - \nabla_k H_{\rm pot}^{\rm n},
\end{equation}
where $\nabla_k$ is the gradient component along the nuclear coordinate ${\bf r}_k$.
The first two terms in Eq.~\eqref{eq:Ehrenfest_Hamiltonian_functional} describe the purely classical nuclear component; the third term contains the semi-classical interaction between the electronic and nuclear degrees of freedom; the final three terms are the quantum electronic components of Ehrenfest TDDFT, respectively given by Eqs.~\eqref{eq:kinetic-energy_functional} and~\eqref{eq:Hartree-energy_functional}, and discussed in section~\ref{sec:exchange-correlation-energy_functional}.
Here as well, the Hamiltonian structure of Ehrenfest TDDFT may guide the design of symplectic schemes for accurate and efficient propagation of the coupled electron-nucleus dynamics.



\begin{thebibliography}{54}%
\makeatletter
\providecommand \@ifxundefined [1]{%
 \@ifx{#1\undefined}
}%
\providecommand \@ifnum [1]{%
 \ifnum #1\expandafter \@firstoftwo
 \else \expandafter \@secondoftwo
 \fi
}%
\providecommand \@ifx [1]{%
 \ifx #1\expandafter \@firstoftwo
 \else \expandafter \@secondoftwo
 \fi
}%
\providecommand \natexlab [1]{#1}%
\providecommand \enquote  [1]{``#1''}%
\providecommand \bibnamefont  [1]{#1}%
\providecommand \bibfnamefont [1]{#1}%
\providecommand \citenamefont [1]{#1}%
\providecommand \href@noop [0]{\@secondoftwo}%
\providecommand \href [0]{\begingroup \@sanitize@url \@href}%
\providecommand \@href[1]{\@@startlink{#1}\@@href}%
\providecommand \@@href[1]{\endgroup#1\@@endlink}%
\providecommand \@sanitize@url [0]{\catcode `\\12\catcode `\$12\catcode
  `\&12\catcode `\#12\catcode `\^12\catcode `\_12\catcode `\%12\relax}%
\providecommand \@@startlink[1]{}%
\providecommand \@@endlink[0]{}%
\providecommand \url  [0]{\begingroup\@sanitize@url \@url }%
\providecommand \@url [1]{\endgroup\@href {#1}{\urlprefix }}%
\providecommand \urlprefix  [0]{URL }%
\providecommand \Eprint [0]{\href }%
\providecommand \doibase [0]{https://doi.org/}%
\providecommand \selectlanguage [0]{\@gobble}%
\providecommand \bibinfo  [0]{\@secondoftwo}%
\providecommand \bibfield  [0]{\@secondoftwo}%
\providecommand \translation [1]{[#1]}%
\providecommand \BibitemOpen [0]{}%
\providecommand \bibitemStop [0]{}%
\providecommand \bibitemNoStop [0]{.\EOS\space}%
\providecommand \EOS [0]{\spacefactor3000\relax}%
\providecommand \BibitemShut  [1]{\csname bibitem#1\endcsname}%
\let\auto@bib@innerbib\@empty
\bibitem [{\citenamefont {Marques}\ \emph {et~al.}(2006)\citenamefont
  {Marques}, \citenamefont {Ullrich}, \citenamefont {Nogueira}, \citenamefont
  {Rubio}, \citenamefont {Burke},\ and\ \citenamefont
  {Gross}}]{Marques_2006_book}%
  \BibitemOpen
  \bibinfo {editor} {\bibfnamefont {M.~A.}\ \bibnamefont {Marques}}, \bibinfo
  {editor} {\bibfnamefont {C.~A.}\ \bibnamefont {Ullrich}}, \bibinfo {editor}
  {\bibfnamefont {F.}~\bibnamefont {Nogueira}}, \bibinfo {editor}
  {\bibfnamefont {A.}~\bibnamefont {Rubio}}, \bibinfo {editor} {\bibfnamefont
  {K.}~\bibnamefont {Burke}},\ and\ \bibinfo {editor} {\bibfnamefont
  {E.~K.~U.}\ \bibnamefont {Gross}},\ eds.,\ \href@noop {} {\emph {\bibinfo
  {title} {Time-Dependent Density Functional Theory}}},\ Lecture Notes in
  Physics\ (\bibinfo  {publisher} {Springer Berlin},\ \bibinfo {address}
  {Heidelberg},\ \bibinfo {year} {2006})\BibitemShut {NoStop}%
\bibitem [{\citenamefont {Marques}\ \emph
  {et~al.}(2012{\natexlab{a}})\citenamefont {Marques}, \citenamefont {Maitra},
  \citenamefont {Nogueira}, \citenamefont {Gross},\ and\ \citenamefont
  {Rubio}}]{Marques_2012_book}%
  \BibitemOpen
  \bibinfo {editor} {\bibfnamefont {M.~A.}\ \bibnamefont {Marques}}, \bibinfo
  {editor} {\bibfnamefont {N.~T.}\ \bibnamefont {Maitra}}, \bibinfo {editor}
  {\bibfnamefont {F.~M.}\ \bibnamefont {Nogueira}}, \bibinfo {editor}
  {\bibfnamefont {E.}~\bibnamefont {Gross}},\ and\ \bibinfo {editor}
  {\bibfnamefont {A.}~\bibnamefont {Rubio}},\ eds.,\ \href@noop {} {\emph
  {\bibinfo {title} {Fundamentals of Time-Dependent Density Functional
  Theory}}},\ Lecture Notes in Physics\ (\bibinfo  {publisher} {Springer
  Berlin},\ \bibinfo {address} {Heidelberg},\ \bibinfo {year}
  {2012})\BibitemShut {NoStop}%
\bibitem [{\citenamefont {Maitra}(2016)}]{Maitra_2016}%
  \BibitemOpen
  \bibfield  {author} {\bibinfo {author} {\bibfnamefont {N.~T.}\ \bibnamefont
  {Maitra}},\ }\bibfield  {title} {\bibinfo {title} {Perspective: Fundamental
  aspects of time-dependent density functional theory},\ }\href@noop {}
  {\bibfield  {journal} {\bibinfo  {journal} {J. Chem. Phys.}\ }\textbf
  {\bibinfo {volume} {144}},\ \bibinfo {pages} {220901} (\bibinfo {year}
  {2016})}\BibitemShut {NoStop}%
\bibitem [{\citenamefont {Apr\`{a}}\ \emph {et~al.}(2020)\citenamefont
  {Apr\`{a}}, \citenamefont {Bylaska}, \citenamefont {de~Jong}, \citenamefont
  {Govind}, \citenamefont {Kowalski}, \citenamefont {Straatsma}, \citenamefont
  {Valiev}, \citenamefont {van Dam}, \citenamefont {Alexeev}, \citenamefont
  {Anchell}, \citenamefont {Anisimov}, \citenamefont {Aquino}, \citenamefont
  {Atta-Fynn}, \citenamefont {Autschbach}, \citenamefont {Bauman},
  \citenamefont {Becca}, \citenamefont {Bernholdt}, \citenamefont
  {Bhaskaran-Nair}, \citenamefont {Bogatko}, \citenamefont {Borowski},
  \citenamefont {Boschen}, \citenamefont {Brabec}, \citenamefont {Bruner},
  \citenamefont {Cau\"{e}t}, \citenamefont {Chen}, \citenamefont {Chuev},
  \citenamefont {Cramer}, \citenamefont {Daily}, \citenamefont {Deegan},
  \citenamefont {Dunning}, \citenamefont {Dupuis}, \citenamefont {Dyall},
  \citenamefont {Fann}, \citenamefont {Fischer}, \citenamefont {Fonari},
  \citenamefont {Fr\"{u}chtl}, \citenamefont {Gagliardi}, \citenamefont
  {Garza}, \citenamefont {Gawande}, \citenamefont {Ghosh}, \citenamefont
  {Glaesemann}, \citenamefont {G\"{o}tz}, \citenamefont {Hammond},
  \citenamefont {Helms}, \citenamefont {Hermes}, \citenamefont {Hirao},
  \citenamefont {Hirata}, \citenamefont {Jacquelin}, \citenamefont {Jensen},
  \citenamefont {Johnson}, \citenamefont {J\'{o}nsson}, \citenamefont
  {Kendall}, \citenamefont {Klemm}, \citenamefont {Kobayashi}, \citenamefont
  {Konkov}, \citenamefont {Krishnamoorthy}, \citenamefont {Krishnan},
  \citenamefont {Lin}, \citenamefont {Lins}, \citenamefont {Littlefield},
  \citenamefont {Logsdail}, \citenamefont {Lopata}, \citenamefont {Ma},
  \citenamefont {Marenich}, \citenamefont {Martin~del Campo}, \citenamefont
  {Mejia-Rodriguez}, \citenamefont {Moore}, \citenamefont {Mullin},
  \citenamefont {Nakajima}, \citenamefont {Nascimento}, \citenamefont
  {Nichols}, \citenamefont {Nichols}, \citenamefont {Nieplocha}, \citenamefont
  {Otero-de-la Roza}, \citenamefont {Palmer}, \citenamefont {Panyala},
  \citenamefont {Pirojsirikul}, \citenamefont {Peng}, \citenamefont {Peverati},
  \citenamefont {Pittner}, \citenamefont {Pollack}, \citenamefont {Richard},
  \citenamefont {Sadayappan}, \citenamefont {Schatz}, \citenamefont {Shelton},
  \citenamefont {Silverstein}, \citenamefont {Smith}, \citenamefont {Soares},
  \citenamefont {Song}, \citenamefont {Swart}, \citenamefont {Taylor},
  \citenamefont {Thomas}, \citenamefont {Tipparaju}, \citenamefont {Truhlar},
  \citenamefont {Tsemekhman}, \citenamefont {Van~Voorhis}, \citenamefont
  {V\'{a}zquez-Mayagoitia}, \citenamefont {Verma}, \citenamefont {Villa},
  \citenamefont {Vishnu}, \citenamefont {Vogiatzis}, \citenamefont {Wang},
  \citenamefont {Weare}, \citenamefont {Williamson}, \citenamefont {Windus},
  \citenamefont {Woli\'{n}ski}, \citenamefont {Wong}, \citenamefont {Wu},
  \citenamefont {Yang}, \citenamefont {Yu}, \citenamefont {Zacharias},
  \citenamefont {Zhang}, \citenamefont {Zhao},\ and\ \citenamefont
  {Harrison}}]{Arpa_2020}%
  \BibitemOpen
  \bibfield  {author} {\bibinfo {author} {\bibfnamefont {E.}~\bibnamefont
  {Apr\`{a}}}, \bibinfo {author} {\bibfnamefont {E.}~\bibnamefont {Bylaska}},
  \bibinfo {author} {\bibfnamefont {W.}~\bibnamefont {de~Jong}}, \bibinfo
  {author} {\bibfnamefont {N.}~\bibnamefont {Govind}}, \bibinfo {author}
  {\bibfnamefont {K.}~\bibnamefont {Kowalski}}, \bibinfo {author}
  {\bibfnamefont {T.}~\bibnamefont {Straatsma}}, \bibinfo {author}
  {\bibfnamefont {M.}~\bibnamefont {Valiev}}, \bibinfo {author} {\bibfnamefont
  {H.}~\bibnamefont {van Dam}}, \bibinfo {author} {\bibfnamefont
  {Y.}~\bibnamefont {Alexeev}}, \bibinfo {author} {\bibfnamefont
  {J.}~\bibnamefont {Anchell}}, \bibinfo {author} {\bibfnamefont
  {V.}~\bibnamefont {Anisimov}}, \bibinfo {author} {\bibfnamefont
  {F.}~\bibnamefont {Aquino}}, \bibinfo {author} {\bibfnamefont
  {R.}~\bibnamefont {Atta-Fynn}}, \bibinfo {author} {\bibfnamefont
  {J.}~\bibnamefont {Autschbach}}, \bibinfo {author} {\bibfnamefont
  {N.}~\bibnamefont {Bauman}}, \bibinfo {author} {\bibfnamefont
  {J.}~\bibnamefont {Becca}}, \bibinfo {author} {\bibfnamefont
  {D.}~\bibnamefont {Bernholdt}}, \bibinfo {author} {\bibfnamefont
  {K.}~\bibnamefont {Bhaskaran-Nair}}, \bibinfo {author} {\bibfnamefont
  {S.}~\bibnamefont {Bogatko}}, \bibinfo {author} {\bibfnamefont
  {P.}~\bibnamefont {Borowski}}, \bibinfo {author} {\bibfnamefont
  {J.}~\bibnamefont {Boschen}}, \bibinfo {author} {\bibfnamefont
  {J.}~\bibnamefont {Brabec}}, \bibinfo {author} {\bibfnamefont
  {A.}~\bibnamefont {Bruner}}, \bibinfo {author} {\bibfnamefont
  {E.}~\bibnamefont {Cau\"{e}t}}, \bibinfo {author} {\bibfnamefont
  {Y.}~\bibnamefont {Chen}}, \bibinfo {author} {\bibfnamefont {G.}~\bibnamefont
  {Chuev}}, \bibinfo {author} {\bibfnamefont {C.}~\bibnamefont {Cramer}},
  \bibinfo {author} {\bibfnamefont {J.}~\bibnamefont {Daily}}, \bibinfo
  {author} {\bibfnamefont {M.}~\bibnamefont {Deegan}}, \bibinfo {author}
  {\bibfnamefont {T.}~\bibnamefont {Dunning}}, \bibinfo {author} {\bibfnamefont
  {M.}~\bibnamefont {Dupuis}}, \bibinfo {author} {\bibfnamefont
  {K.}~\bibnamefont {Dyall}}, \bibinfo {author} {\bibfnamefont
  {G.}~\bibnamefont {Fann}}, \bibinfo {author} {\bibfnamefont {S.}~\bibnamefont
  {Fischer}}, \bibinfo {author} {\bibfnamefont {A.}~\bibnamefont {Fonari}},
  \bibinfo {author} {\bibfnamefont {H.}~\bibnamefont {Fr\"{u}chtl}}, \bibinfo
  {author} {\bibfnamefont {L.}~\bibnamefont {Gagliardi}}, \bibinfo {author}
  {\bibfnamefont {J.}~\bibnamefont {Garza}}, \bibinfo {author} {\bibfnamefont
  {N.}~\bibnamefont {Gawande}}, \bibinfo {author} {\bibfnamefont
  {S.}~\bibnamefont {Ghosh}}, \bibinfo {author} {\bibfnamefont
  {K.}~\bibnamefont {Glaesemann}}, \bibinfo {author} {\bibfnamefont
  {A.}~\bibnamefont {G\"{o}tz}}, \bibinfo {author} {\bibfnamefont
  {J.}~\bibnamefont {Hammond}}, \bibinfo {author} {\bibfnamefont
  {V.}~\bibnamefont {Helms}}, \bibinfo {author} {\bibfnamefont
  {E.}~\bibnamefont {Hermes}}, \bibinfo {author} {\bibfnamefont
  {K.}~\bibnamefont {Hirao}}, \bibinfo {author} {\bibfnamefont
  {S.}~\bibnamefont {Hirata}}, \bibinfo {author} {\bibfnamefont
  {M.}~\bibnamefont {Jacquelin}}, \bibinfo {author} {\bibfnamefont
  {L.}~\bibnamefont {Jensen}}, \bibinfo {author} {\bibfnamefont
  {B.}~\bibnamefont {Johnson}}, \bibinfo {author} {\bibfnamefont
  {H.}~\bibnamefont {J\'{o}nsson}}, \bibinfo {author} {\bibfnamefont
  {R.}~\bibnamefont {Kendall}}, \bibinfo {author} {\bibfnamefont
  {M.}~\bibnamefont {Klemm}}, \bibinfo {author} {\bibfnamefont
  {R.}~\bibnamefont {Kobayashi}}, \bibinfo {author} {\bibfnamefont
  {V.}~\bibnamefont {Konkov}}, \bibinfo {author} {\bibfnamefont
  {S.}~\bibnamefont {Krishnamoorthy}}, \bibinfo {author} {\bibfnamefont
  {M.}~\bibnamefont {Krishnan}}, \bibinfo {author} {\bibfnamefont
  {Z.}~\bibnamefont {Lin}}, \bibinfo {author} {\bibfnamefont {R.}~\bibnamefont
  {Lins}}, \bibinfo {author} {\bibfnamefont {R.}~\bibnamefont {Littlefield}},
  \bibinfo {author} {\bibfnamefont {A.}~\bibnamefont {Logsdail}}, \bibinfo
  {author} {\bibfnamefont {K.}~\bibnamefont {Lopata}}, \bibinfo {author}
  {\bibfnamefont {W.}~\bibnamefont {Ma}}, \bibinfo {author} {\bibfnamefont
  {A.}~\bibnamefont {Marenich}}, \bibinfo {author} {\bibfnamefont
  {J.}~\bibnamefont {Martin~del Campo}}, \bibinfo {author} {\bibfnamefont
  {D.}~\bibnamefont {Mejia-Rodriguez}}, \bibinfo {author} {\bibfnamefont
  {J.}~\bibnamefont {Moore}}, \bibinfo {author} {\bibfnamefont
  {J.}~\bibnamefont {Mullin}}, \bibinfo {author} {\bibfnamefont
  {T.}~\bibnamefont {Nakajima}}, \bibinfo {author} {\bibfnamefont
  {D.}~\bibnamefont {Nascimento}}, \bibinfo {author} {\bibfnamefont
  {J.}~\bibnamefont {Nichols}}, \bibinfo {author} {\bibfnamefont
  {P.}~\bibnamefont {Nichols}}, \bibinfo {author} {\bibfnamefont
  {J.}~\bibnamefont {Nieplocha}}, \bibinfo {author} {\bibfnamefont
  {A.}~\bibnamefont {Otero-de-la Roza}}, \bibinfo {author} {\bibfnamefont
  {B.}~\bibnamefont {Palmer}}, \bibinfo {author} {\bibfnamefont
  {A.}~\bibnamefont {Panyala}}, \bibinfo {author} {\bibfnamefont
  {T.}~\bibnamefont {Pirojsirikul}}, \bibinfo {author} {\bibfnamefont
  {B.}~\bibnamefont {Peng}}, \bibinfo {author} {\bibfnamefont {R.}~\bibnamefont
  {Peverati}}, \bibinfo {author} {\bibfnamefont {J.}~\bibnamefont {Pittner}},
  \bibinfo {author} {\bibfnamefont {L.}~\bibnamefont {Pollack}}, \bibinfo
  {author} {\bibfnamefont {R.}~\bibnamefont {Richard}}, \bibinfo {author}
  {\bibfnamefont {P.}~\bibnamefont {Sadayappan}}, \bibinfo {author}
  {\bibfnamefont {G.}~\bibnamefont {Schatz}}, \bibinfo {author} {\bibfnamefont
  {W.}~\bibnamefont {Shelton}}, \bibinfo {author} {\bibfnamefont
  {D.}~\bibnamefont {Silverstein}}, \bibinfo {author} {\bibfnamefont
  {D.}~\bibnamefont {Smith}}, \bibinfo {author} {\bibfnamefont
  {T.}~\bibnamefont {Soares}}, \bibinfo {author} {\bibfnamefont
  {D.}~\bibnamefont {Song}}, \bibinfo {author} {\bibfnamefont {M.}~\bibnamefont
  {Swart}}, \bibinfo {author} {\bibfnamefont {H.}~\bibnamefont {Taylor}},
  \bibinfo {author} {\bibfnamefont {G.}~\bibnamefont {Thomas}}, \bibinfo
  {author} {\bibfnamefont {V.}~\bibnamefont {Tipparaju}}, \bibinfo {author}
  {\bibfnamefont {D.}~\bibnamefont {Truhlar}}, \bibinfo {author} {\bibfnamefont
  {K.}~\bibnamefont {Tsemekhman}}, \bibinfo {author} {\bibfnamefont
  {T.}~\bibnamefont {Van~Voorhis}}, \bibinfo {author} {\bibfnamefont
  {A.}~\bibnamefont {V\'{a}zquez-Mayagoitia}}, \bibinfo {author} {\bibfnamefont
  {P.}~\bibnamefont {Verma}}, \bibinfo {author} {\bibfnamefont
  {O.}~\bibnamefont {Villa}}, \bibinfo {author} {\bibfnamefont
  {A.}~\bibnamefont {Vishnu}}, \bibinfo {author} {\bibfnamefont
  {K.}~\bibnamefont {Vogiatzis}}, \bibinfo {author} {\bibfnamefont
  {D.}~\bibnamefont {Wang}}, \bibinfo {author} {\bibfnamefont {J.}~\bibnamefont
  {Weare}}, \bibinfo {author} {\bibfnamefont {M.}~\bibnamefont {Williamson}},
  \bibinfo {author} {\bibfnamefont {T.}~\bibnamefont {Windus}}, \bibinfo
  {author} {\bibfnamefont {K.}~\bibnamefont {Woli\'{n}ski}}, \bibinfo {author}
  {\bibfnamefont {A.}~\bibnamefont {Wong}}, \bibinfo {author} {\bibfnamefont
  {Q.}~\bibnamefont {Wu}}, \bibinfo {author} {\bibfnamefont {C.}~\bibnamefont
  {Yang}}, \bibinfo {author} {\bibfnamefont {Q.}~\bibnamefont {Yu}}, \bibinfo
  {author} {\bibfnamefont {M.}~\bibnamefont {Zacharias}}, \bibinfo {author}
  {\bibfnamefont {Z.}~\bibnamefont {Zhang}}, \bibinfo {author} {\bibfnamefont
  {Y.}~\bibnamefont {Zhao}},\ and\ \bibinfo {author} {\bibfnamefont
  {R.}~\bibnamefont {Harrison}},\ }\bibfield  {title} {\bibinfo {title}
  {{NWC}hem: Past, present, and future},\ }\href@noop {} {\bibfield  {journal}
  {\bibinfo  {journal} {J. Chem. Phys.}\ }\textbf {\bibinfo {volume} {152}},\
  \bibinfo {pages} {184102} (\bibinfo {year} {2020})}\BibitemShut {NoStop}%
\bibitem [{\citenamefont {Tancogne-Dejean}\ \emph {et~al.}(2020)\citenamefont
  {Tancogne-Dejean}, \citenamefont {Oliveira}, \citenamefont {Andrade},
  \citenamefont {Appel}, \citenamefont {Borca}, \citenamefont {Le~Breton},
  \citenamefont {Buchholz}, \citenamefont {Castro}, \citenamefont {Corni},
  \citenamefont {Correa}, \citenamefont {De~Giovannini}, \citenamefont
  {Delgado}, \citenamefont {Eich}, \citenamefont {Flick}, \citenamefont {Gil},
  \citenamefont {Gomez}, \citenamefont {Helbig}, \citenamefont {H\"{u}bener},
  \citenamefont {Jest\"{a}dt}, \citenamefont {Jornet-Somoza}, \citenamefont
  {Larsen}, \citenamefont {Lebedeva}, \citenamefont {L\"{u}ders}, \citenamefont
  {Marques}, \citenamefont {Ohlmann}, \citenamefont {Pipolo}, \citenamefont
  {Rampp}, \citenamefont {Rozzi}, \citenamefont {Strubbe}, \citenamefont
  {Sato}, \citenamefont {Sch\"{a}fer}, \citenamefont {Theophilou},
  \citenamefont {Welden},\ and\ \citenamefont {Rubio}}]{Tancogne-Dejean_2020}%
  \BibitemOpen
  \bibfield  {author} {\bibinfo {author} {\bibfnamefont {N.}~\bibnamefont
  {Tancogne-Dejean}}, \bibinfo {author} {\bibfnamefont {M.~J.}\ \bibnamefont
  {Oliveira}}, \bibinfo {author} {\bibfnamefont {X.}~\bibnamefont {Andrade}},
  \bibinfo {author} {\bibfnamefont {H.}~\bibnamefont {Appel}}, \bibinfo
  {author} {\bibfnamefont {C.~H.}\ \bibnamefont {Borca}}, \bibinfo {author}
  {\bibfnamefont {G.}~\bibnamefont {Le~Breton}}, \bibinfo {author}
  {\bibfnamefont {F.}~\bibnamefont {Buchholz}}, \bibinfo {author}
  {\bibfnamefont {A.}~\bibnamefont {Castro}}, \bibinfo {author} {\bibfnamefont
  {S.}~\bibnamefont {Corni}}, \bibinfo {author} {\bibfnamefont {A.~A.}\
  \bibnamefont {Correa}}, \bibinfo {author} {\bibfnamefont {U.}~\bibnamefont
  {De~Giovannini}}, \bibinfo {author} {\bibfnamefont {A.}~\bibnamefont
  {Delgado}}, \bibinfo {author} {\bibfnamefont {F.~G.}\ \bibnamefont {Eich}},
  \bibinfo {author} {\bibfnamefont {J.}~\bibnamefont {Flick}}, \bibinfo
  {author} {\bibfnamefont {G.}~\bibnamefont {Gil}}, \bibinfo {author}
  {\bibfnamefont {A.}~\bibnamefont {Gomez}}, \bibinfo {author} {\bibfnamefont
  {N.}~\bibnamefont {Helbig}}, \bibinfo {author} {\bibfnamefont
  {H.}~\bibnamefont {H\"{u}bener}}, \bibinfo {author} {\bibfnamefont
  {R.}~\bibnamefont {Jest\"{a}dt}}, \bibinfo {author} {\bibfnamefont
  {J.}~\bibnamefont {Jornet-Somoza}}, \bibinfo {author} {\bibfnamefont {A.~H.}\
  \bibnamefont {Larsen}}, \bibinfo {author} {\bibfnamefont {I.~V.}\
  \bibnamefont {Lebedeva}}, \bibinfo {author} {\bibfnamefont {M.}~\bibnamefont
  {L\"{u}ders}}, \bibinfo {author} {\bibfnamefont {M.~A.}\ \bibnamefont
  {Marques}}, \bibinfo {author} {\bibfnamefont {S.~T.}\ \bibnamefont
  {Ohlmann}}, \bibinfo {author} {\bibfnamefont {S.}~\bibnamefont {Pipolo}},
  \bibinfo {author} {\bibfnamefont {M.}~\bibnamefont {Rampp}}, \bibinfo
  {author} {\bibfnamefont {C.~A.}\ \bibnamefont {Rozzi}}, \bibinfo {author}
  {\bibfnamefont {D.~A.}\ \bibnamefont {Strubbe}}, \bibinfo {author}
  {\bibfnamefont {S.~A.}\ \bibnamefont {Sato}}, \bibinfo {author}
  {\bibfnamefont {C.}~\bibnamefont {Sch\"{a}fer}}, \bibinfo {author}
  {\bibfnamefont {I.}~\bibnamefont {Theophilou}}, \bibinfo {author}
  {\bibfnamefont {A.}~\bibnamefont {Welden}},\ and\ \bibinfo {author}
  {\bibfnamefont {A.}~\bibnamefont {Rubio}},\ }\bibfield  {title} {\bibinfo
  {title} {Octopus, a computational framework for exploring light-driven
  phenomena and quantum dynamics in extended and finite systems},\ }\href@noop
  {} {\bibfield  {journal} {\bibinfo  {journal} {J. Chem. Phys.}\ }\textbf
  {\bibinfo {volume} {152}},\ \bibinfo {pages} {124119} (\bibinfo {year}
  {2020})}\BibitemShut {NoStop}%
\bibitem [{\citenamefont {Smith}\ \emph {et~al.}(2020)\citenamefont {Smith},
  \citenamefont {Burns}, \citenamefont {Simmonett}, \citenamefont {Parrish},
  \citenamefont {Schieber}, \citenamefont {Galvelis}, \citenamefont {Kraus},
  \citenamefont {Kruse}, \citenamefont {Di~Remigio}, \citenamefont {Alenaizan},
  \citenamefont {James}, \citenamefont {Lehtola}, \citenamefont {Misiewicz},
  \citenamefont {Scheurer}, \citenamefont {Shaw}, \citenamefont {Schriber},
  \citenamefont {Xie}, \citenamefont {Glick}, \citenamefont {Sirianni},
  \citenamefont {O’Brien}, \citenamefont {Waldrop}, \citenamefont {Kumar},
  \citenamefont {Hohenstein}, \citenamefont {Pritchard}, \citenamefont
  {Brooks}, \citenamefont {Schaefer}, \citenamefont {Sokolov}, \citenamefont
  {Patkowski}, \citenamefont {DePrince}, \citenamefont {Bozkaya}, \citenamefont
  {King}, \citenamefont {Evangelista}, \citenamefont {Turney}, \citenamefont
  {Crawford},\ and\ \citenamefont {Sherrill}}]{Smith_2022}%
  \BibitemOpen
  \bibfield  {author} {\bibinfo {author} {\bibfnamefont {D.~G.}\ \bibnamefont
  {Smith}}, \bibinfo {author} {\bibfnamefont {L.~A.}\ \bibnamefont {Burns}},
  \bibinfo {author} {\bibfnamefont {A.~C.}\ \bibnamefont {Simmonett}}, \bibinfo
  {author} {\bibfnamefont {R.~M.}\ \bibnamefont {Parrish}}, \bibinfo {author}
  {\bibfnamefont {M.~C.}\ \bibnamefont {Schieber}}, \bibinfo {author}
  {\bibfnamefont {R.}~\bibnamefont {Galvelis}}, \bibinfo {author}
  {\bibfnamefont {P.}~\bibnamefont {Kraus}}, \bibinfo {author} {\bibfnamefont
  {H.}~\bibnamefont {Kruse}}, \bibinfo {author} {\bibfnamefont
  {R.}~\bibnamefont {Di~Remigio}}, \bibinfo {author} {\bibfnamefont
  {A.}~\bibnamefont {Alenaizan}}, \bibinfo {author} {\bibfnamefont {A.~M.}\
  \bibnamefont {James}}, \bibinfo {author} {\bibfnamefont {S.}~\bibnamefont
  {Lehtola}}, \bibinfo {author} {\bibfnamefont {J.~P.}\ \bibnamefont
  {Misiewicz}}, \bibinfo {author} {\bibfnamefont {M.}~\bibnamefont {Scheurer}},
  \bibinfo {author} {\bibfnamefont {R.~A.}\ \bibnamefont {Shaw}}, \bibinfo
  {author} {\bibfnamefont {J.~B.}\ \bibnamefont {Schriber}}, \bibinfo {author}
  {\bibfnamefont {Y.}~\bibnamefont {Xie}}, \bibinfo {author} {\bibfnamefont
  {Z.~L.}\ \bibnamefont {Glick}}, \bibinfo {author} {\bibfnamefont {D.~A.}\
  \bibnamefont {Sirianni}}, \bibinfo {author} {\bibfnamefont {J.~S.}\
  \bibnamefont {O’Brien}}, \bibinfo {author} {\bibfnamefont {J.~M.}\
  \bibnamefont {Waldrop}}, \bibinfo {author} {\bibfnamefont {A.}~\bibnamefont
  {Kumar}}, \bibinfo {author} {\bibfnamefont {E.~G.}\ \bibnamefont
  {Hohenstein}}, \bibinfo {author} {\bibfnamefont {B.~P.}\ \bibnamefont
  {Pritchard}}, \bibinfo {author} {\bibfnamefont {B.~R.}\ \bibnamefont
  {Brooks}}, \bibinfo {author} {\bibfnamefont {H.~F.}\ \bibnamefont
  {Schaefer}}, \bibinfo {author} {\bibfnamefont {A.~Y.}\ \bibnamefont
  {Sokolov}}, \bibinfo {author} {\bibfnamefont {K.}~\bibnamefont {Patkowski}},
  \bibinfo {author} {\bibfnamefont {A.~E.}\ \bibnamefont {DePrince}}, \bibinfo
  {author} {\bibfnamefont {U.}~\bibnamefont {Bozkaya}}, \bibinfo {author}
  {\bibfnamefont {R.~A.}\ \bibnamefont {King}}, \bibinfo {author}
  {\bibfnamefont {F.~A.}\ \bibnamefont {Evangelista}}, \bibinfo {author}
  {\bibfnamefont {J.~M.}\ \bibnamefont {Turney}}, \bibinfo {author}
  {\bibfnamefont {T.~D.}\ \bibnamefont {Crawford}},\ and\ \bibinfo {author}
  {\bibfnamefont {C.~D.}\ \bibnamefont {Sherrill}},\ }\bibfield  {title}
  {\bibinfo {title} {{PSI}4 1.4: Open-source software for high-throughput
  quantum chemistry},\ }\href@noop {} {\bibfield  {journal} {\bibinfo
  {journal} {J. Chem. Phys.}\ }\textbf {\bibinfo {volume} {152}},\ \bibinfo
  {pages} {184108} (\bibinfo {year} {2020})}\BibitemShut {NoStop}%
\bibitem [{\citenamefont {Castro}\ \emph {et~al.}(2004)\citenamefont {Castro},
  \citenamefont {Marques},\ and\ \citenamefont {Rubio}}]{Castro_2004}%
  \BibitemOpen
  \bibfield  {author} {\bibinfo {author} {\bibfnamefont {A.}~\bibnamefont
  {Castro}}, \bibinfo {author} {\bibfnamefont {M.~A.}\ \bibnamefont
  {Marques}},\ and\ \bibinfo {author} {\bibfnamefont {A.}~\bibnamefont
  {Rubio}},\ }\bibfield  {title} {\bibinfo {title} {Propagators for the
  time-dependent {K}ohn–{S}ham equations},\ }\href@noop {} {\bibfield
  {journal} {\bibinfo  {journal} {J. Chem. Phys.}\ }\textbf {\bibinfo {volume}
  {121}},\ \bibinfo {pages} {3425} (\bibinfo {year} {2004})}\BibitemShut
  {NoStop}%
\bibitem [{\citenamefont {G\'{o}mez~Pueyo}\ \emph {et~al.}(2018)\citenamefont
  {G\'{o}mez~Pueyo}, \citenamefont {Marques}, \citenamefont {Rubio},\ and\
  \citenamefont {Castro}}]{Gomez_Pueyo_2018}%
  \BibitemOpen
  \bibfield  {author} {\bibinfo {author} {\bibfnamefont {A.}~\bibnamefont
  {G\'{o}mez~Pueyo}}, \bibinfo {author} {\bibfnamefont {M.~A.}\ \bibnamefont
  {Marques}}, \bibinfo {author} {\bibfnamefont {A.}~\bibnamefont {Rubio}},\
  and\ \bibinfo {author} {\bibfnamefont {A.}~\bibnamefont {Castro}},\
  }\bibfield  {title} {\bibinfo {title} {Propagators for the time-dependent
  {K}ohn-{S}ham equations: Multistep, {R}unge-{K}utta, exponential
  {R}unge-{K}utta, and commutator free {M}agnus methods},\ }\href@noop {}
  {\bibfield  {journal} {\bibinfo  {journal} {J. Chem. Theory Comput.}\
  }\textbf {\bibinfo {volume} {14}},\ \bibinfo {pages} {3040} (\bibinfo {year}
  {2018})}\BibitemShut {NoStop}%
\bibitem [{\citenamefont {Moler}\ and\ \citenamefont
  {Van~Loan}(2003)}]{Moler_2003}%
  \BibitemOpen
  \bibfield  {author} {\bibinfo {author} {\bibfnamefont {C.}~\bibnamefont
  {Moler}}\ and\ \bibinfo {author} {\bibfnamefont {C.}~\bibnamefont
  {Van~Loan}},\ }\bibfield  {title} {\bibinfo {title} {Nineteen dubious ways to
  compute the exponential of a matrix, twenty-five years later},\ }\href@noop
  {} {\bibfield  {journal} {\bibinfo  {journal} {SIAM Review}\ }\textbf
  {\bibinfo {volume} {45}},\ \bibinfo {pages} {3} (\bibinfo {year}
  {2003})}\BibitemShut {NoStop}%
\bibitem [{\citenamefont {G\'{o}mez~Pueyo}\ \emph
  {et~al.}(2020{\natexlab{a}})\citenamefont {G\'{o}mez~Pueyo}, \citenamefont
  {Blanes},\ and\ \citenamefont {Castro}}]{Gomez_Pueyo_2020}%
  \BibitemOpen
  \bibfield  {author} {\bibinfo {author} {\bibfnamefont {A.}~\bibnamefont
  {G\'{o}mez~Pueyo}}, \bibinfo {author} {\bibfnamefont {S.}~\bibnamefont
  {Blanes}},\ and\ \bibinfo {author} {\bibfnamefont {A.}~\bibnamefont
  {Castro}},\ }\bibfield  {title} {\bibinfo {title} {Propagators for
  quantum-classical models: Commutator-free {M}agnus methods},\ }\href@noop {}
  {\bibfield  {journal} {\bibinfo  {journal} {J. Chem. Theory Comput.}\
  }\textbf {\bibinfo {volume} {16}},\ \bibinfo {pages} {1420} (\bibinfo {year}
  {2020}{\natexlab{a}})}\BibitemShut {NoStop}%
\bibitem [{\citenamefont {Kohn}\ and\ \citenamefont {Sham}(1965)}]{Kohn_1965}%
  \BibitemOpen
  \bibfield  {author} {\bibinfo {author} {\bibfnamefont {W.}~\bibnamefont
  {Kohn}}\ and\ \bibinfo {author} {\bibfnamefont {L.}~\bibnamefont {Sham}},\
  }\bibfield  {title} {\bibinfo {title} {Self-consistent equations including
  exchange and correlation effects},\ }\href@noop {} {\bibfield  {journal}
  {\bibinfo  {journal} {Phys.~Rev.}\ }\textbf {\bibinfo {volume} {140}},\
  \bibinfo {pages} {A1133} (\bibinfo {year} {1965})}\BibitemShut {NoStop}%
\bibitem [{\citenamefont {Runge}\ and\ \citenamefont
  {Gross}(1984)}]{Runge_1984}%
  \BibitemOpen
  \bibfield  {author} {\bibinfo {author} {\bibfnamefont {E.}~\bibnamefont
  {Runge}}\ and\ \bibinfo {author} {\bibfnamefont {E.~K.~U.}\ \bibnamefont
  {Gross}},\ }\bibfield  {title} {\bibinfo {title} {Density-functional theory
  for time-dependent systems},\ }\href@noop {} {\bibfield  {journal} {\bibinfo
  {journal} {Phys. Rev. Lett.}\ }\textbf {\bibinfo {volume} {52}},\ \bibinfo
  {pages} {997} (\bibinfo {year} {1984})}\BibitemShut {NoStop}%
\bibitem [{\citenamefont {Yoshida}(1990)}]{Yoshida_1990}%
  \BibitemOpen
  \bibfield  {author} {\bibinfo {author} {\bibfnamefont {H.}~\bibnamefont
  {Yoshida}},\ }\bibfield  {title} {\bibinfo {title} {Construction of higher
  order symplectic integrators},\ }\href@noop {} {\bibfield  {journal}
  {\bibinfo  {journal} {Phys. Lett. A}\ }\textbf {\bibinfo {volume} {150}},\
  \bibinfo {pages} {262} (\bibinfo {year} {1990})}\BibitemShut {NoStop}%
\bibitem [{\citenamefont {Bandrauk}\ and\ \citenamefont
  {Shen}(1992)}]{Bandrauk_1992}%
  \BibitemOpen
  \bibfield  {author} {\bibinfo {author} {\bibfnamefont {A.~D.}\ \bibnamefont
  {Bandrauk}}\ and\ \bibinfo {author} {\bibfnamefont {H.}~\bibnamefont
  {Shen}},\ }\bibfield  {title} {\bibinfo {title} {Higher order exponential
  split operator method for solving time-dependent schrödinger equations},\
  }\href@noop {} {\bibfield  {journal} {\bibinfo  {journal} {Can. J. Chem.}\
  }\textbf {\bibinfo {volume} {70}},\ \bibinfo {pages} {555} (\bibinfo {year}
  {1992})}\BibitemShut {NoStop}%
\bibitem [{\citenamefont {Blanes}\ and\ \citenamefont
  {Moan}(2002)}]{Blanes_2002}%
  \BibitemOpen
  \bibfield  {author} {\bibinfo {author} {\bibfnamefont {S.}~\bibnamefont
  {Blanes}}\ and\ \bibinfo {author} {\bibfnamefont {P.}~\bibnamefont {Moan}},\
  }\bibfield  {title} {\bibinfo {title} {Practical symplectic partitioned
  {R}unge-{K}utta and {R}unge-{K}utta-{N}ystr\"{o}m methods},\ }\href@noop {}
  {\bibfield  {journal} {\bibinfo  {journal} {J. Comp. Appl. Math.}\ }\textbf
  {\bibinfo {volume} {142}},\ \bibinfo {pages} {313} (\bibinfo {year}
  {2002})}\BibitemShut {NoStop}%
\bibitem [{\citenamefont {McLachlan}(2022)}]{McLachlan_2022}%
  \BibitemOpen
  \bibfield  {author} {\bibinfo {author} {\bibfnamefont {R.~I.}\ \bibnamefont
  {McLachlan}},\ }\bibfield  {title} {\bibinfo {title} {Tuning symplectic
  integrators is easy and worthwhile},\ }\href@noop {} {\bibfield  {journal}
  {\bibinfo  {journal} {Commun. Comput. Phys.}\ }\textbf {\bibinfo {volume}
  {31}},\ \bibinfo {pages} {987} (\bibinfo {year} {2022})}\BibitemShut
  {NoStop}%
\bibitem [{\citenamefont {Xiao}\ \emph {et~al.}(2015)\citenamefont {Xiao},
  \citenamefont {Qin}, \citenamefont {Liu}, \citenamefont {He}, \citenamefont
  {Zhang},\ and\ \citenamefont {Sun}}]{Xiao_2015}%
  \BibitemOpen
  \bibfield  {author} {\bibinfo {author} {\bibfnamefont {J.}~\bibnamefont
  {Xiao}}, \bibinfo {author} {\bibfnamefont {H.}~\bibnamefont {Qin}}, \bibinfo
  {author} {\bibfnamefont {J.}~\bibnamefont {Liu}}, \bibinfo {author}
  {\bibfnamefont {Y.}~\bibnamefont {He}}, \bibinfo {author} {\bibfnamefont
  {R.}~\bibnamefont {Zhang}},\ and\ \bibinfo {author} {\bibfnamefont
  {Y.}~\bibnamefont {Sun}},\ }\bibfield  {title} {\bibinfo {title} {Explicit
  high-order non-canonical symplectic particle-in-cell algorithms for
  vlasov-maxwell systems},\ }\href@noop {} {\bibfield  {journal} {\bibinfo
  {journal} {Phys. Plasmas}\ }\textbf {\bibinfo {volume} {22}},\ \bibinfo
  {pages} {112504} (\bibinfo {year} {2015})}\BibitemShut {NoStop}%
\bibitem [{\citenamefont {He}\ \emph {et~al.}(2015)\citenamefont {He},
  \citenamefont {Qin}, \citenamefont {Sun}, \citenamefont {Xiao}, \citenamefont
  {Zhang},\ and\ \citenamefont {Liu}}]{He_2015}%
  \BibitemOpen
  \bibfield  {author} {\bibinfo {author} {\bibfnamefont {Y.}~\bibnamefont
  {He}}, \bibinfo {author} {\bibfnamefont {H.}~\bibnamefont {Qin}}, \bibinfo
  {author} {\bibfnamefont {Y.}~\bibnamefont {Sun}}, \bibinfo {author}
  {\bibfnamefont {J.}~\bibnamefont {Xiao}}, \bibinfo {author} {\bibfnamefont
  {R.}~\bibnamefont {Zhang}},\ and\ \bibinfo {author} {\bibfnamefont
  {J.}~\bibnamefont {Liu}},\ }\bibfield  {title} {\bibinfo {title} {Hamiltonian
  time integrators for vlasov-maxwell equations},\ }\href@noop {} {\bibfield
  {journal} {\bibinfo  {journal} {Phys. Plasmas}\ }\textbf {\bibinfo {volume}
  {22}},\ \bibinfo {pages} {124503} (\bibinfo {year} {2015})}\BibitemShut
  {NoStop}%
\bibitem [{\citenamefont {Xiao}\ \emph {et~al.}(2016)\citenamefont {Xiao},
  \citenamefont {Qin}, \citenamefont {Morrison}, \citenamefont {Liu},
  \citenamefont {Yu}, \citenamefont {Zhang},\ and\ \citenamefont
  {He}}]{Xiao_2016}%
  \BibitemOpen
  \bibfield  {author} {\bibinfo {author} {\bibfnamefont {J.}~\bibnamefont
  {Xiao}}, \bibinfo {author} {\bibfnamefont {H.}~\bibnamefont {Qin}}, \bibinfo
  {author} {\bibfnamefont {P.~J.}\ \bibnamefont {Morrison}}, \bibinfo {author}
  {\bibfnamefont {J.}~\bibnamefont {Liu}}, \bibinfo {author} {\bibfnamefont
  {Z.}~\bibnamefont {Yu}}, \bibinfo {author} {\bibfnamefont {R.}~\bibnamefont
  {Zhang}},\ and\ \bibinfo {author} {\bibfnamefont {Y.}~\bibnamefont {He}},\
  }\bibfield  {title} {\bibinfo {title} {Explicit high-order noncanonical
  symplectic algorithms for ideal two-fluid systems},\ }\href@noop {}
  {\bibfield  {journal} {\bibinfo  {journal} {Phys. Plasmas}\ }\textbf
  {\bibinfo {volume} {23}},\ \bibinfo {pages} {112107} (\bibinfo {year}
  {2016})}\BibitemShut {NoStop}%
\bibitem [{\citenamefont {Kraus}\ \emph {et~al.}(2017)\citenamefont {Kraus},
  \citenamefont {Kormann}, \citenamefont {Morrison},\ and\ \citenamefont
  {Sonnendrücker}}]{Kraus_2017}%
  \BibitemOpen
  \bibfield  {author} {\bibinfo {author} {\bibfnamefont {M.}~\bibnamefont
  {Kraus}}, \bibinfo {author} {\bibfnamefont {K.}~\bibnamefont {Kormann}},
  \bibinfo {author} {\bibfnamefont {P.~J.}\ \bibnamefont {Morrison}},\ and\
  \bibinfo {author} {\bibfnamefont {E.}~\bibnamefont {Sonnendrücker}},\
  }\bibfield  {title} {\bibinfo {title} {Gempic: geometric electromagnetic
  particle-in-cell methods},\ }\href@noop {} {\bibfield  {journal} {\bibinfo
  {journal} {J. Plasma Phys.}\ }\textbf {\bibinfo {volume} {83}},\ \bibinfo
  {pages} {905830401} (\bibinfo {year} {2017})}\BibitemShut {NoStop}%
\bibitem [{\citenamefont {Szabo}\ and\ \citenamefont
  {Ostlund}(1996)}]{Szabo_1996}%
  \BibitemOpen
  \bibfield  {author} {\bibinfo {author} {\bibfnamefont {A.}~\bibnamefont
  {Szabo}}\ and\ \bibinfo {author} {\bibfnamefont {N.~S.}\ \bibnamefont
  {Ostlund}},\ }\href@noop {} {\emph {\bibinfo {title} {Modern quantum
  chemistry: introduction to advanced electronic structure theory}}}\ (\bibinfo
   {publisher} {Courier Corporation},\ \bibinfo {address} {New York},\ \bibinfo
  {year} {1996})\BibitemShut {NoStop}%
\bibitem [{\citenamefont {Ishikawa}\ and\ \citenamefont
  {Sato}(2015)}]{Ishikawa_2015}%
  \BibitemOpen
  \bibfield  {author} {\bibinfo {author} {\bibfnamefont {K.~L.}\ \bibnamefont
  {Ishikawa}}\ and\ \bibinfo {author} {\bibfnamefont {T.}~\bibnamefont
  {Sato}},\ }\bibfield  {title} {\bibinfo {title} {A review on ab initio
  approaches for multielectron dynamics},\ }\href@noop {} {\bibfield  {journal}
  {\bibinfo  {journal} {IEEE J. Sel. Top. Quantum Electron.}\ }\textbf
  {\bibinfo {volume} {21}},\ \bibinfo {pages} {1} (\bibinfo {year}
  {2015})}\BibitemShut {NoStop}%
\bibitem [{\citenamefont {van Leeuwen}(1999)}]{van_Leeuwen_1999}%
  \BibitemOpen
  \bibfield  {author} {\bibinfo {author} {\bibfnamefont {R.}~\bibnamefont {van
  Leeuwen}},\ }\bibfield  {title} {\bibinfo {title} {Mapping from densities to
  potentials in time-dependent density-functional theory},\ }\href@noop {}
  {\bibfield  {journal} {\bibinfo  {journal} {Phys. Rev. Lett.}\ }\textbf
  {\bibinfo {volume} {82}},\ \bibinfo {pages} {3863} (\bibinfo {year}
  {1999})}\BibitemShut {NoStop}%
\bibitem [{\citenamefont {Marques}\ and\ \citenamefont
  {Gross}(2004)}]{Marques_2004}%
  \BibitemOpen
  \bibfield  {author} {\bibinfo {author} {\bibfnamefont {M.}~\bibnamefont
  {Marques}}\ and\ \bibinfo {author} {\bibfnamefont {E.}~\bibnamefont
  {Gross}},\ }\bibfield  {title} {\bibinfo {title} {Time-dependent density
  functional theory},\ }\href@noop {} {\bibfield  {journal} {\bibinfo
  {journal} {Annu. Rev. Phys. Chem.}\ }\textbf {\bibinfo {volume} {55}},\
  \bibinfo {pages} {427} (\bibinfo {year} {2004})}\BibitemShut {NoStop}%
\bibitem [{\citenamefont {Ullrich}\ and\ \citenamefont
  {Yang}(2014)}]{Ullrich_2014}%
  \BibitemOpen
  \bibfield  {author} {\bibinfo {author} {\bibfnamefont {C.~A.}\ \bibnamefont
  {Ullrich}}\ and\ \bibinfo {author} {\bibfnamefont {Z.-h.}\ \bibnamefont
  {Yang}},\ }\bibfield  {title} {\bibinfo {title} {A brief compendium of
  time-dependent density functional theory},\ }\href@noop {} {\bibfield
  {journal} {\bibinfo  {journal} {Braz. J. Phys.}\ }\textbf {\bibinfo {volume}
  {44}},\ \bibinfo {pages} {154} (\bibinfo {year} {2014})}\BibitemShut
  {NoStop}%
\bibitem [{\citenamefont {Vignale}\ \emph {et~al.}(1997)\citenamefont
  {Vignale}, \citenamefont {Ullrich},\ and\ \citenamefont
  {Conti}}]{Vignale_1997}%
  \BibitemOpen
  \bibfield  {author} {\bibinfo {author} {\bibfnamefont {G.}~\bibnamefont
  {Vignale}}, \bibinfo {author} {\bibfnamefont {C.~A.}\ \bibnamefont
  {Ullrich}},\ and\ \bibinfo {author} {\bibfnamefont {S.}~\bibnamefont
  {Conti}},\ }\bibfield  {title} {\bibinfo {title} {Time-dependent density
  functional theory beyond the adiabatic local density approximation},\
  }\href@noop {} {\bibfield  {journal} {\bibinfo  {journal} {Phys. Rev. Lett.}\
  }\textbf {\bibinfo {volume} {79}},\ \bibinfo {pages} {4878} (\bibinfo {year}
  {1997})}\BibitemShut {NoStop}%
\bibitem [{\citenamefont {Maitra}\ \emph {et~al.}(2002)\citenamefont {Maitra},
  \citenamefont {Burke},\ and\ \citenamefont {Woodward}}]{Maitra_2002}%
  \BibitemOpen
  \bibfield  {author} {\bibinfo {author} {\bibfnamefont {N.~T.}\ \bibnamefont
  {Maitra}}, \bibinfo {author} {\bibfnamefont {K.}~\bibnamefont {Burke}},\ and\
  \bibinfo {author} {\bibfnamefont {C.}~\bibnamefont {Woodward}},\ }\bibfield
  {title} {\bibinfo {title} {Memory in time-dependent density functional
  theory},\ }\href@noop {} {\bibfield  {journal} {\bibinfo  {journal} {Phys.
  Rev. Lett.}\ }\textbf {\bibinfo {volume} {89}},\ \bibinfo {pages} {023002}
  (\bibinfo {year} {2002})}\BibitemShut {NoStop}%
\bibitem [{\citenamefont {Wijewardane}\ and\ \citenamefont
  {Ullrich}(2005)}]{Wijewardane_2005}%
  \BibitemOpen
  \bibfield  {author} {\bibinfo {author} {\bibfnamefont {H.~O.}\ \bibnamefont
  {Wijewardane}}\ and\ \bibinfo {author} {\bibfnamefont {C.~A.}\ \bibnamefont
  {Ullrich}},\ }\bibfield  {title} {\bibinfo {title} {Time-dependent
  {K}ohn-{S}ham theory with memory},\ }\href@noop {} {\bibfield  {journal}
  {\bibinfo  {journal} {Phys. Rev. Lett.}\ }\textbf {\bibinfo {volume} {95}},\
  \bibinfo {pages} {086401} (\bibinfo {year} {2005})}\BibitemShut {NoStop}%
\bibitem [{\citenamefont {Lacombe}\ and\ \citenamefont
  {Maitra}(2020)}]{Lacombe_2020}%
  \BibitemOpen
  \bibfield  {author} {\bibinfo {author} {\bibfnamefont {L.}~\bibnamefont
  {Lacombe}}\ and\ \bibinfo {author} {\bibfnamefont {N.~T.}\ \bibnamefont
  {Maitra}},\ }\bibfield  {title} {\bibinfo {title} {Developing new and
  understanding old approximations in {TDDFT}},\ }\href@noop {} {\bibfield
  {journal} {\bibinfo  {journal} {Faraday Discuss.}\ }\textbf {\bibinfo
  {volume} {224}},\ \bibinfo {pages} {382} (\bibinfo {year}
  {2020})}\BibitemShut {NoStop}%
\bibitem [{\citenamefont {Bandrauk}\ \emph {et~al.}(2013)\citenamefont
  {Bandrauk}, \citenamefont {Fillion-Gourdeau},\ and\ \citenamefont
  {Lorin}}]{Bandrauk_2013}%
  \BibitemOpen
  \bibfield  {author} {\bibinfo {author} {\bibfnamefont {A.}~\bibnamefont
  {Bandrauk}}, \bibinfo {author} {\bibfnamefont {F.}~\bibnamefont
  {Fillion-Gourdeau}},\ and\ \bibinfo {author} {\bibfnamefont {E.}~\bibnamefont
  {Lorin}},\ }\bibfield  {title} {\bibinfo {title} {Atoms and molecules in
  intense laser fields: gauge invariance of theory and models},\ }\href@noop {}
  {\bibfield  {journal} {\bibinfo  {journal} {J. Phys. B: At. Mol. Opt. Phys.}\
  }\textbf {\bibinfo {volume} {46}},\ \bibinfo {pages} {153001} (\bibinfo
  {year} {2013})}\BibitemShut {NoStop}%
\bibitem [{\citenamefont {Seidl}\ \emph {et~al.}(1996)\citenamefont {Seidl},
  \citenamefont {G\"orling}, \citenamefont {Vogl}, \citenamefont {Majewski},\
  and\ \citenamefont {Levy}}]{Seidl_1996}%
  \BibitemOpen
  \bibfield  {author} {\bibinfo {author} {\bibfnamefont {A.}~\bibnamefont
  {Seidl}}, \bibinfo {author} {\bibfnamefont {A.}~\bibnamefont {G\"orling}},
  \bibinfo {author} {\bibfnamefont {P.}~\bibnamefont {Vogl}}, \bibinfo {author}
  {\bibfnamefont {J.~A.}\ \bibnamefont {Majewski}},\ and\ \bibinfo {author}
  {\bibfnamefont {M.}~\bibnamefont {Levy}},\ }\bibfield  {title} {\bibinfo
  {title} {Generalized {K}ohn-{S}ham schemes and the band-gap problem},\
  }\href@noop {} {\bibfield  {journal} {\bibinfo  {journal} {Phys. Rev. B}\
  }\textbf {\bibinfo {volume} {53}},\ \bibinfo {pages} {3764} (\bibinfo {year}
  {1996})}\BibitemShut {NoStop}%
\bibitem [{\citenamefont {Ullrich}\ \emph {et~al.}(1995)\citenamefont
  {Ullrich}, \citenamefont {Gossmann},\ and\ \citenamefont
  {Gross}}]{Ullrich_1995}%
  \BibitemOpen
  \bibfield  {author} {\bibinfo {author} {\bibfnamefont {C.~A.}\ \bibnamefont
  {Ullrich}}, \bibinfo {author} {\bibfnamefont {U.~J.}\ \bibnamefont
  {Gossmann}},\ and\ \bibinfo {author} {\bibfnamefont {E.~K.~U.}\ \bibnamefont
  {Gross}},\ }\bibfield  {title} {\bibinfo {title} {Time-dependent optimized
  effective potential},\ }\href@noop {} {\bibfield  {journal} {\bibinfo
  {journal} {Phys. Rev. Lett.}\ }\textbf {\bibinfo {volume} {74}},\ \bibinfo
  {pages} {872} (\bibinfo {year} {1995})}\BibitemShut {NoStop}%
\bibitem [{\citenamefont {Wijewardane}\ and\ \citenamefont
  {Ullrich}(2008)}]{Wijewardane_2008}%
  \BibitemOpen
  \bibfield  {author} {\bibinfo {author} {\bibfnamefont {H.~O.}\ \bibnamefont
  {Wijewardane}}\ and\ \bibinfo {author} {\bibfnamefont {C.~A.}\ \bibnamefont
  {Ullrich}},\ }\bibfield  {title} {\bibinfo {title} {Real-time electron
  dynamics with exact-exchange time-dependent density-functional theory},\
  }\href@noop {} {\bibfield  {journal} {\bibinfo  {journal} {Phys. Rev. Lett.}\
  }\textbf {\bibinfo {volume} {100}},\ \bibinfo {pages} {056404} (\bibinfo
  {year} {2008})}\BibitemShut {NoStop}%
\bibitem [{\citenamefont {Morrison}(1998)}]{Morrison_1998}%
  \BibitemOpen
  \bibfield  {author} {\bibinfo {author} {\bibfnamefont {P.~J.}\ \bibnamefont
  {Morrison}},\ }\bibfield  {title} {\bibinfo {title} {Hamiltonian description
  of the ideal fluid},\ }\href@noop {} {\bibfield  {journal} {\bibinfo
  {journal} {Rev. Mod. Phys.}\ }\textbf {\bibinfo {volume} {70}},\ \bibinfo
  {pages} {467} (\bibinfo {year} {1998})}\BibitemShut {NoStop}%
\bibitem [{\citenamefont {Morrison}(2005)}]{Morrison_2005}%
  \BibitemOpen
  \bibfield  {author} {\bibinfo {author} {\bibfnamefont {P.~J.}\ \bibnamefont
  {Morrison}},\ }\bibfield  {title} {\bibinfo {title} {Hamiltonian and action
  principle formulations of plasma physics},\ }\href@noop {} {\bibfield
  {journal} {\bibinfo  {journal} {Phys. Plasmas}\ }\textbf {\bibinfo {volume}
  {12}},\ \bibinfo {pages} {058102} (\bibinfo {year} {2005})}\BibitemShut
  {NoStop}%
\bibitem [{\citenamefont {Morrison}(2006)}]{Morrison_2006}%
  \BibitemOpen
  \bibfield  {author} {\bibinfo {author} {\bibfnamefont {P.}~\bibnamefont
  {Morrison}},\ }\bibfield  {title} {\bibinfo {title} {Hamiltonian fluid
  dynamics},\ }in\ \href
  {https://doi.org/https://doi.org/10.1016/B0-12-512666-2/00246-7} {\emph
  {\bibinfo {booktitle} {Encyclopedia of Mathematical Physics}}},\ \bibinfo
  {editor} {edited by\ \bibinfo {editor} {\bibfnamefont {J.-P.}\ \bibnamefont
  {Fran\c{c}oise}}, \bibinfo {editor} {\bibfnamefont {G.~L.}\ \bibnamefont
  {Naber}},\ and\ \bibinfo {editor} {\bibfnamefont {T.~S.}\ \bibnamefont
  {Tsun}}}\ (\bibinfo  {publisher} {Academic Press},\ \bibinfo {address}
  {Oxford},\ \bibinfo {year} {2006})\ p.\ \bibinfo {pages} {593}\BibitemShut
  {NoStop}%
\bibitem [{\citenamefont {Perdew}\ \emph {et~al.}(1996)\citenamefont {Perdew},
  \citenamefont {Burke},\ and\ \citenamefont {Ernzerhof}}]{Perdew_1996}%
  \BibitemOpen
  \bibfield  {author} {\bibinfo {author} {\bibfnamefont {J.~P.}\ \bibnamefont
  {Perdew}}, \bibinfo {author} {\bibfnamefont {K.}~\bibnamefont {Burke}},\ and\
  \bibinfo {author} {\bibfnamefont {M.}~\bibnamefont {Ernzerhof}},\ }\bibfield
  {title} {\bibinfo {title} {Generalized gradient approximation made simple},\
  }\href@noop {} {\bibfield  {journal} {\bibinfo  {journal} {Phys. Rev. Lett.}\
  }\textbf {\bibinfo {volume} {77}},\ \bibinfo {pages} {3865} (\bibinfo {year}
  {1996})}\BibitemShut {NoStop}%
\bibitem [{\citenamefont {Perdew}\ and\ \citenamefont
  {Schmidt}(2001)}]{Perdew_2001}%
  \BibitemOpen
  \bibfield  {author} {\bibinfo {author} {\bibfnamefont {J.~P.}\ \bibnamefont
  {Perdew}}\ and\ \bibinfo {author} {\bibfnamefont {K.}~\bibnamefont
  {Schmidt}},\ }\bibfield  {title} {\bibinfo {title} {Jacob’s ladder of
  density functional approximations for the exchange-correlation energy},\
  }\href@noop {} {\bibfield  {journal} {\bibinfo  {journal} {AIP Conf. Proc.}\
  }\textbf {\bibinfo {volume} {577}},\ \bibinfo {pages} {1} (\bibinfo {year}
  {2001})}\BibitemShut {NoStop}%
\bibitem [{\citenamefont {Marques}\ \emph
  {et~al.}(2012{\natexlab{b}})\citenamefont {Marques}, \citenamefont
  {Oliveira},\ and\ \citenamefont {Burnus}}]{Marques_2012}%
  \BibitemOpen
  \bibfield  {author} {\bibinfo {author} {\bibfnamefont {M.~A.}\ \bibnamefont
  {Marques}}, \bibinfo {author} {\bibfnamefont {M.~J.}\ \bibnamefont
  {Oliveira}},\ and\ \bibinfo {author} {\bibfnamefont {T.}~\bibnamefont
  {Burnus}},\ }\bibfield  {title} {\bibinfo {title} {Libxc: A library of
  exchange and correlation functionals for density functional theory},\
  }\href@noop {} {\bibfield  {journal} {\bibinfo  {journal} {Comput. Phys.
  Commun.}\ }\textbf {\bibinfo {volume} {183}},\ \bibinfo {pages} {2272}
  (\bibinfo {year} {2012}{\natexlab{b}})}\BibitemShut {NoStop}%
\bibitem [{Note1()}]{Note1}%
  \BibitemOpen
  \bibinfo {note} {The kinetic-energy density is often denoted with the letter
  $\tau $. Here we use $\kappa $ to distinguish it from the evolution parameter
  used in the propagation schemes discussed in section~\ref
  {sec:Symplectic_schemes}.}\BibitemShut {Stop}%
\bibitem [{\citenamefont {Lehtola}\ \emph {et~al.}(2018)\citenamefont
  {Lehtola}, \citenamefont {Steigemann}, \citenamefont {Oliveira},\ and\
  \citenamefont {Marques}}]{Lehtola_2018}%
  \BibitemOpen
  \bibfield  {author} {\bibinfo {author} {\bibfnamefont {S.}~\bibnamefont
  {Lehtola}}, \bibinfo {author} {\bibfnamefont {C.}~\bibnamefont {Steigemann}},
  \bibinfo {author} {\bibfnamefont {M.~J.}\ \bibnamefont {Oliveira}},\ and\
  \bibinfo {author} {\bibfnamefont {M.~A.}\ \bibnamefont {Marques}},\
  }\bibfield  {title} {\bibinfo {title} {Recent developments in libxc -- a
  comprehensive library of functionals for density functional theory},\
  }\href@noop {} {\bibfield  {journal} {\bibinfo  {journal} {Software X}\
  }\textbf {\bibinfo {volume} {7}},\ \bibinfo {pages} {1} (\bibinfo {year}
  {2018})}\BibitemShut {NoStop}%
\bibitem [{\citenamefont {Gaarde}\ \emph {et~al.}(2011)\citenamefont {Gaarde},
  \citenamefont {Buth}, \citenamefont {Tate},\ and\ \citenamefont
  {Schafer}}]{Gaarde_2011}%
  \BibitemOpen
  \bibfield  {author} {\bibinfo {author} {\bibfnamefont {M.~B.}\ \bibnamefont
  {Gaarde}}, \bibinfo {author} {\bibfnamefont {C.}~\bibnamefont {Buth}},
  \bibinfo {author} {\bibfnamefont {J.~L.}\ \bibnamefont {Tate}},\ and\
  \bibinfo {author} {\bibfnamefont {K.~J.}\ \bibnamefont {Schafer}},\
  }\bibfield  {title} {\bibinfo {title} {Transient absorption and reshaping of
  ultrafast xuv light by laser-dressed helium},\ }\href@noop {} {\bibfield
  {journal} {\bibinfo  {journal} {Phys. Rev. A}\ }\textbf {\bibinfo {volume}
  {83}},\ \bibinfo {pages} {013419} (\bibinfo {year} {2011})}\BibitemShut
  {NoStop}%
\bibitem [{\citenamefont {Legrand}\ \emph {et~al.}(2002)\citenamefont
  {Legrand}, \citenamefont {Suraud},\ and\ \citenamefont
  {Reinhard}}]{Legrand_2002}%
  \BibitemOpen
  \bibfield  {author} {\bibinfo {author} {\bibfnamefont {C.}~\bibnamefont
  {Legrand}}, \bibinfo {author} {\bibfnamefont {E.}~\bibnamefont {Suraud}},\
  and\ \bibinfo {author} {\bibfnamefont {P.-G.}\ \bibnamefont {Reinhard}},\
  }\bibfield  {title} {\bibinfo {title} {Comparison of
  self-interaction-corrections for metal clusters},\ }\href@noop {} {\bibfield
  {journal} {\bibinfo  {journal} {J.~Phys.~B: At.~Mol.~Opt.~Phys.}\ }\textbf
  {\bibinfo {volume} {35}},\ \bibinfo {pages} {1115} (\bibinfo {year}
  {2002})}\BibitemShut {NoStop}%
\bibitem [{\citenamefont {Bruner}\ \emph {et~al.}(2017)\citenamefont {Bruner},
  \citenamefont {Hernandez}, \citenamefont {Mauger}, \citenamefont {Abanador},
  \citenamefont {LaMaster}, \citenamefont {Gaarde}, \citenamefont {Schafer},\
  and\ \citenamefont {Lopata}}]{Bruner_2017}%
  \BibitemOpen
  \bibfield  {author} {\bibinfo {author} {\bibfnamefont {A.}~\bibnamefont
  {Bruner}}, \bibinfo {author} {\bibfnamefont {S.}~\bibnamefont {Hernandez}},
  \bibinfo {author} {\bibfnamefont {F.}~\bibnamefont {Mauger}}, \bibinfo
  {author} {\bibfnamefont {P.~M.}\ \bibnamefont {Abanador}}, \bibinfo {author}
  {\bibfnamefont {D.~J.}\ \bibnamefont {LaMaster}}, \bibinfo {author}
  {\bibfnamefont {M.~B.}\ \bibnamefont {Gaarde}}, \bibinfo {author}
  {\bibfnamefont {K.~J.}\ \bibnamefont {Schafer}},\ and\ \bibinfo {author}
  {\bibfnamefont {K.}~\bibnamefont {Lopata}},\ }\bibfield  {title} {\bibinfo
  {title} {Attosecond charge migration with {TDDFT}: Accurate dynamics from a
  well-defined initial state},\ }\href@noop {} {\bibfield  {journal} {\bibinfo
  {journal} {J. Phys. Chem. Lett.}\ }\textbf {\bibinfo {volume} {8}},\ \bibinfo
  {pages} {3991} (\bibinfo {year} {2017})}\BibitemShut {NoStop}%
\bibitem [{\citenamefont {Tuthill}\ \emph {et~al.}(2020)\citenamefont
  {Tuthill}, \citenamefont {Mauger}, \citenamefont {Scarborough}, \citenamefont
  {Jones}, \citenamefont {Gaarde}, \citenamefont {Lopata}, \citenamefont
  {Schafer},\ and\ \citenamefont {DiMauro}}]{Tuthill_2020}%
  \BibitemOpen
  \bibfield  {author} {\bibinfo {author} {\bibfnamefont {D.~R.}\ \bibnamefont
  {Tuthill}}, \bibinfo {author} {\bibfnamefont {F.}~\bibnamefont {Mauger}},
  \bibinfo {author} {\bibfnamefont {T.~D.}\ \bibnamefont {Scarborough}},
  \bibinfo {author} {\bibfnamefont {R.~R.}\ \bibnamefont {Jones}}, \bibinfo
  {author} {\bibfnamefont {M.~B.}\ \bibnamefont {Gaarde}}, \bibinfo {author}
  {\bibfnamefont {K.}~\bibnamefont {Lopata}}, \bibinfo {author} {\bibfnamefont
  {K.~J.}\ \bibnamefont {Schafer}},\ and\ \bibinfo {author} {\bibfnamefont
  {L.~F.}\ \bibnamefont {DiMauro}},\ }\bibfield  {title} {\bibinfo {title}
  {Multidimensional molecular high-harmonic spectroscopy: A road map for charge
  migration studies},\ }\href@noop {} {\bibfield  {journal} {\bibinfo
  {journal} {J. Mol. Spectrosc.}\ }\textbf {\bibinfo {volume} {372}},\ \bibinfo
  {pages} {111353} (\bibinfo {year} {2020})}\BibitemShut {NoStop}%
\bibitem [{\citenamefont {Folorunso}\ \emph {et~al.}(2021)\citenamefont
  {Folorunso}, \citenamefont {Bruner}, \citenamefont {Mauger}, \citenamefont
  {Hamer}, \citenamefont {Hernandez}, \citenamefont {Jones}, \citenamefont
  {DiMauro}, \citenamefont {Gaarde}, \citenamefont {Schafer},\ and\
  \citenamefont {Lopata}}]{Folorunso_2021}%
  \BibitemOpen
  \bibfield  {author} {\bibinfo {author} {\bibfnamefont {A.~S.}\ \bibnamefont
  {Folorunso}}, \bibinfo {author} {\bibfnamefont {A.}~\bibnamefont {Bruner}},
  \bibinfo {author} {\bibfnamefont {F.}~\bibnamefont {Mauger}}, \bibinfo
  {author} {\bibfnamefont {K.~A.}\ \bibnamefont {Hamer}}, \bibinfo {author}
  {\bibfnamefont {S.}~\bibnamefont {Hernandez}}, \bibinfo {author}
  {\bibfnamefont {R.~R.}\ \bibnamefont {Jones}}, \bibinfo {author}
  {\bibfnamefont {L.~F.}\ \bibnamefont {DiMauro}}, \bibinfo {author}
  {\bibfnamefont {M.~B.}\ \bibnamefont {Gaarde}}, \bibinfo {author}
  {\bibfnamefont {K.~J.}\ \bibnamefont {Schafer}},\ and\ \bibinfo {author}
  {\bibfnamefont {K.}~\bibnamefont {Lopata}},\ }\bibfield  {title} {\bibinfo
  {title} {Molecular modes of attosecond charge migration},\ }\href@noop {}
  {\bibfield  {journal} {\bibinfo  {journal} {Phys. Rev. Lett.}\ }\textbf
  {\bibinfo {volume} {126}},\ \bibinfo {pages} {133002} (\bibinfo {year}
  {2021})}\BibitemShut {NoStop}%
\bibitem [{\citenamefont {Mauger}\ \emph {et~al.}(2022)\citenamefont {Mauger},
  \citenamefont {Folorunso}, \citenamefont {Hamer}, \citenamefont {Chandre},
  \citenamefont {Gaarde}, \citenamefont {Lopata},\ and\ \citenamefont
  {Schafer}}]{Mauger_2022}%
  \BibitemOpen
  \bibfield  {author} {\bibinfo {author} {\bibfnamefont {F.}~\bibnamefont
  {Mauger}}, \bibinfo {author} {\bibfnamefont {A.~S.}\ \bibnamefont
  {Folorunso}}, \bibinfo {author} {\bibfnamefont {K.~A.}\ \bibnamefont
  {Hamer}}, \bibinfo {author} {\bibfnamefont {C.}~\bibnamefont {Chandre}},
  \bibinfo {author} {\bibfnamefont {M.~B.}\ \bibnamefont {Gaarde}}, \bibinfo
  {author} {\bibfnamefont {K.}~\bibnamefont {Lopata}},\ and\ \bibinfo {author}
  {\bibfnamefont {K.~J.}\ \bibnamefont {Schafer}},\ }\bibfield  {title}
  {\bibinfo {title} {Charge migration and attosecond solitons in conjugated
  organic molecules},\ }\href@noop {} {\bibfield  {journal} {\bibinfo
  {journal} {Phys. Rev. Research}\ }\textbf {\bibinfo {volume} {4}},\ \bibinfo
  {pages} {013073} (\bibinfo {year} {2022})}\BibitemShut {NoStop}%
\bibitem [{\citenamefont {Serre}(1992)}]{Serre_1992_book}%
  \BibitemOpen
  \bibfield  {author} {\bibinfo {author} {\bibfnamefont {J.~P.}\ \bibnamefont
  {Serre}},\ }\href@noop {} {\emph {\bibinfo {title} {Lie Algebras and Lie
  Groups}}},\ Lecture Notes in Mathematics\ (\bibinfo  {publisher} {Springer
  Berlin},\ \bibinfo {address} {Heidelberg},\ \bibinfo {year}
  {1992})\BibitemShut {NoStop}%
\bibitem [{\citenamefont {Strang}(1968)}]{Strang_1968}%
  \BibitemOpen
  \bibfield  {author} {\bibinfo {author} {\bibfnamefont {G.}~\bibnamefont
  {Strang}},\ }\bibfield  {title} {\bibinfo {title} {On the construction and
  comparison of difference schemes},\ }\href@noop {} {\bibfield  {journal}
  {\bibinfo  {journal} {SIAM J. Numer. Analysis}\ }\textbf {\bibinfo {volume}
  {5}},\ \bibinfo {pages} {506} (\bibinfo {year} {1968})}\BibitemShut {NoStop}%
\bibitem [{\citenamefont {Forest}\ and\ \citenamefont
  {Ruth}(1990)}]{Forest_1990}%
  \BibitemOpen
  \bibfield  {author} {\bibinfo {author} {\bibfnamefont {E.}~\bibnamefont
  {Forest}}\ and\ \bibinfo {author} {\bibfnamefont {R.~D.}\ \bibnamefont
  {Ruth}},\ }\bibfield  {title} {\bibinfo {title} {Fourth-order symplectic
  integration},\ }\href@noop {} {\bibfield  {journal} {\bibinfo  {journal}
  {Physica D}\ }\textbf {\bibinfo {volume} {43}},\ \bibinfo {pages} {105}
  (\bibinfo {year} {1990})}\BibitemShut {NoStop}%
\bibitem [{\citenamefont {Fass\`{o}}(2003)}]{Fasso_2003}%
  \BibitemOpen
  \bibfield  {author} {\bibinfo {author} {\bibfnamefont {F.}~\bibnamefont
  {Fass\`{o}}},\ }\bibfield  {title} {\bibinfo {title} {Comparison of splitting
  algorithms for the rigid body},\ }\href@noop {} {\bibfield  {journal}
  {\bibinfo  {journal} {J. Comput. Phys.}\ }\textbf {\bibinfo {volume} {189}},\
  \bibinfo {pages} {527} (\bibinfo {year} {2003})}\BibitemShut {NoStop}%
\bibitem [{\citenamefont {G\'{o}mez~Pueyo}\ \emph
  {et~al.}(2020{\natexlab{b}})\citenamefont {G\'{o}mez~Pueyo}, \citenamefont
  {Blanes},\ and\ \citenamefont {Castro}}]{Gomez_Pueyo_2020_2}%
  \BibitemOpen
  \bibfield  {author} {\bibinfo {author} {\bibfnamefont {A.}~\bibnamefont
  {G\'{o}mez~Pueyo}}, \bibinfo {author} {\bibfnamefont {S.}~\bibnamefont
  {Blanes}},\ and\ \bibinfo {author} {\bibfnamefont {A.}~\bibnamefont
  {Castro}},\ }\bibfield  {title} {\bibinfo {title} {Performance of fourth and
  sixth-order commutator-free {M}agnus expansion integrators for {E}hrenfest
  dynamics},\ }\href@noop {} {\bibfield  {journal} {\bibinfo  {journal} {Comp.
  and Math. Methods}\ }\textbf {\bibinfo {volume} {3}},\ \bibinfo {pages}
  {e1100} (\bibinfo {year} {2020}{\natexlab{b}})}\BibitemShut {NoStop}%
\bibitem [{\citenamefont {Castro}\ and\ \citenamefont
  {Gross}(2013)}]{Castro_2013}%
  \BibitemOpen
  \bibfield  {author} {\bibinfo {author} {\bibfnamefont {A.}~\bibnamefont
  {Castro}}\ and\ \bibinfo {author} {\bibfnamefont {E.}~\bibnamefont {Gross}},\
  }\bibfield  {title} {\bibinfo {title} {Optimal control theory for
  quantum-classical systems: Ehrenfest molecular dynamics based on
  time-dependent density-functional theory},\ }\href@noop {} {\bibfield
  {journal} {\bibinfo  {journal} {J. Phys. A: Math. Theor.}\ }\textbf {\bibinfo
  {volume} {47}},\ \bibinfo {pages} {025204} (\bibinfo {year}
  {2013})}\BibitemShut {NoStop}%
\bibitem [{\citenamefont {Javanainen}\ \emph {et~al.}(1988)\citenamefont
  {Javanainen}, \citenamefont {Eberly},\ and\ \citenamefont
  {Su}}]{Javanainen_1988}%
  \BibitemOpen
  \bibfield  {author} {\bibinfo {author} {\bibfnamefont {J.}~\bibnamefont
  {Javanainen}}, \bibinfo {author} {\bibfnamefont {J.~H.}\ \bibnamefont
  {Eberly}},\ and\ \bibinfo {author} {\bibfnamefont {Q.}~\bibnamefont {Su}},\
  }\bibfield  {title} {\bibinfo {title} {Numerical simulations of multiphoton
  ionization and above-threshold electron spectra},\ }\href@noop {} {\bibfield
  {journal} {\bibinfo  {journal} {Phys.~Rev.~A}\ }\textbf {\bibinfo {volume}
  {38}},\ \bibinfo {pages} {3430} (\bibinfo {year} {1988})}\BibitemShut
  {NoStop}%
\end{thebibliography}

%

\end{document}